\documentclass[useAMS,usenatbib]{mn2e}
\usepackage{graphicx}
\usepackage{pdflscape}  % Landscape pages
\title[Young L dwarf spectral sequence of IMPOs in UpSco]
{The optical$+$infrared L dwarf spectral sequence of young planetary-mass objects in the Upper Scorpius association
\thanks{Based on observations collected with the Gran Telescopio Canarias (programmes 
GTC4-14A, GTC25-16A, and GTC40-17A)
installed in the Spanish Observatorio del Roque de los Muchachos of the Instituto
de Astrof\'isica de Canarias, in the island of La Palma and the ESO VLT telescope
under programmes 095.C-0812(A) and 089-C.0102(ABC).}}

\author[N. Lodieu et al.]{
N.\ Lodieu$^{1,2}$\thanks{E-mail: nlodieu@iac.es},
M.\ R.\ Zapatero Osorio$^{3}$, V.\ J.\ S.\ B\'ejar$^{1,2}$, K.\ Pe\~na Ram\'irez$^{4}$
\\
$^{1}$Instituto de Astrof\'isica de Canarias (IAC), C/ V\'ia L\'actea s/n, E-38200 La Laguna, Tenerife, Spain \\
$^{2}$Departamento de Astrof\'isica, Universidad de La Laguna (ULL), E-38206 La Laguna, Tenerife, Spain \\
$^{3}$Centro de Astrobiolog\'ia (CSIC-INTA), Carretera de Ajalvir km 4, 28850 Torrej\'on de Ardoz, Madrid \\
$^{4}$Unidad de Astronom\'ia, Facultad de Cs\. B\'asicas, Universidad de Antofagasta, Av.\ U.\ de Antofagasta 02800, Antofagasta, Chile \\
}

\begin{document}

\date{Accepted \today. Received \today; in original form \today}

\pagerange{\pageref{firstpage}--\pageref{lastpage}} \pubyear{2017}

\maketitle

\label{firstpage}

%
%%%%%%%%%%%%%%%%%%%%%%%%%%%%%%%%%%%%%%%%%%
%%%%%%%%  Abstract  %%%%%%%%
%%%%%%%%%%%%%%%%%%%%%%%%%%%%%%%%%%%%%%%%%%
%
\begin{abstract}
We present the results of photometric and spectroscopic follow-ups of the lowest mass member candidates 
in the nearest OB association, Upper Scorpius ($\sim$5--10 Myr; 145$\pm$17 pc), with the Gran Telescopio de 
Canarias (GTC) and European Southern Observatory (ESO) Very Large Telescope (VLT). We confirm the 
membership of the large majority ($>$\,80\%) of the candidates selected originally photometrically 
and astrometrically based on their spectroscopic features, weak equivalent widths of gravity-sensitive
doublets, and radial velocities. Confirmed members follow a sequence over a wide magnitude range
($J$\,=\,17.0--19.3 mag) in several colour-magnitude diagrams with optical, near-, and mid-infrared 
photometry, and have near-infrared spectral types in the L1--L7 interval with likely masses below 
15 Jupiter masses. We find that optical spectral types tend to be earlier than near-infrared spectral 
types by a few subclasses for spectral types later than M9\@. 
We investigate the behaviour of spectral indices defined in the literature as a function of spectral
type and gravity by comparison with values reported in the literature for young and old  dwarfs.
{\bf{We also derive effective temperatures in the 1900--1600\,K from fits of synthetic model-atmosphere 
spectra to the observed photometry but we caution the procedure carries large uncertainties.}}
We determine bolometric corrections for young L dwarfs with ages of $\sim$5--10 Myr (Upper Sco 
association) and find them similar in the $J$-band but larger by 0.1-0.4 mag in the $K$-band 
{\bf{with respect to field L dwarfs}}.
Finally, we discovered two faint young L dwarfs, VISTA\,J1607$-$2146 (L4.5) and 
VISTA\,J1611$-$2215 (L5) that have H$\alpha$ emission and possible flux excesses at 4.5 $\mu$m, pointing towards the presence of accretion from a disk onto the central objects of mass below 
$\sim$15 M$_{\rm Jup}$ at the age of 5--10 Myr.
\end{abstract}

\begin{keywords}
Stars: low-mass stars and brown dwarfs --- techniques: photometric --- techniques: spectroscopic
--- surveys --- stars: luminosity function, mass function
\end{keywords}

%
%%%%%%%%%%%%%%%%%%%%%%%%%%%%%%%%%%%%%%%%%%
%%%%%%%%  Introduction  %%%%%%%%
%%%%%%%%%%%%%%%%%%%%%%%%%%%%%%%%%%%%%%%%%%
%
\section{Introduction}
\label{USco_GTC_XSH:intro}

Various independent groups explored the very low-mass end of the initial mass function
\citep[IMF;][]{salpeter55,miller79,scalo86,chabrier05a,kroupa13} in young clusters and
star-forming regions following L to T-type spectral sequences. 
\citet{barrado07b} conducted a deep ($K$\,$\sim$\,17 mag) 1-deg$^{2}$ survey in 
$\lambda$\,Orionis down to the low-mass and substellar regimes with subsequent optical 
and near-infrared spectroscopy \citep{bayo11a} and in-depth study of the physical properties 
of the confirmed members \citep{bayo12a}. \citet{scholz09a} reported a possible mass cut-off 
in NGC\,1333 in the range 0.02--0.012 M$_{\odot}$ from a deep photometric and spectroscopic 
survey of 0.25 deg$^{2}$ down to $J$\,=\,20.8 mag with spectroscopic follow-up
\citep{scholz12b}. The census of spectroscopic members in this region is now complete
down to $\sim$0.01 M$_{\odot}$ at an age of 3 Myr \citep{luhman16b}. In IC\,348, \citet{alves13a}
presented optical and near-infrared spectroscopy of late-M/early-L dwarf candidates while
\citet{burgess09} reported one high-probability T-type candidate.
The current census of spectroscopic members in IC\,348 
amounts for 478 objects down to $K$\,=\,16.2 mag and A$_{J}$\,$<$\,3 mag \citep{luhman16b}.
In $\sigma$ Orionis, \citet{penya12a} reported 23 new candidates extending the previous
mass functions \citep[e.g.][]{zapatero00,bejar01,caballero07d,bihain09,lodieu09e,bejar11} 
down to a completeness of 0.006 M$_{\odot}$.  The membership of T-type sources 
\citep{zapatero00,zapatero02b,burgasser04b,zapatero08a} towards the cluster was revised 
in \citet{penya11a} and \citet{penya15}.
Many independent surveys investigated spectroscopically the substellar population of the 
Orion Nebula Cluster \citep{lucas01a,weights09,hillenbrand13,ingraham14,suenaga14}.
In $\rho$\,Oph, \citet{geers11} selected member candidates from a deep ($J$\,=\,20.6 mag)
multi-band survey of 0.25 deg$^{2}$ with near-infrared spectroscopic follow-up down to
the deuterium-burning limit \citep{muzic12a}. In the same region, \citet{alves12} presented 
optical spectroscopy of brown dwarf candidates as faint $J$\,$\sim$\,20.5 mag \citep{alves10}, 
casting doubts about the membership of the T2-type member \citep{marsh10a} identified in a 
deep near-infrared survey \citep{marsh10b}. \citet*{chiang15b} presented two young
early T-type candidates as well as a L-type dwarf identified in their methane and mid-infrared
survey \citep{chiang15a}. In Serpens, \citet{spezzi12b} reported a few T-type candidates 
that require additional follow-up before deriving the IMF below the deuterium-burning limit.
In Lupus-3, \citet{muzic14} conducted an optical and near-infrared survey of 1.4 deg$^{2}$
down to planetary masses (0.009 M$_{\odot}$ or $J$\,=\,18.3 mag) with one probable new brown
dwarf \citep{muzic15}.
In more mature clusters, \citet{zapatero14b} confirmed seven early-L and L/T transition 
members in the Pleiades with masses in the 0.035--0.010 M$_{\odot}$ range from a deep $JH$ 
0.8-deg$^{2}$ survey \citep{zapatero14a}.
\citet{bouvier08a} confirmed astrometrically and spectroscopically two early-T dwarfs
with masses around 0.05 M$_{\odot}$ in a deep optical survey of 16 deg${^2}$ in the 
625 Myr-old Hyades cluster.

The spectroscopic characterisation of L dwarf members of young open clusters and
star-forming regions has revealed key features sensitive to gravity. After photometric
selection and astrometric confirmation (when possible depending on the mean proper motion
of the cluster/region), weak gravity-sensitive features such as alkali lines (e.g.\ the 
potassium and sodium doublets) at optical and infrared wavelengths add strong credit to 
the membership of candidates \citep{martin96,luhman98,mclean00a,martin01b,gorlova03}. 
Other criteria include weaker hydride (e.g.\ FeH and CrH) and stronger oxide (e.g.\ VO, TiO) bands 
\citep{kirkpatrick06} and the presence of the triangular or peaked-shape of the $H$-band 
spectra \citep{lucas01a}, which together with the aforementioned features, are well-accepted 
low surface gravity features
\citep{zapatero00,martin03b,lodieu08a,cruz09,scholz12c,allers13,bonnefoy14a,manjavacas14,muzic14,muzic15,martin17a}.

Upper Scorpius (hereafter UpSco) is part of the nearest OB association to the Sun, Scorpius Centaurus. 
The region is nearby with a distance of $\sim$145 pc from Hipparcos \citep{deBruijne97} and a recent update 
from Gaia \citep[144.2$\pm$17.6 pc;][]{fang17a}. UpSco is young with different age determinations and
a possible spread among its members \citep{preibisch99,preibisch01,pecaut12,song12,pecaut16a,rizzuto16}.
The members of UpSco exhibit a significant mean proper motion
\citep[$\mu_\alpha\cos\delta$\,=\,$-$10.5 and $\mu_\delta$\,=\,$-$23.2 mas/yr with a standard dispersion of about 6.4 mas/yr][]{deBruijne97,deZeeuw99,fang17a}.
The bright end of the UpSco population has been examined in X-rays
\citep{walter94,kunkel99,preibisch98}, astrometrically \citep{deBruijne97,deZeeuw99},
and spectroscopically \citep{preibisch02}.
The low-mass and substellar population has been investigated over the past
decade in great detail with the advent of modern detectors permitting wide
and/or deep surveys
\citep{ardila00,martin04,slesnick06,lodieu06,lodieu07a,kraus08a,bejar08,slesnick08,lafreniere10a,dawson11,lodieu11c,lafreniere11,dawson13,lodieu13c,lafreniere14,dawson14,penya16a,best17a}.
The first transiting systems in the association have been announced in the last years thanks to 
the Kepler K2 mission \citep{borucki10,lissauer14,batalha14}, including a triple system 
made of a F star and two solar-type stars \citep{alonso15a} and several M dwarf eclisping 
binaries \citep{kraus15a,lodieu15c,david16a}. These systems are of prime 
importance because they provide the first mass and radius measurements independent of models in 
this region. Finally, we should note the existence of a transiting Neptune-size planet candidate 
around a M3 member of UpSco \citep{mann16b,david16b}.

Our group has recently identified several tens of substellar and isolated planetary-mass
candidates from a deep infrared survey of 13.5 square degrees in the central region
of UpSco \citep[][hereafter L13]{lodieu13d}, which represents the basis of the work presented in
this paper. This photometric survey is among the widest and deepest in a young
star-forming region to investigate the shape of the IMF at very low masses.
Here, we present optical photometry in the Sloan $i$ filter obtained with the
OSIRIS instrument \citep[Optical System for Imaging and low Resolution Integrated
Spectroscopy;][]{cepa00} on the 10.4-m Gran Telescopio de Canarias (GTC).
We also present optical and near-infrared spectroscopy of photometrically-confirmed member 
candidates with GTC/OSIRIS and the European Southern Observatory (ESO) Very Large Telescope (VLT)
X-shooter spectrograph \citep{vernet11}.
In Section \ref{USco_GTC_XSH:GTC_phot} we describe the optical photometric observations
obtained with the GTC and its associated data reduction.
In Section \ref{USco_GTC_XSH:XSH_spec} we detail the spectroscopic follow-up conducted 
with ESO VLT for 15 member candidates in UpSco, including optical spectroscopic of a sub-sample
with GTC/OSIRIS\@.
In Section \ref{USco_GTC_XSH:WISE_midIR} we show mid-infrared photometry
from the AllWISE public catalogue for the coolest UpSco members.
In Section \ref{USco_GTC_XSH:properties} we present an analysis of the photometric
and spectroscopic properties of the L-type members in UpSco, discuss their membership 
and physical properties. We define the first sequence for young L dwarf members
of the UpSco association with an age of $\sim$5--10 Myr.
{\bf{In the literature, it has been shown that the low gravity sequence of L dwarfs 
does not logically flow from the high gravity sequence. In this paper, we are looking to create a uniform 
sample out of the UpSco objects, which have the same metallicity and age.}}

%
%%%%%%%%%%%%%%%%%%%%%%%%%%%%%%%%%%%%%%%%%%
%%%%% Table: Log of GTC observations %%%%%
%%%%%%%%%%%%%%%%%%%%%%%%%%%%%%%%%%%%%%%%%%
%
\begin{table*}
 \centering
 \caption[]{Logs of the GTC OSIRIS photometric observations. We list the names of the target according to
 the IAU nomenclature, coordinates in sexagesimal format (J2000), $J$ magnitudes from VISTA (L13; MKO system), 
the dates of observations,
the numbers of repeated on-source integration times, the seeing rounded to the nearest digit,
the mean airmass, the Sloan $i$-band magnitudes (AB system) measured on the GTC/OSIRIS images 
with their uncertainties (values quoted without errors are 3$\sigma$ limits), and the derived 
$i-J$ colours (or lower limits).
 }
{\normalsize
 \begin{tabular}{@{\hspace{0mm}}c c c c c c c c c c@{\hspace{0mm}}}
 \hline
 \hline
Name & R.A.    &     Dec       &  ExpT & Date & Seeing & Airmass & SDSS$i$ & $i-J$ \cr
 \hline
VISTA\,J  & hh:mm:ss.ss & ${^\circ}$:$'$:$''$ & sec & yyyy-mm-dd & arcsec &   & mag & mag   \cr
 \hline
16073161$-$2146544 & 16:07:31.61   & $-$21:46:54.4 & 11$\times$60 & 2014-08-27 & 0.7 & 1.87 & 23.42$\pm$0.21 & 5.38$\pm$0.27 \cr % USco\_ZJY\_001
16041304$-$2241034 & 16:04:13.04   & $-$22:41:03.4 & 11$\times$60 & 2014-08-28 & 0.8 & 1.87 & 22.16$\pm$0.22 & 3.99$\pm$0.23 \cr % USco\_ZJY\_002
16140756$-$2211522 & 16:14:07.56   & $-$22:11:52.2 & 11$\times$60 & 2014-08-23 & 1.3 & 1.92 & 22.65$\pm$0.28 & 4.39$\pm$0.28 \cr % USco\_ZJY\_003
16084343$-$2245162 & 16:08:43.43   & $-$22:45:16.2 & 11$\times$60 & 2014-08-23 & 1.3 & 1.86 & 23.08$\pm$0.21 & 3.48$\pm$0.64 \cr % USco\_ZJY\_004
16053909$-$2403328 & 16:05:39.09   & $-$24:03:32.8 & 11$\times$60 & 2014-08-23 & 1.3 & 1.84 & 23.18$\pm$0.26 & 4.86$\pm$0.26 \cr % USco\_ZJY\_005
16042042$-$2134530 & 16:04:20.42   & $-$21:34:53.0 & 11$\times$60 & 2014-08-31 & 1.2 & 1.89 & 22.32$\pm$0.22 & 3.91$\pm$0.23 \cr % USco\_ZJY\_006 && USco\_cand75
16020000$-$2057341 & 16:02:00.00   & $-$20:57:34.1 & 11$\times$60 & 2014-08-21 & 0.9 & 1.80 & 22.77$\pm$0.21 & 4.27$\pm$0.21 \cr % USco\_ZJY\_007
16041234$-$2127470 & 16:04:12.34   & $-$21:27:47.0 & 11$\times$60 & 2014-08-24 & 1.3 & 1.79 & 22.69$\pm$0.21 & 4.17$\pm$0.22 \cr % USco\_ZJY\_008
16082915$-$2116296 & 16:08:29.15   & $-$21:16:29.6 & 11$\times$60 & 2014-08-21 & 0.9 & 1.70 & 23.24$\pm$0.22 & 4.55$\pm$0.22 \cr % USco\_ZJY\_009
16013692$-$2212027 & 16:01:36.92   & $-$22:12:02.7 & 30$\times$60 & 2014-08-01 & 1.2 & 1.80 & 23.67$\pm$0.23 & 4.40$\pm$0.24 \cr % USco\_ZJY\_010_exp1 && USco\_cand96
16151270$-$2229492 & 16:15:12.70   & $-$22:29:49.2 & 30$\times$60 & 2014-08-17 & 1.5 & 1.79 & 23.50$\pm$0.22 & 4.16$\pm$0.23 \cr % USco\_ZJY\_011_exp1 && USco\_cand97
16114892$-$2105286 & 16:11:48.92      & $-$21:05:28.6 & 30$\times$60 & 2014-08-18 & 1.6 & 1.69 & $>$24.05          & $>$4.38 \cr % USco\_IPMO1
16162104$-$2355201 & 16:16:21.04      & $-$23:55:20.1 & 30$\times$60 & 2014-08-12 & 0.9 & 1.62 & $>$24.05          & $>$4.05 \cr % USco\_IPMO4
16125183$-$2316500 & 16:12:51.83      & $-$23:16:50.0 & 30$\times$60 & 2014-08-16 & 0.9 & 1.74 & 23.91$\pm$0.23 & 4.07$\pm$0.25 \cr % USco\_IPMO3
16130655$-$2255327 & 16:13:06.55      & $-$22:55:32.7 & 30$\times$60 & 2014-08-03 & 0.9 & 1.68 & 22.45$\pm$0.21 & 2.65$\pm$0.23 \cr % USco\_IPMO2
16102763$-$2305543 & 16:10:27.63      & $-$23:05:54.3 & 30$\times$60 & 2014-08-11 & 0.9 & 1.71 & 21.26$\pm$0.20 & 1.26$\pm$0.25 \cr % USco\_IPMO5
16091868$-$2229239 & 16:09:18.68  & $-$22:29:23.9 & 11$\times$60 & 2014-08-26 & 0.9 & 1.70 & 22.87$\pm$0.21 & 4.93$\pm$0.25 \cr % USco\_ZJY\_010
16114437$-$2215446 & 16:11:44.37   & $-$22:15:44.6 & 11$\times$60 & 2014-08-24 & 1.0 & 1.71 & 22.45$\pm$0.21 & 4.59$\pm$0.26 \cr % USco\_ZJY\_012
16051705$-$2130449 & 16:05:17.05   & $-$21:30:44.9 & 10$\times$60 & 2014-08-28 & 0.8 & 1.88 & 22.33$\pm$0.21 & 3.95$\pm$0.21 \cr % USco\_ZJY\_013
16142256$-$2331178 & 16:14:22.56   & $-$23:31:17.8 &  2$\times$60 & 2014-08-28 & 0.8 & 1.91 & 21.74$\pm$0.21 & 4.29$\pm$0.21 \cr % USco\_ZJY\_014
16130231$-$2124285 & 16:13:02.31   & $-$21:24:28.5 &  1$\times$65 & 2014-08-28 & 0.8 & 2.05 & 21.96$\pm$0.21 & 4.78$\pm$0.21 \cr % USco\_ZJY\_015 && USco\_cand55
15593638$-$2214159 & 15:59:36.38   & $-$22:14:15.9 &  1$\times$65 & 2014-08-31 & 1.7 & 2.00 & 21.80$\pm$0.21 & 4.83$\pm$0.21 \cr % USco\_ZJY\_016 && USco\_cand2
16072782$-$2239042 & 16:07:27.82   & $-$22:39:04.2 &  1$\times$70 & 2014-08-28 & 0.8 & 2.20 & 21.14$\pm$0.21 & 4.28$\pm$0.21 \cr % USco\_ZJY\_018
16161064$-$2151243 & 16:16:10.64   & $-$21:51:24.3 &  1$\times$65 & 2014-08-28 & 0.8 & 2.17 & 20.32$\pm$0.20 & 3.69$\pm$0.20 \cr % USco\_ZJY\_019
16071478$-$2321013 & 16:07:14.78   & $-$23:21:01.3 &  2$\times$60 & 2014-08-28 & 0.8 & 2.33 & 20.89$\pm$0.21 & 4.33$\pm$0.21 \cr % USco\_ZJY\_020
16130140$-$2142547 & 16:13:01.40   & $-$21:42:54.7 &  1$\times$60 & 2014-08-28 & 0.8 & 2.07 & 19.94$\pm$0.20 & 3.62$\pm$0.20 \cr % USco\_ZJY\_021
16060629$-$2335135 & 16:06:06.29   & $-$23:35:13.5 &  2$\times$60 & 2014-08-31 & 1.6 & 2.26 & under spike    &     ---       \cr % USco\_ZJY\_022
16064554$-$2121593 & 16:06:45.54   & $-$21:21:59.3 &  2$\times$60 & 2014-08-31 & 1.6 & 2.12 & 20.45$\pm$0.20 & 4.31$\pm$0.20 \cr % USco\_ZJY\_023
16081842$-$2232252 & 16:08:18.42   & $-$22:32:25.2 &  2$\times$60 & 2014-08-28 & 0.8 & 2.25 & 20.26$\pm$0.20 & 4.17$\pm$0.20 \cr % USco\_ZJY\_024
  \hline
 \label{tab_USco_GTC_XSH:log_obs_GTC}
 \end{tabular}
}
\end{table*}
%

%
%%%%%%%%%%%%%%%%%%%%%%%%%%%%%%%%%%%%%%%%%%
%%%%% Optical photometry %%%%%
%%%%%%%%%%%%%%%%%%%%%%%%%%%%%%%%%%%%%%%%%%
%
\section{Optical photometry with GTC}
\label{USco_GTC_XSH:GTC_phot}
\subsection{SDSS$i$ photometry}
\label{USco_GTC_XSH:GTC_phot_obs}

The OSIRIS instrument \citep{cepa00} is mounted on the 10.4-m GTC operating at the Observatory
del Roque de Los Muchachos (La Palma, Canary Islands).
The detector consists of two 2048$\times$4096 Marconi CCD42-82 with a 8 arcsec gap
between them and operates at optical wavelengths from 0.365 to 1.0\,$\mu$m. The unvignetted
instrument field-of-view is about 7$'$\,$\times$\,7$'$ with an unbinned pixel scale of 0.125 arcsec.
We employed a 2$\times$2 binning because it is currently the standard mode of observations.
We placed all targets in the CCD2 because it provides the largest unvignetted field of view.

The observations were conducted in service during the month of August 2014 over several
days (programme GTC4-14A; PI Lodieu). We list the dates of observations, exposure times, mean seeing 
at the time of observations, and airmass in Table \ref{tab_USco_GTC_XSH:log_obs_GTC}. 
The conditions at the time of the observations were grey and clear or photometric except on 28 August 2014\@.
However, the exposure times, seeing conditions, and airmass at the time of observations span wide ranges,
resulting in a globally inhomogeneous photometric dataset.
Bias and skyflats were obtained during the afternoon or morning of the respective nights.
A photometric standard star was also observed with the same instrumental configuration as the science
targets as part of the GTC calibration plan, except for the nights of 1, 11, 17, 27, and 28 August 2014\@. 
Four different standards were taken over the
full period of observations for our program (SA\,110-232; G\,158-100, PG\,1528, SA\,109-381). 

We imaged 31 member candidates in the UpSco association published by L13\@.
We used mainly individual on-source integrations of 60\,sec repeated many times to detect (or
attempt a detection) of all UpSco candidates. The 60\,sec integration is set by the brightness
of the sky in the Sloan $i$ filter, where exposures longer than 120\,sec usually saturate
completely the detector. Table \ref{tab_USco_GTC_XSH:log_obs_GTC} provides the coordinates
of the UpSco targets along with the measured $i$-band magnitudes and inferred $i-J$ colours. 
We plot the positions of the GTC OSIRIS pointings in Fig.\ \ref{fig_USco_GTC_XSH:radec_cover}.

\subsection{Data reduction and photometry}
\label{USco_GTC_XSH:GTC_phot_DR}

\begin{figure}
  \centering
  \includegraphics[width=\linewidth, angle=0]{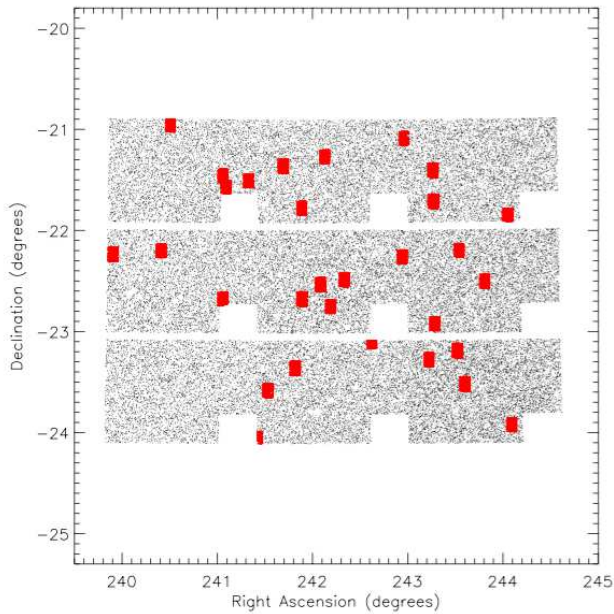}
  \caption{Location of the GTC OSIRIS CCD2 pointing (red regions) overlapping with the deep VISTA 
$ZYJ$ survey (L13; light grey; 1 every 20 sources plotted only). Some of the GTC fields do not 
fully overlap with the VISTA survey.
}
  \label{fig_USco_GTC_XSH:radec_cover}
\end{figure}

We reduced the OSIRIS optical images in a standard manner under the IRAF environment\footnote{IRAF 
is distributed by National Optical Astronomy Observatory, which is operated by the Association of 
Universities for Research in Astronomy, Inc., under contract with the National Science Foundation.}
\citep{tody86,tody93}. We considered only the CCD2 of the detector where the target
was located. First, we subtracted the median-combined bias and divided by the median-combined 
flat-field every single raw frame. Then, we averaged all frames to produce a final 
combined image for each target. All OSIRIS pointings contain one single UpSco target keeping in
mind that  the VISTA survey covered 13.5 deg$^{2}$ of UpSco (Fig.\ \ref{fig_USco_GTC_XSH:radec_cover}),
except in one case where two nearby sources were placed in the same OSIRIS field-of-view to save 
telescope time.

First, we calibrated astrometrically the combined images with a sequence of IRAF tasks: 
{\tt{ccxymatch}}, {\tt{ccmap}}, and {\tt{mscimage}}. We identified bright sources with 
their pixel coordinates using {\tt{daofind}} with a 10$\sigma$ detection threshold and used 
the 2MASS point-source catalogue \citep{cutri03,skrutskie06} as reference to convert pixels
into the world coordinate system. We included non-linear distortion terms to the tangent 
plane projection due to the slight rotation of the detector on the focal plane creating
distortions at the edges of the images. We found 
that the x axis is rotated by 180 degrees towards the East and the y axis by one-tenth of a degree.
The precision of final astrometry solution is of the order of or better than 0.15 arcsec in both axis. 

\begin{table*}
 \centering
 \caption[]{Cross-match between the catalogue of GTC OSIRIS $i$-band (SDSS AB system) 
with the deep $ZYJ$ (MKO system) VISTA catalogue. 
Columns 1 and 2 give the coordinates (J2000) on the GTC OSIRIS images in
sexagesimal format; columns 3 and 4 give the coordinates on the GTC OSIRIS images in pixels;
column 5 provides the $i$-band photometry and its uncertainty; columns 6 and 7 give the coordinates
on the VISTA images in sexagesimal format; columns 8--10 provides the VISTA $ZYJ$ photometry and
their error bars; columns 11--13 give the classification of the source according to the UKIDSS
scheme \citep[negative and positive values indicate point sources and galaxies, respectively;][]{lawrence07};
column 14 gives the separation between the positions on the GTC and VISTA images.
The full version of this table will be available in the electronic version of the paper.
 }
\scriptsize
 \begin{tabular}{@{\hspace{0mm}}c @{\hspace{1mm}}c @{\hspace{2mm}}c @{\hspace{1mm}}c @{\hspace{1mm}}c @{\hspace{1mm}}c @{\hspace{1mm}}c @{\hspace{1mm}}c @{\hspace{1mm}}c @{\hspace{1mm}}c @{\hspace{1mm}}c @{\hspace{2mm}}c @{\hspace{1mm}}c @{\hspace{1mm}}c@{\hspace{0mm}}}
 \hline
 \hline
R.A. (GTC) &  Dec (GTC)  & xc & yc & $i$ & R.A. (VISTA) & dec (VISTA) & $Z$ & $Y$ & $J$ & zClass & yClass & jClass & Sep \cr
 \hline
hh:mm:ss.ss & ${^\circ}$:$'$:$''$ & pix & pix & mag & hh:mm:ss.ss & ${^\circ}$:$'$:$''$ & mag & mag & mag & & & & arcsec \cr
 \hline
15:59:22.34 & -22:14:26.9 &  975.417 &  936.077 & 20.188$\pm$0.047 & 15:59:22.34 & -22:14:27.3 & 20.007$\pm$0.020 & 19.703$\pm$0.046 & 19.482$\pm$0.081 & -1 & -1 & -1 & 0.382 \cr
\ldots{} & \ldots{} & \ldots{} & \ldots{} & \ldots{} & \ldots{} & \ldots{} & \ldots{} & \ldots{} & \ldots{} & \ldots{} & \ldots{} & \ldots{} & \ldots{} \cr
16:16:24.47 & -23:51:50.9 &   21.189 & 1776.340 & 17.189$\pm$0.002 & 16:16:24.48 & -23:51:51.0 & 16.948$\pm$0.004 & 16.699$\pm$0.004 & 16.335$\pm$0.005 & -1 & -1 & -1 & 0.119 \cr
 \hline
 \label{tab_USco_GTC_XSH:Xmatch_GTC_VISTA}
 \end{tabular}
\end{table*}

We performed aperture and point-spread function (PSF) photometry with the task {\tt{phot}} in IRAF\@. 
The full catalogue of sources in all OSIRIS pointings contains 15,354 objects.
The depths of the combined images vary from pointing to pointing because of the observing strategy,
which was strongly dependent on the on-source integration times (see Table \ref{tab_USco_GTC_XSH:log_obs_GTC}) 
and the distinct airmass and seeing conditions at the time of observations. We adapted the
photometric aperture to the seeing conditions of each combined image, choosing
apertures of the order of 4 times the full-width-half-maximum.

\begin{figure}
  \centering
  \includegraphics[width=\linewidth, angle=0]{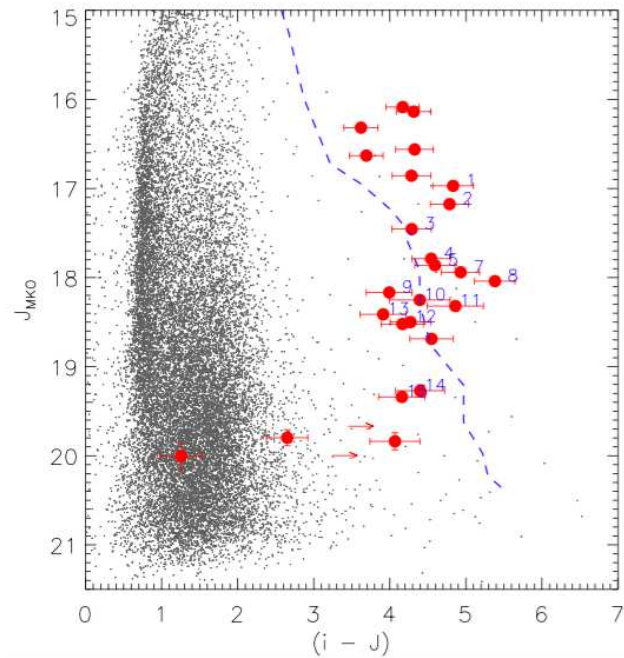}
  \caption{($i-J$,$J$) colour-magnitude diagram for all sources in the GTC OSIRIS CCD2 
field-of-view (grey dots) with infrared photometry from the deep $ZYJ$ VISTA survey
(L13). Overplotted as red dots are the $ZYJ$ candidates detected on the GTC OSIRIS\@.
We give lower limits for two objects (VISTA\,J16114892$-$2105286 and VISTA\,J16162104$-$2355201) 
The two objects with the bluest $i-J$ colours are VISTA\,J16130655$-$2255327 and 
VISTA\,J16102763$-$2305543,
marked as right-pointing red arrow. Blue numbers correspond to the IDs as in 
Table \ref{tab_USco_GTC_XSH:log_obs_XSH}. Overplotted as a blue dashed line is the sequence 
of M and L dwarfs (M5--L8) from SDSS shifted to the distance of UpSco.
}
  \label{fig_USco_GTC_XSH:CMD_iJJ}
\end{figure}

We divided our candidates into two groups to perform aperture photometry:
\begin{itemize}
\item[$\bullet$] For candidates clearly detected on the OSIRIS images, we directly
extracted the photometry from the full catalogue (red filled dots in Fig.\ \ref{fig_USco_GTC_XSH:CMD_iJJ}).
\item[$\bullet$] For the two remaining candidates below the 3$\sigma$ limit of detection
(VISTA\,J16114892$-$2105286 and VISTA\,J16162104$-$2355201), we estimated the limited magnitudes
as three times the root-mean-square at the expected location of the candidates as their flux peak 
and compared with the peak fluxes of three nearby isolated and well-detected stars with
measured photometry (red arrows in Fig.\ \ref{fig_USco_GTC_XSH:CMD_iJJ}).
\end{itemize}

To calibrate photometrically the OSIRIS images, we proceeded as follows: we obtained
PSF photometry for both the target data and the standard stars using a zero point of 25 mag 
as starting point. The derived photometric zero points from the standard stars agree
within $\pm$0.05 mag with those published by the observatory.
For the targets observed on the nights with no photometric standard star available,
we calibrate photometrically using the following procedure. First, we cross-matched the catalogue of
OSIRIS sources with the full VISTA$+$UKIDSS GCS catalogue. We produced a histogram of the 
$i-H$ colours for all the sources in the non-calibrated fields using the instrumental
magnitudes in the $i$-band and compared them with the combination of the same histograms
of the UpSco fields with already calibrated photometry using standard stars. We shifted
instrumental magnitudes in the $i$-band to match the distribution of both histograms and
convert them into apparent magnitudes. The uncertainty of this procedure 
is limited to the dispersion of the histograms and can be as high as 0.15--0.20 mag.
We added this calibration error in quadrature to the photometric uncertainties derived
from {\tt{phot}} in IRAF\@.

%
%%%%%%%%%%%%%%%%%%%%%%%%%%%%%%%%%%%%%%%%%%%%%%%
%%%%% Figure: (RA,dec) coverage %%%%%
%%%%%%%%%%%%%%%%%%%%%%%%%%%%%%%%%%%%%%%%%%%%%%%
%
%
%

%
%%%%%%%%%%%%%%%%%%%%%%%%%%%%%%%%%%%%%%%%%%%%%%%
%%%%% Figure: (i-J,J) CMD %%%%%
%%%%%%%%%%%%%%%%%%%%%%%%%%%%%%%%%%%%%%%%%%%%%%%
%

The final $i$-band magnitudes and their uncertainties (in the case of detections)
are listed in Table \ref{tab_USco_GTC_XSH:log_obs_GTC} along with the associated
$i-J$ colours. The error bars are dominated by the photometric calibration in most cases.
We display the ($i-J$,$J$) colour-magnitude diagram with all optical detections with
infrared counterparts in Fig.\ \ref{fig_USco_GTC_XSH:CMD_iJJ}.
We highlighted our candidates with larger symbols and 3$\sigma$ non-detections with right-pointing
arrows. We conclude that most of our photometric candidates remain as potential UpSco members
based on optical-to-near-infrared indices. Only two sources, VISTA\,J16130655$-$2255327 and 
J16102763$-$2305543, have blue $i-J$ colours that might indicate that they are field 
contaminants. This suggests that the minimum level of contamination in our original photometric
sample is of the order of $\sim$7\%.
For comparison, we overplotted the sequence of field M and L dwarfs from M5 to L8
from the Sloan team \citep{west08,schmidt10b} shifted to the distance of 145 pc.

%
%%%%%%%%%%%%%%%%%%%%%%%%%%%%%%%%%%%%%%%%%%
%%%%% Spectroscopy %%%%%
%%%%%%%%%%%%%%%%%%%%%%%%%%%%%%%%%%%%%%%%%%
%
\section{Spectroscopic datasets}
\label{USco_GTC_XSH:XSH_spec}
%

%
%%%%%%%%%%%%%%%%%%%%%%%%%%%%%%%%%%%%%
%%%%% Table: Known members with SpTypes %%%%%
%%%%%%%%%%%%%%%%%%%%%%%%%%%%%%%%%%%%%
%
\begin{table*}
 \centering
 \caption[]{Coordinates (J2000), magnitudes ($ZYJ$ from VISTA and $HK$ from the UKIDSS Galactic Clusters Survey,
all in the MKO system), and spectral types of candidates previously confirmed
 as members of the UpSco association. LM12 stands for \citet{luhman12c}, L08a for \citet{lodieu08a},
H09 for \citet{herczeg09}, and L11a for \citet{lodieu11a}. Objects
 observed with VLT/X-shooter and publicly available from the ESO archive are highlighted
 in the last column with "XSH".
 }
{\small
 \begin{tabular}{@{\hspace{0mm}}c @{\hspace{1mm}}c @{\hspace{2mm}}c @{\hspace{2mm}}c @{\hspace{2mm}}c @{\hspace{2mm}}c @{\hspace{2mm}}c @{\hspace{2mm}}c @{\hspace{1mm}}c@{\hspace{0mm}}}
 \hline
 \hline
R.A.    &     Dec       &   $Z$ & $Y$ & $J$ & $H$ & $K$ & SpT & References \cr
 \hline
          &                   &   mag & mag & mag & mag & mag &     &       \cr
\hline
16:00:43.17 & $-$22:29:14.5 & 17.112$\pm$0.004 & 16.051$\pm$0.003 & 15.197$\pm$0.003 & 14.646$\pm$0.006 & 14.174$\pm$0.004 & M6.5 & LM12 \cr
16:06:03.75 & $-$22:19:30.2 & 18.292$\pm$0.007 & 16.911$\pm$0.004 & 15.811$\pm$0.004 & 15.112$\pm$0.008 & 14.416$\pm$0.005 & M8.75,L2 & LM12,L08a \cr
16:06:06.29 & $-$23:35:13.5 &      ---                      & 17.220$\pm$0.007 & 16.231$\pm$0.006 & 15.619$\pm$0.012 & 14.999$\pm$0.010 & L0,M8.5 & L08a,LM12,XSH \cr
16:07:14.78 & $-$23:21:01.3 & 19.090$\pm$0.015 & 17.677$\pm$0.009 & 16.561$\pm$0.007 & 15.807$\pm$0.016 & 15.103$\pm$0.009 & M9.25,L0 & LM12,L08a \cr
16:07:23.81 & $-$22:11:02.1 & 17.270$\pm$0.004 & 16.045$\pm$0.003 & 15.145$\pm$0.003 & 14.565$\pm$0.006 & 13.995$\pm$0.004 & M7.5,M8.5,L1 & H09,L11a,L08a,XSH \cr
16:07:27.82 & $-$22:39:04.2 & 19.478$\pm$0.015 & 18.014$\pm$0.009 & 16.856$\pm$0.008 & 16.153$\pm$0.021 & 15.471$\pm$0.013 & M9,L1 & LM12,L08a \cr
16:07:37.98 & $-$22:42:47.1 & 19.291$\pm$0.013 & 17.875$\pm$0.008 & 16.781$\pm$0.007 & 16.091$\pm$0.020 & 15.403$\pm$0.012 & M8.75,L0 & LM12,L08a,XSH \cr
16:08:18.42 & $-$22:32:25.2 & 18.555$\pm$0.008 & 17.157$\pm$0.005 & 16.086$\pm$0.004 & 15.387$\pm$0.011 & 14.699$\pm$0.007 & M9.25,L0 & LM12,L08a,XSH \cr
16:08:28.47 & $-$23:15:10.5 & 17.724$\pm$0.006 & 16.424$\pm$0.004 & 15.423$\pm$0.003 & 14.795$\pm$0.007 & 14.137$\pm$0.004 & M9,L1 & LM12,L08a,XSH \cr
16:08:47.44 & $-$22:35:48.1 & 17.870$\pm$0.006 & 16.639$\pm$0.004 & 15.722$\pm$0.003 & 15.080$\pm$0.009 & 14.526$\pm$0.006 & M8,M9 & LM12,L08a \cr
16:10:47.13 & $-$22:39:49.6 & 17.483$\pm$0.005 & 16.205$\pm$0.003 & 15.236$\pm$0.003 & 14.631$\pm$0.006 & 14.027$\pm$0.004 & M8.5,M9 & L11a,L08a,XSH \cr
16:14:41.68 & $-$23:51:06.0 & 18.589$\pm$0.012 & 17.157$\pm$0.006 & 16.078$\pm$0.004 & 15.357$\pm$0.010 & 14.643$\pm$0.006 & M9.25,L1 & LM12,L08a \cr
16:15:16.66 & $-$23:40:46.5 & 17.989$\pm$0.008 & 16.674$\pm$0.004 & 15.631$\pm$0.003 & 14.935$\pm$0.006 & 14.288$\pm$0.004 & M9 & L08a \cr
 \hline
 \label{tab_USco_GTC_XSH:spectro_known_Members}
 \end{tabular}
}
\end{table*}
\subsection{VLT/X-shooter spectroscopy}
\label{USco_GTC_XSH:XSH_spec_obs}

We carried out spectroscopy from the UV- to the $K$-band with the X-shooter
spectrograph \citep{dOdorico06,vernet11} mounted on the Cassegrain focus of the
Very Large Telescope (VLT) Unit 2\@.
X-shooter is a multi wavelength cross--dispersed echelle spectrograph made of
three independent arms covering simultaneously the ultraviolet (UVB;
0.3--0.56 $\mu$m), visible (VIS; 0.56--1.02 $\mu$m), and near--infrared
(NIR; 1.02--2.48 $\mu$m) wavelength ranges thanks to the presence of two
dichroics splitting the light. The spectrograph is equipped with three
detectors: a 4096$\times$2048 E2V CCD44-82, a 4096$\times$2048 MIT/LL
CCID\,20, and a 2096$\times$2096 Hawaii 2RG for the UVB, VIS, and NIR arms,
respectively.

\subsubsection{New X-shooter near-infrared spectroscopy}
\label{USco_GTC_XSH:XSH_spec_obs_NEW}

We performed our observations in visitor mode during the second half of four
nights on 11--14 April 2015.
We set the read-out mode to 400k and low gain without binning. We used the 
1.3 arcsec slit in the UVB and 1.2 arcsec slits in the VIS and NIR, yielding 
nominal resolving powers of 4000 (8.1 pixels per full--width--half--maximum),
6700 (7.9 pixels per full--width--half--maximum), and 3900 (5.8 pixels per
full--width--half--maximum) in the UVB, VIS, and NIR arms, respectively.
Conditions during the first part of the second half on the night of 11 April
were affected by thin cirrus but the seeing remained around 1.0 arcsec or better.
The third target, VISTA\,J16020000$-$2057341 turns out to be the most affected
by the cirrus that increased in size and numbers during the observing blocks.
The sky was clear and the seeing sub-arcsec during the remaining half nights.
In addition, we retrieved the X-shooter spectra of six previously-known UpSco members
with optical spectral types as well as another seven M/L-type transition members
confirmed spectroscopically (Table \ref{tab_USco_GTC_XSH:spectro_known_Members}), 
all of them belonging to the sequence of VISTA candidates (L13).
We report the list of targets and detailed logs of observations in
Table \ref{tab_USco_GTC_XSH:log_obs_XSH}.

We set the individual on-source integration times to 300 sec in the NIR arm
and used multiple AB patterns to correct for the sky contribution
(mainly) in the near-infrared. All observations were conducted with the slit
oriented at parallactic angle. We acquired the faintest targets ($J$\,$\geq$\,17.5 mag)
with blind offsets, i.e.\ we pointed to a bright star within 30 arcsec or less and applied
an offset using the positions of the bright star and the target measured on the deep VISTA
images. We provide the coordinates of the targets observed with X-shooter and
their $J$-band magnitudes in Table \ref{tab_USco_GTC_XSH:log_obs_XSH}.
We also give the date of the observations, the number of exposures and individual
on-source integrations in each arm.
We did not detect any of our targets in the UVB region and little flux in the
VIS arm even for the brightest sources so we focus on the data reduction and analysis
of the NIR spectra only in the rest of the paper.

\subsubsection{Data reduction of X-shooter spectra}
\label{USco_GTC_XSH:XSH_spec_DR}

We reduced the X-shooter NIR spectra of 15 targets listed in 
Table \ref{tab_USco_GTC_XSH:log_obs_XSH} with the {\tt{esoreflex}} pipeline version 2.8.1\@.
It is a graphical interface using a sequence of command lines from the {\tt{esorex}}
package that produces fully calibrated 2D and 1D spectra.
The pipeline includes the following steps: first a bad pixel map is created using a set of 
40 linearity frames taken within 10 days of our run to identify non-linear pixels. Next, 
a master bias and master dark are created for the optical and near-infrared data, respectively. 
Afterwards, a first guess order and arc line tables
created by illuminating the X-shooter pinhole mask with a continuum and arc lamp are produced.
Later, a master flat and an order table tracing the flat edges before establishing the 2D map 
of the instrument are constructed. Subsequently, the efficiency of the whole system made of the
telescope, instrument, and detector is determined. Finally, 2D spectra for the targets,
telluric standards, and flux standards are generated. We used the final 2D spectra produced
by {\tt{esoreflex}} rather than the final 1D spectra because of the faintness of our 
targets. We should also emphasise that the pipeline does not correct for telluric bands.
Moreover, the ESO staff has informed us about a problem with the sky illumination of the 
last order of X-shooter which leads to a jump in the spectra at $\sim$2.27 $\mu$m where the 
last and the next to last order are joined. This issue affects solely the determination of 
the CO spectral index (Section \ref{USco_GTC_XSH:spectral_indices}) by making it larger; 
because of this artefact, we do not use the CO spectral index \citep{burgasser02} in our analysis.
We also note the presence of a possible artefact around 2.1 microns in some of our targets, which 
affects indices including the $K$-band region of the spectra (CH4-K, KH, H2O-D2, H2O-2, sH2OK).

\begin{figure*}
  \centering
  \includegraphics[width=0.98\linewidth, angle=0]{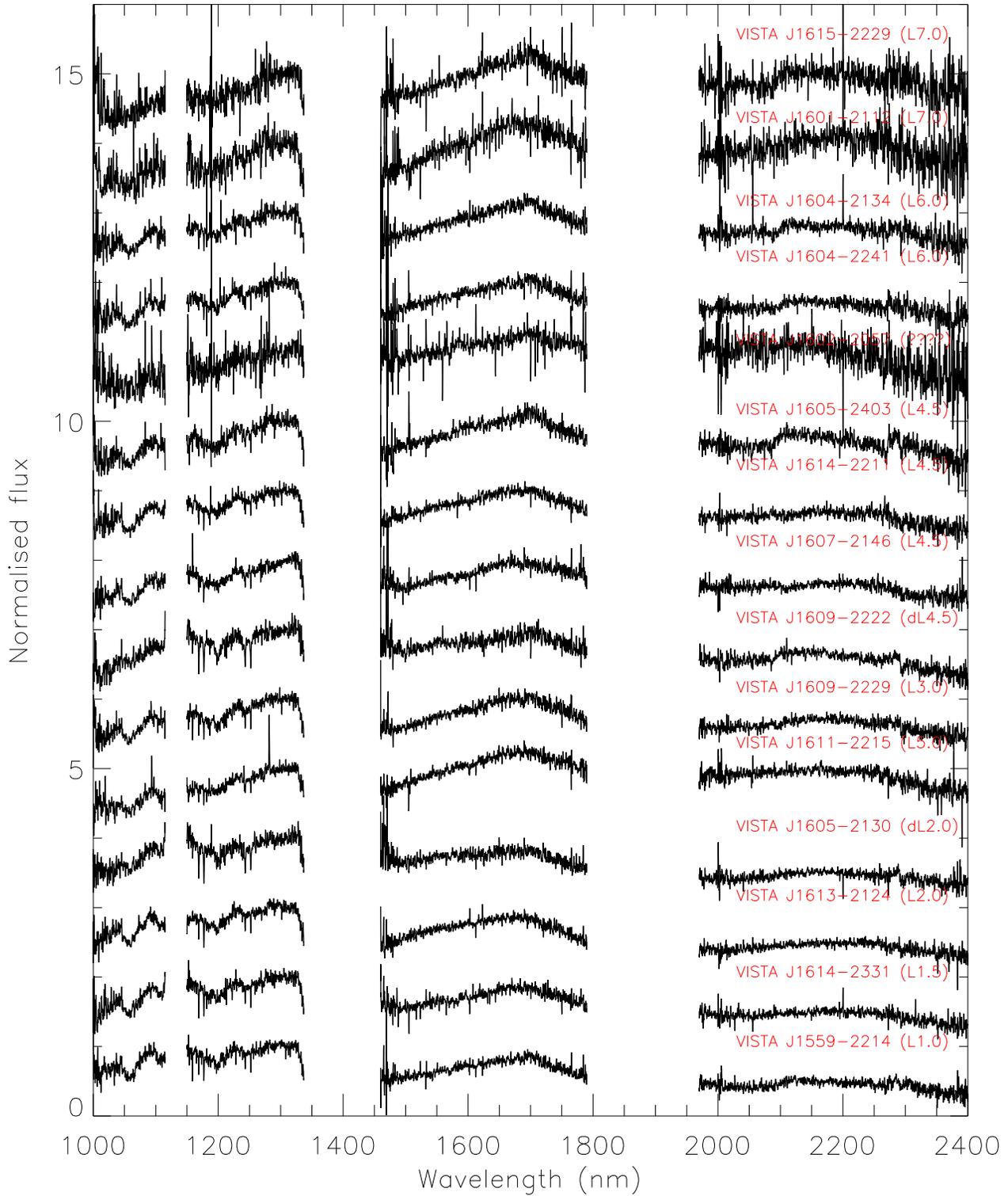}
  \caption{VLT X-shooter spectra of 15 UpSco member candidates (this work).
We note that 1611$-$2215 might be of earlier type with an infrared excess (L2.0exc).
{\bf{All spectra are normalised at 1.30--1.32 microns.}}
}

  \label{fig_USco_GTC_XSH:spectra_XSH}
\end{figure*}

\begin{figure*}
  \centering
  \includegraphics[width=\linewidth, angle=0]{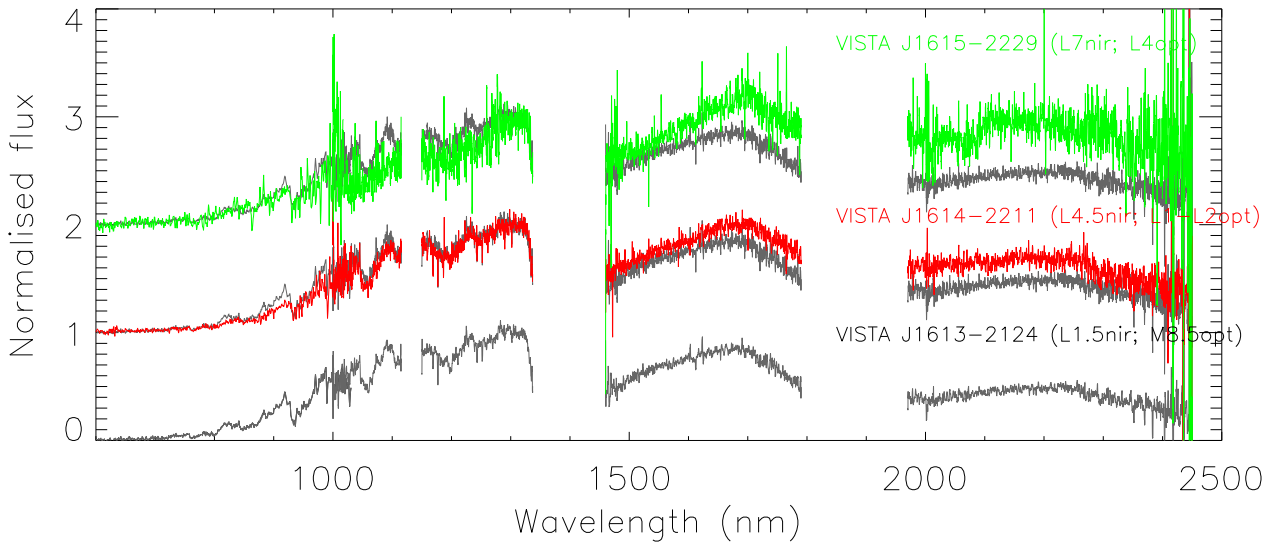}
  \caption{Optical and near-infrared spectra of VISTA\,J1613$-$2124 (grey; L1.5nir; M8.5opt),
VISTA\,J1614$-$2211 (red; L4.5nir; L2opt), and VISTA\,J1615$-$2229 (green; L7nir; L4opt) showing the
evolution as a function of spectral types among our UpSco sequence.
{\bf{All spectra are normalised at 1.30--1.32 microns.}}
}

  \label{fig_USco_GTC_XSH:spectra_XSH_internal_compare}
\end{figure*}

We carried out the next steps under the IRAF environment.
First, we extracted a 1D spectrum for our targets and their associated telluric standard stars 
with the IRAF routine {\tt{apsum}}. To correct for the telluric bands and lines, we fit the 
1D spectrum of the telluric stars taken each night within {\tt{splot}} and normalised the 
continuum to one. We removed the strongest Paschen and Bracket lines in
the spectrum of the telluric late-B star in the resulting spectrum. Finally, we divided
the spectrum of our targets by this spectrum of the telluric standard and multiplied by
a blackbody with the effective temperature of the telluric standard. The final NIR
spectra of the UpSco member candidates are displayed in Fig.\ \ref{fig_USco_GTC_XSH:spectra_XSH}.
We compare the X-shooter spectra of our candidates with known low- and high-gravity dwarfs
in Fig.\ \ref{fig_USco_GTC_XSH:spectra_XSH}. We present the internal UpSco comparison in
Fig.\ \ref{fig_USco_GTC_XSH:spectra_XSH_internal_compare}, where the spectral change of some key 
features can be seen from earlier to later spectral types. We use this 
figure \ref{fig_USco_GTC_XSH:spectra_XSH_internal_compare} to establish a relative spectral 
sequence of our data in Section \ref{USco_GTC_XSH:spectral_classification}.

\subsubsection{Known UpSco members with X-shooter spectra}
\label{USco_GTC_XSH:XSH_spec_obs_KNOWN}

We cross-matched our list of VISTA candidates with the literature and found that several
candidates have been previously confirmed as members of the UpSco association in the
literature. A few of them (6) have also been observed with VLT/Xshooter by other groups.
We list these objects with their infrared magnitudes and spectral types in
Table \ref{tab_USco_GTC_XSH:spectro_known_Members}.

\begin{figure*}
  \centering
  \includegraphics[width=\linewidth, angle=0]{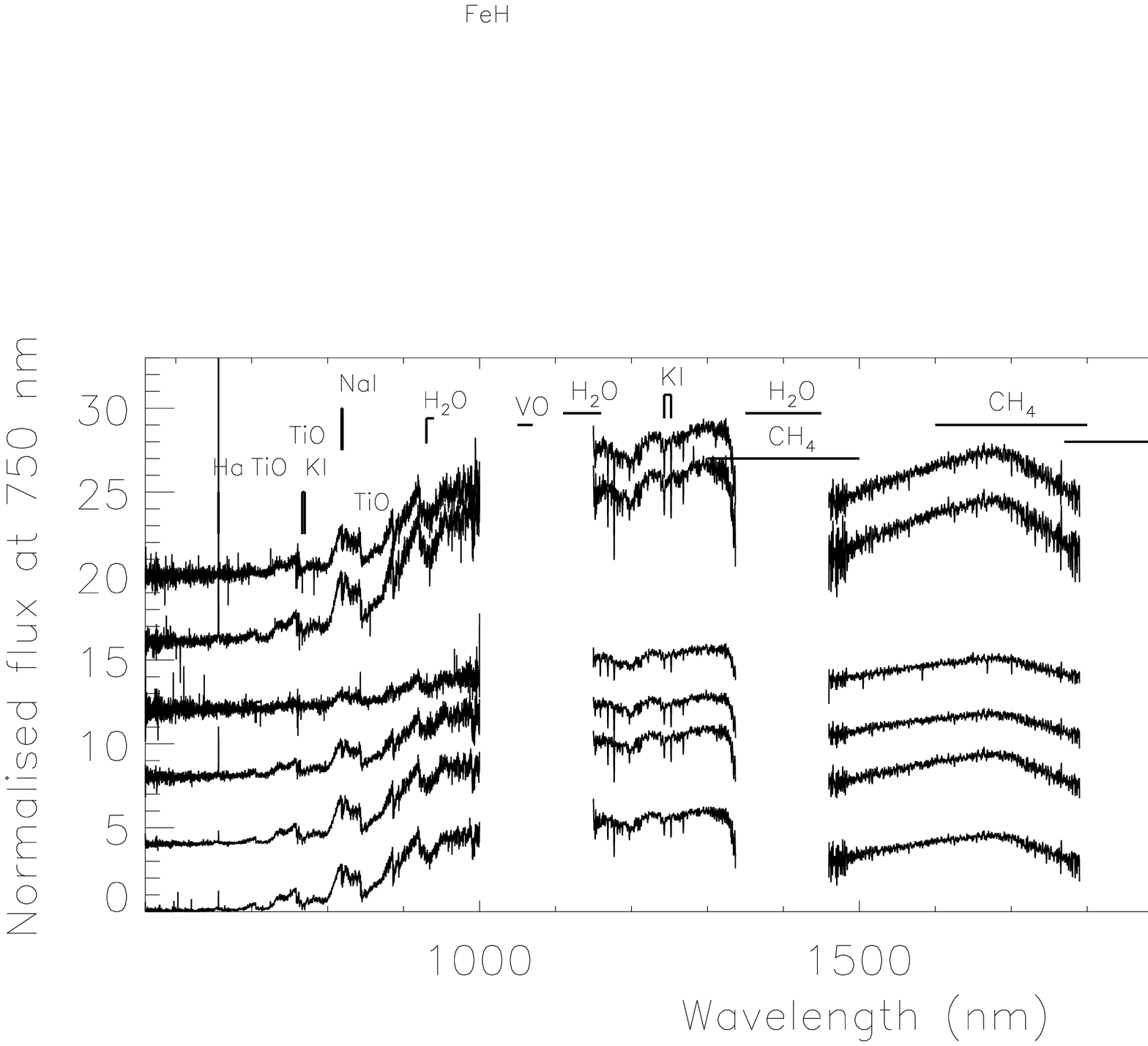}
  \caption{VLT X-shooter spectra of known UpSco members in the ESO archive with
their optical (first value) and near-infrared (second value) spectral classification.
All spectra are corrected for telluric bands.
From bottom to top: VISTA\,J16072381$-$2211021 (M8.5, L1),
USco16104713$-$2239496 (M8.5,M9), USco16060629$-$2335135 (M8.5,L0)
USco16073798$-$2242471 (M8.75,L0), USco16082847$-$2315105 (M9.0,L1),
USco16081842$-$2232252 (M9.25,L0).}
  \label{fig_USco_GTC_XSH:spectra_XSH_archive}
\end{figure*}

We reduced the six X-shooter spectra downloaded from the ESO archive with the same version of
the pipeline. We extracted the 1D spectra from the 2D images created by the pipeline.
We corrected those spectra for telluric bands with the {\tt{molecfit}} package distributed
by ESO \citep{kausch15,smette15}\footnote{http://www.eso.org/sci/software/pipelines/skytools/molecfit}
mainly because no specific telluric standard was taken at the same time and airmass as these targets. 
The 0.55--2.4\,$\mu$m optical-to-infrared spectra of these six M8.5--M9.25 UpSco members are 
displayed in Fig.\ \ref{fig_USco_GTC_XSH:spectra_XSH_archive}. They represent benchmark 
5--10 Myr-old objects for future discoveries of young brown dwarfs in moving groups due 
to the high quality of their spectra and the intermediate spectral resolution. 

%
%%%%%%%%%%%%%%%%%%%%%%%%%%%%%%%%%%%%%
%%%%% Table: Log of VLT/X-shooter observations %%%%%
%%%%%%%%%%%%%%%%%%%%%%%%%%%%%%%%%%%%%
%
\begin{table}
 \centering
 \caption[]{Logs of the VLT X-shooter spectroscopic observations.
 We provide the coordinates (J2000) of the targets with ID used in figures, their $J$-band magnitudes,
the date of observations, and the exposure times for the NIR arm. The exposure 
times chosen for the NIR arm are used for the UVB and VIS arms to optimise overheads.
 }
\scriptsize
 \begin{tabular}{@{\hspace{0mm}}c @{\hspace{2mm}}c @{\hspace{2mm}}c @{\hspace{2mm}}c @{\hspace{2mm}}c @{\hspace{0mm}}}
 \hline
 \hline
ID & Name      &   $J$ & Date  & ExpT  \cr
 \hline
   &  VISTA\,J &  mag  &  yyyy-mm-dd  & sec   \cr
 \hline
 5 & 16114437$-$2215446 & 17.861$\pm$0.016 & 2015-04-11 & 10$\times$300 \cr
10 & 16140756$-$2211522 & 18.251$\pm$0.022 & 2015-04-11 & 12$\times$300 \cr
12 & 16020000$-$2057341 & 18.500$\pm$0.032 & 2015-04-11 & 12$\times$300 \cr
11 & 16053909$-$2403328 & 18.500$\pm$0.032 & 2015-04-11 & 12$\times$300 \cr
 2 & 16130231$-$2124285 & 17.175$\pm$0.011 & 2015-04-12 &  4$\times$300 \cr
14 & 16013692$-$2212027 & 19.270$\pm$0.066 & 2015-04-12 & 12$\times$300 \cr
 9 & 16041304$-$2241034 & 18.167$\pm$0.025 & 2015-04-12 & 10$\times$300 \cr
13 & 16042042$-$2134530 & 18.413$\pm$0.029 & 2015-04-12 & 12$\times$300 \cr
 7 & 16091868$-$2229239 & 17.940$\pm$0.017 & 2015-04-12 & 10$\times$300 \cr
 1 & 15593638$-$2214159 & 16.970$\pm$0.009 & 2015-04-12 &  4$\times$300 \cr
15 & 16151270$-$2229492 & 19.339$\pm$0.057 & 2015-04-13 & 12$\times$300 \cr
 8 & 16073161$-$2146544 & 18.040$\pm$0.022 & 2015-04-13 & 10$\times$300 \cr
 4 & 16051705$-$2130449 & 17.788$\pm$0.017 & 2015-04-13 & 10$\times$300 \cr
 6 & 16095636$-$2222457 & 17.867$\pm$0.016 & 2015-04-13 & 10$\times$300 \cr
 3 & 16142256$-$2331178 & 17.456$\pm$0.012 & 2015-04-13 &  6$\times$300 \cr
 \hline
 \label{tab_USco_GTC_XSH:log_obs_XSH}
 \end{tabular}
\end{table}
%

%
%%%%%%%%%%%%%%%%%%%%%%%%%%%%%%%%%%%%%%%%%%%%%%%
%%%%% Figure: XSH spectra %%%%%
%%%%%%%%%%%%%%%%%%%%%%%%%%%%%%%%%%%%%%%%%%%%%%%
%
%

%
%%%%%%%%%%%%%%%%%%%%%%%%%%%%%%%%%%%%%%%%%%%%%%%%%%%%%%
%%%%% Figure: XSH spectra compared to templates %%%%%
%%%%%%%%%%%%%%%%%%%%%%%%%%%%%%%%%%%%%%%%%%%%%%%%%%%%%%
\begin{figure*}
  \centering
  \includegraphics[width=\linewidth, angle=0]{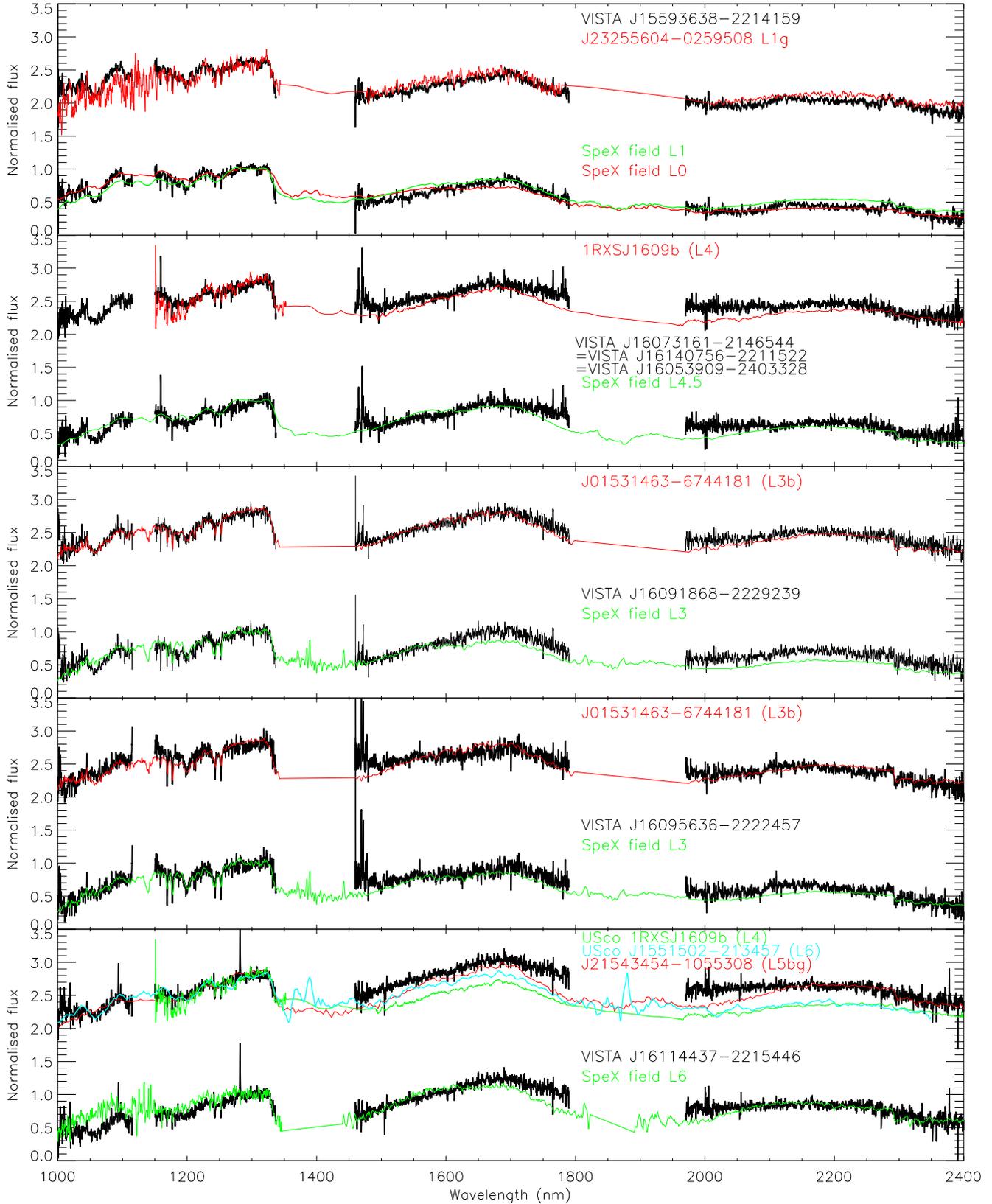}
  \caption{VLT X-shooter spectra of all UpSco member candidates presented in this work
compared to smoothed spectra of field L dwarf templates of similar spectral types 
and known young objects. The names of our targets and the templates are given on each 
plot with their spectral types (b\,=\,$\beta$, g\,=\,$\gamma$).
We note that 1611$-$2215 might be of earlier type with an infrared excess (L2.0exc).
All spectra are normalised at 1.30--1.32 microns.
}
  \label{fig_USco_GTC_XSH:spectra_XSH_compare}
\end{figure*}

\begin{figure*}
  \centering
  \includegraphics[width=\linewidth, angle=0]{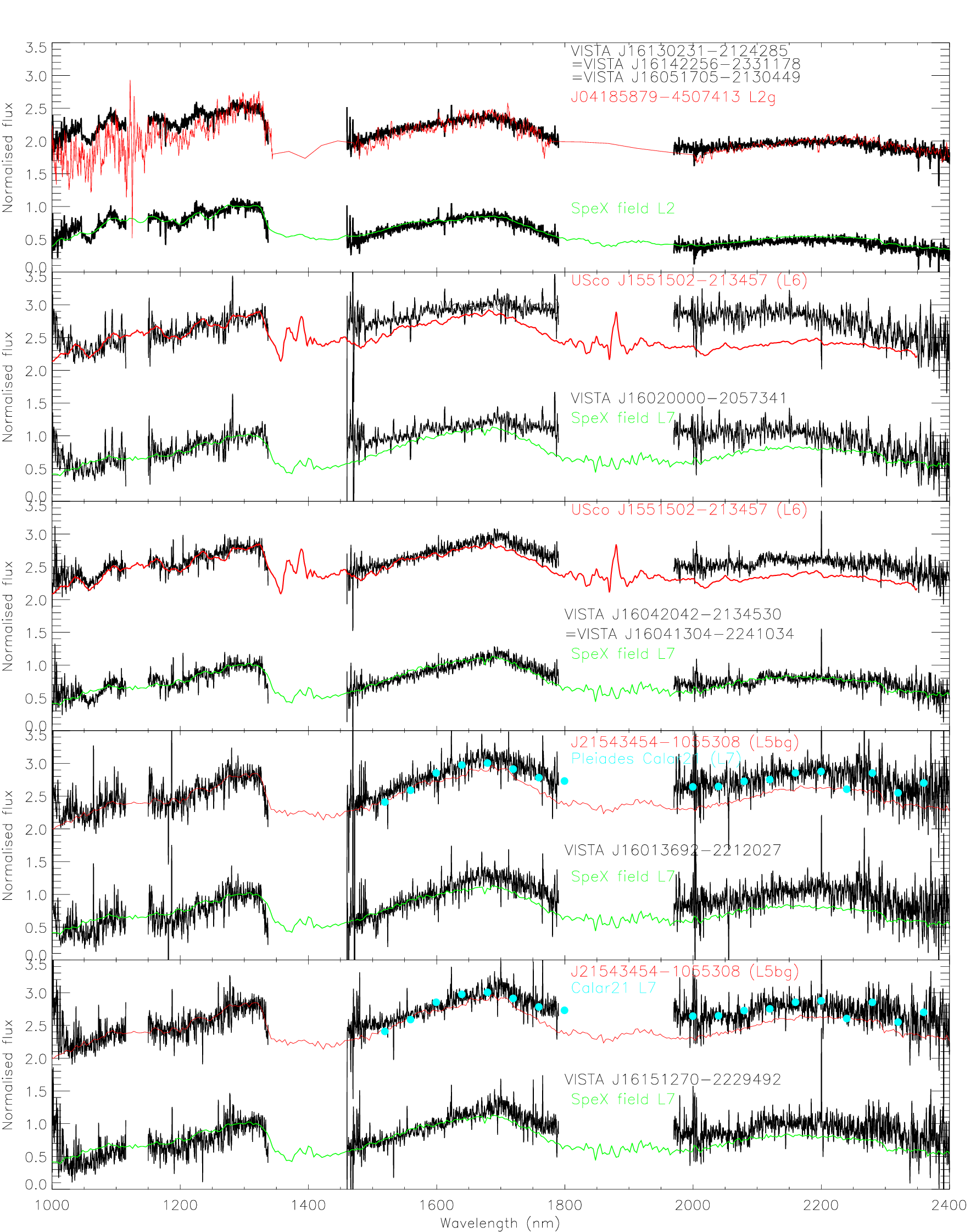}
  \contcaption{}
\end{figure*}

%
%%%%%%%%%%%%%%%%%%%%%%%%%%%%%%%%%%%%%%%%%%%%%%%
%%%%% Figure: XSH spectra: internal comparison %%%%%
%%%%%%%%%%%%%%%%%%%%%%%%%%%%%%%%%%%%%%%%%%%%%%%
%
%

%
%%%%%%%%%%%%%%%%%%%%%%%%%%%%%%%%%%%%%%%%%%%%%%%
%%%%% Figure: XSH spectra: ESO archive %%%%%
%%%%%%%%%%%%%%%%%%%%%%%%%%%%%%%%%%%%%%%%%%%%%%%
%
\subsection{GTC OSIRIS optical spectroscopy}
\label{USco_GTC_XSH:optical_LowRes}

We collected low-resolution optical spectra with GTC OSIRIS for all 15 UpSco targets with X-shooter 
near-infrared spectra (Table \ref{tab_USco_GTC_XSH:logs_spectro_GTC}). 
We carried out the observations in two visitor mode run over six half nights between 7 and 13 June 2016 
(programme GTC25-16A; PI Lodieu) and five half-nights on 25--29 May 2017 (programme GTC40-17A; PI Lodieu). 
In 2016, the nights were dark to grey with the moon increasing 
from 20\% illumination up to 57\% between 7 and 12 June 2016\@. The seeing was typically sub-arcsec during 
the first three nights except during the first 1h on the night of 9 June. The night of 10 June 
was lost due to technical problems and high humidity. The night of 11 June was affected by strong wind altering
the seeing (measured between 1.5 and 2.0 arcsec). The last night was affected by bad seeing ($>$2.5 arcsec).
We collected a spectrum for VISTA\,J16140756$-$2211522 on 2 July 2016 in service mode 
under dark time, clear conditions, and seeing around 0.8 arcsec.
In May 2017, the nights were also dark and the skies clear every night. The seeing was sub-arcsec 
during the first four nights and variable between 1.1 and 1.5 arcsec on the last night (29 May).

We obtained low-resolution optical spectra with the R300R grating covering the 4800--10000\AA{}
and a slit of 1 arcsec resulting in a spectral resolution below 300\@.
We acquired our targets in the Sloan $z$-band filter due to their faintness at optical wavelengths.
We used on-source integrations between 20\,min and 40\,min for the brightest and faintest targets
shifted along the slit several times by a few arcsec (between 5 and 12 arcsec) to avoid nearby stars
in constant AB patterns (a similar technique used in the near-infrared).
We provide the journal of observations in Table \ref{tab_USco_GTC_XSH:logs_spectro_GTC}.
We placed a bright ($Y$\,$\leq$\,18 mag) nearby ($<$50 arcsec) reference star in the slit to
calculate the shifts along the slit with the highest precision possible, implying that our targets
were not observed at parallactic angle. However, prior to three of our targets, we observed the
reference star alone at parallactic angle to evaluate the loss of light. We concluded that the
effect of the angle was negligible redwards of 6500\AA{}. Hence, our targets remain unaffected
because very little or no flux is seen at these blue wavelengths.

\begin{figure}
  \centering
  \includegraphics[width=\linewidth, angle=0]{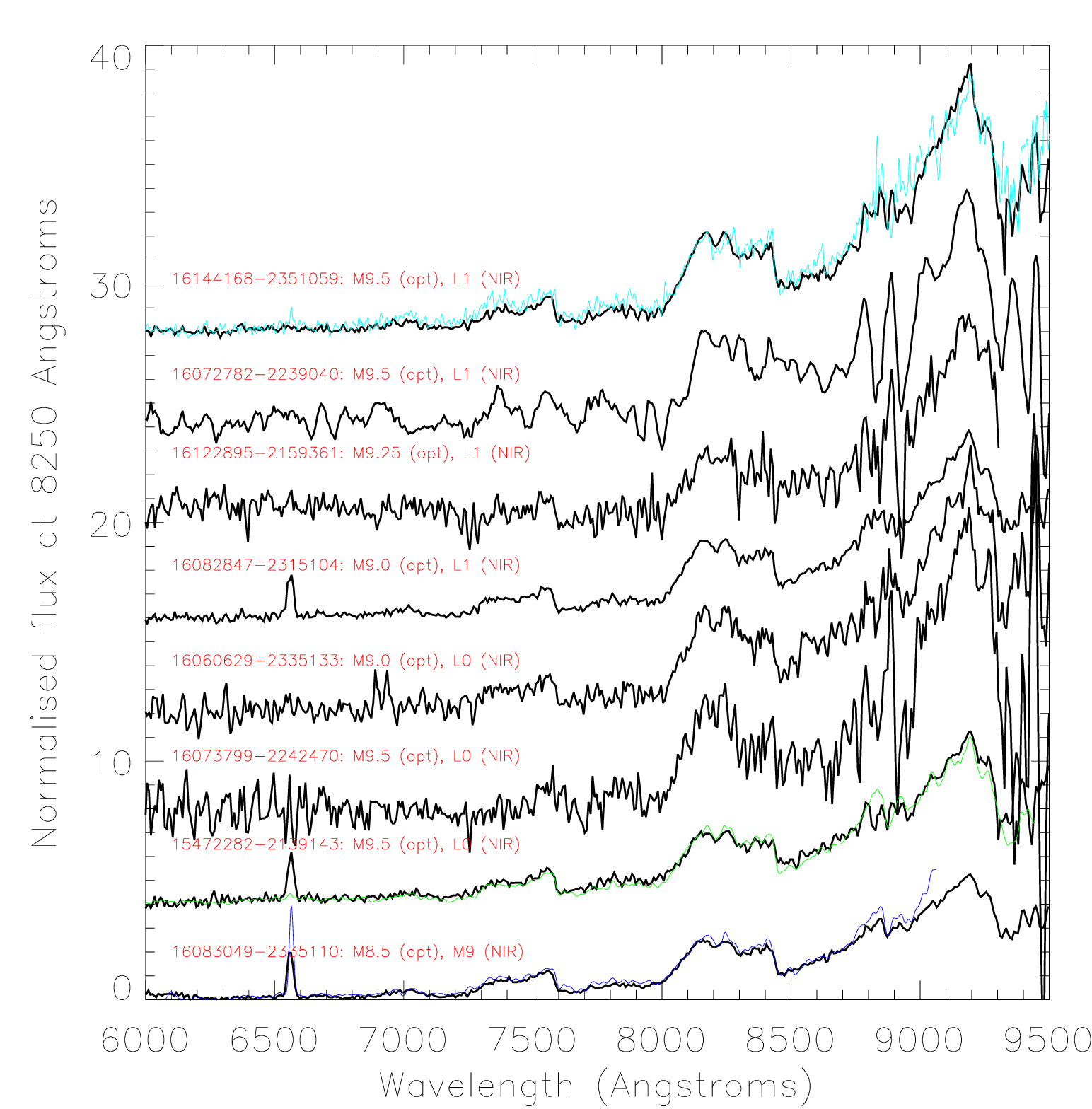}
  \caption{GTC/OSIRIS low-resolution optical spectra of eight M9--L1 UpSco brown dwarf 
members with near-infrared spectral types from \citet{lodieu08a}.
Overplotted in colour are three young spectral templates, members of Taurus 
\citep[KPNO06; M8.5; blue;][]{briceno98,luhman03a}, Chamaeleon 
\citep[11122250$-$7714512; M9; green;][]{luhman04b,luhman07b}, and USco
\citep[UScoCTIO\,108B; M9.5; cyan;][]{bejar08}. Spectra are shifted for clarity.
More details in Section \ref{USco_GTC_XSH:spectral_classification_opt}.
}
  \label{fig_USco_GTC_XSH:plot_spec_optical_backup}
\end{figure}

Because of the poor seeing on 12 June, we obtained optical spectra of a few bright late-M and L dwarf
members of Upper Sco classified in the near-infrared based on Gemini spectra \citep{lodieu08a}.
The targets are listed in the bottom part of Table \ref{tab_USco_GTC_XSH:logs_spectro_GTC}.
We opted for this backup program because of the earlier spectral types reported by \citet{herczeg09}
for two L dwarfs (VISTA\,J160723.82$-$221102.0 and VISTA\,J160603.75$-$221930.0) published in \citet{lodieu08a}.
We plot these objects in Fig.\ \ref{fig_USco_GTC_XSH:plot_spec_optical_backup} and discuss 
the discrepancy between optical and near-infrared classifications in Section \ref{USco_GTC_XSH:spectral_features}.

\begin{figure*}
  \centering
  \includegraphics[width=0.49\linewidth, angle=0]{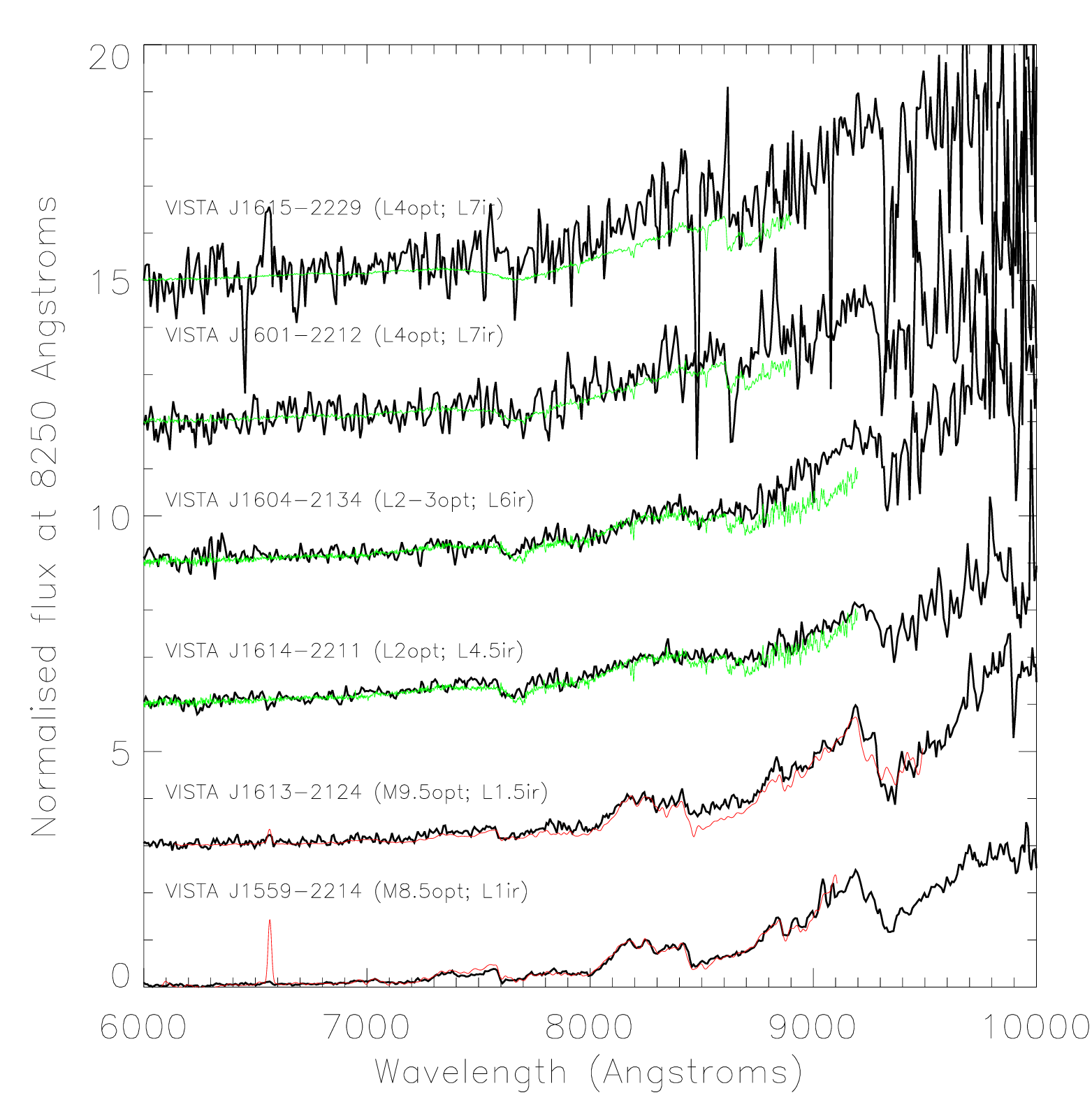}
  \includegraphics[width=0.49\linewidth, angle=0]{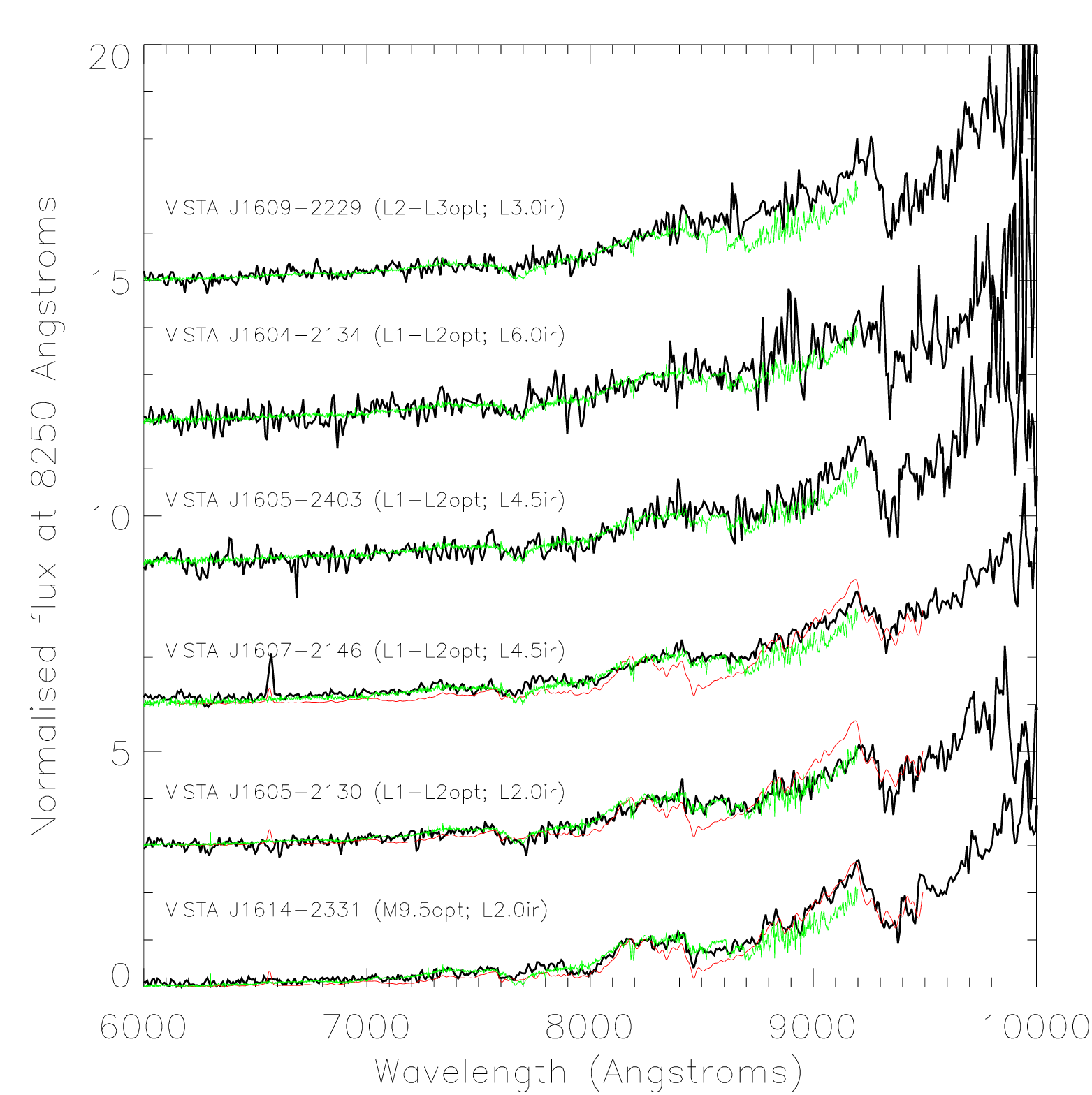}
  \caption{GTC/OSIRIS low-resolution optical spectra of UpSco member candidates 
taken in June 2016 (left) and in May 2017 (right).
The spectral types derived in this work from the VLT/X-shooter near-infrared
spectra and GTC/OSIRIS optical spectra are quoted next to the names.
Overplotted in red are young spectral templates members of the Taurus star-forming 
region \citep[KPNO\,06 (M8.5) and KPNO\,4 (M9.5);][]{briceno98,luhman03a}
and in green Sloan L-type spectral templates marked next to the name of the targets \citep{schmidt10b}.
Spectra are shifted for clarity.
}
  \label{fig_USco_GTC_XSH:plot_spec_optical_LowRes}
\end{figure*}

\begin{figure}
  \centering
  \includegraphics[width=\linewidth, angle=0]{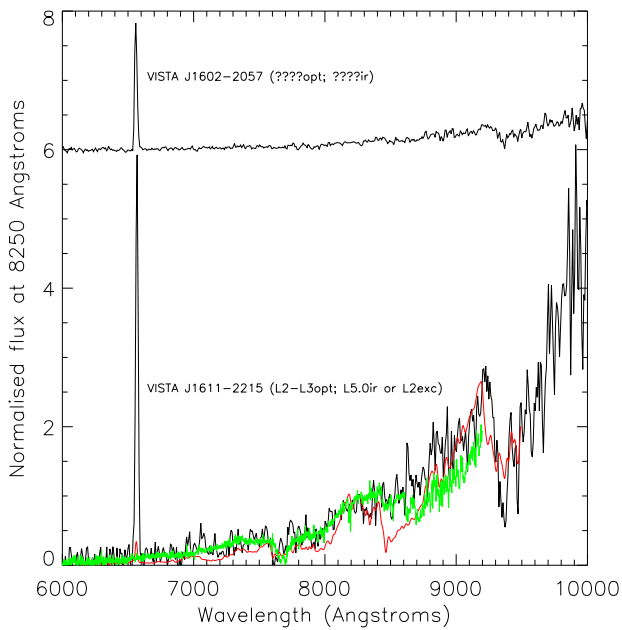}
  \caption{GTC/OSIRIS low-resolution optical spectra (black) of the two sources with strong
H$_{\alpha}$ emission lines: VISTA\,J1611$-$2215 (bottom) and VISTA\,J1602$-$2057 (top).
The coordinates and the spectral types derived from the VLT/X-shooter near-infrared
spectra and optical spectra (Section \ref{USco_GTC_XSH:spectral_classification_opt})
are quoted next to the names. We note that VISTA\,J1602$-$2057 is most likely a non planetary-mass
member of the association because its optical spectrum is not as red the other targets and its 
infrared spectrum does not exhibit strong water bands.
Overplotted in red and green are a member of Taurus classified as M9.5 
\citep[KPNO\,4;][]{briceno98,luhman03a} in red and a Sloan L2 dwarf template 
\citep{schmidt10b}, respectively.
Spectra are shifted for clarity.
}
  \label{fig_USco_GTC_XSH:plot_spec_optical_LowRes_Ha}
\end{figure}

We reduced the optical spectra under the IRAF environment \citep{tody86,tody93} in a standard
manner. First, we median-combined the bias and flat-fields taken during the afternoon which we used
only for the spectrophotometric standard stars and the reference stars.
We calibrated the 1D spectra with the response function derived from two spectrophotometric standard
stars: Ross\,640 classified as DZA5.5 \citep{greenstein67,harrington80,wesemael93,monet03,cutri03,lepine05d} 
and the sdB1 Feige\,66 \citep{reed03,gontcharov06,vanLeeuwen07}. We also corrected for the second order 
contamination by observing the standard star with the R300R grating and the R300R+SDSS$z$ filter.
We calibrated our spectra in wavelength with HgAr$+$Xe$+$Ne lamps taken during the afternoon
with a dispersion of $\sim$7.9\,\AA{} and a rms better than 0.5\,\AA{}.
For the reference stars, we subtracted the mean bias from the raw spectrum of the target and
divided by the normalised flat field. We extracted the one-dimensional spectrum by choosing
optimally the aperture and the background level.
For the targets, we produce images subtracting the ``B'' position from the ``A'' position to
improve the sky subtraction in the red optical part where most of the flux of our targets is
concentrated. Then, we extracted the 1D spectrum optimally and averaged them to produce 
the final spectra displayed in Fig.\ \ref{fig_USco_GTC_XSH:plot_spec_optical_LowRes} 
separating the objects observed in June 2016 (left panel) and May 2017 (right panel).
We display the two strong H$_{\alpha}$ emitters in 
Fig.\ \ref{fig_USco_GTC_XSH:plot_spec_optical_LowRes_Ha}.
We normalised the final OSIRIS optical spectra at 8250\AA{} and compare them to field 
late-M and early-L dwarfs.

%
%%%%%%%%%%%%%%%%%%%%%%%%%%%%%%%%%%%%%%%%%%%%%%%%%
%%%%% Table: Logs of OSIRIS observations  %%%%%
%%%%%%%%%%%%%%%%%%%%%%%%%%%%%%%%%%%%%%%%%%%%%%%%%
%
\begin{table}
\centering
\caption{Journal of observations for the GTC OSIRIS optical spectroscopy 
(Section \ref{USco_GTC_XSH:spectral_classification}): top are
UpSco candidates with X-shooter spectra whereas the bottom part lists backup targets
from \citet{lodieu08a} with their near-infrared spectral types in brackets. 
The photometry comes from our VISTA deep survey (top panel) and UKIDSS Galactic 
Clusters Survey (bottom panel; MKO system).
}
\begin{tabular}{@{\hspace{0mm}}c @{\hspace{2mm}}c @{\hspace{2mm}}c @{\hspace{2mm}}c @{\hspace{2mm}}c@{\hspace{0mm}}}
\hline
\hline
Name      &  $Z$ & Date & ExpT & H$_{\alpha}$ \\
          &  mag & yyyy-mm-dd     & sec  &    \\
\hline
15593638$-$2214159 (L1.0) & 19.523 & 2016-06-07 & 2$\times$1200 &   \cr % cand2 && USco\_ZJY\_016
16130231$-$2124285 (L1.5) & 19.955 & 2016-06-11 & 4$\times$1500 &   \cr % cand55 && USco\_ZJY\_015
16140756$-$2211522 (L4.5) & 21.027 & 2016-07-02 & 4$\times$1800 &   \cr % cand71 && USco\_ZJY\_003
16042042$-$2134530 (L6.0) & 21.138 & 2016-06-07 & 4$\times$1200 &   \cr % cand75 && USco\_ZJY\_006
16013692$-$2212027 (L7.0) & 22.250 & 2016-06-08 & 5$\times$2300 &   \cr % cand96 && USco\_ZJY\_010
16151270$-$2229492 (L7.0) & 22.402 & 2016-06-09 & 6$\times$2300 & Yes \cr % cand97 && USco\_ZJY\_011 
16142256$-$2331178 (L2.0) & 20.236 & 2017-05-25 & 4$\times$1200 & \cr % cand58
16051705$-$2130449 (L2.0) & 20.356 & 2017-05-25 & 4$\times$1200 & \cr % cand19
16091868$-$2229239 (L3.0) & 20.777 & 2017-05-28 & 4$\times$1800 &   \cr % cand41
16073161$-$2146544 (L4.5) & 20.760 & 2017-05-26 & 4$\times$1500 & Yes \cr % cand32
16114437$-$2215446 (L5.0) & 20.946 & 2017-05-27 & 4$\times$1800 & Yes  \cr % cand50
16053909$-$2403328 (L4.5) & 21.103 & 2017-05-27 & 4$\times$1800 &   \cr % cand73
16020000$-$2057341 (????) & 21.372 & 2017-05-28 & 4$\times$1800 & Yes  \cr % cand74
16041304$-$2241034 (L6.0) & 20.652 & 2017-05-26 & 4$\times$1500 &   \cr % cand70
16042042$-$2134530 (L6.0) & 21.138 & 2017-05-29 & 3$\times$1200 &   \cr % cand75 && USco\_ZJY\_006
\hline
15472282$-$2139143 (L0) & 19.94 & 2016-06-12 & 1$\times$1200 & Yes \cr
16060629$-$2335133 (L0) & 18.43 & 2016-06-12 & 1$\times$1200 &   \cr
16072782$-$2239040 (L1) & 19.36 & 2016-06-12 & 1$\times$1200 &   \cr
16073799$-$2242470 (L0) & 19.24 & 2016-06-12 & 1$\times$1800 &   \cr
16082847$-$2315104 (L1) & 17.64 & 2016-06-12 & 1$\times$1200 & Yes \cr
16083049$-$2335110 (M9) & 16.95 & 2016-06-12 & 1$\times$900  & Yes \cr
16122895$-$2159361 (L1) &  ---  & 2016-06-12 & 1$\times$1800 &   \cr
16144168$-$2351059 (L1) & 18.40 & 2016-06-12 & 1$\times$1800 &   \cr
 \hline
 \label{tab_USco_GTC_XSH:logs_spectro_GTC}
\end{tabular}
\end{table}
%

%
%%%%%%%%%%%%%%%%%%%%%%%%%%%%%%%%%%%%%%%%%%%%%%%
%%%%% Figure: OSIRIS Optical spectra %%%%%
%%%%%%%%%%%%%%%%%%%%%%%%%%%%%%%%%%%%%%%%%%%%%%%
%
%

%
%%%%%%%%%%%%%%%%%%%%%%%%%%%%%%%%%%%%%%%%%%%%%%%
%%%%% Figure: OSIRIS Optical spectra: Ha %%%%%
%%%%%%%%%%%%%%%%%%%%%%%%%%%%%%%%%%%%%%%%%%%%%%%
%
%

%
%%%%%%%%%%%%%%%%%%%%%%%%%%%%%%%%%%%%%%%%%%%%%%%%%%%
%%%%% Figure: OSIRIS spectra Backup Targets %%%%%
%%%%%%%%%%%%%%%%%%%%%%%%%%%%%%%%%%%%%%%%%%%%%%%%%%%
%
\subsection{GTC EMIR infrared spectroscopy}
\label{USco_GTC_XSH:Jband_EMIR_LowRes}

We also used the Multi-object Infrared Spectrograph \citep[EMIR;][]{garzon16a} mounted on the 
Naysmith-A focus of the GTC to obtain the near-infrared spectrum of USco\,J155150.2$-$213457. 
We collected these data as part of an EMIR science verification project. This object 
was first identified by \citet{penya16a} using photometric, astrometric, and spectroscopic techniques, and 
was classified as a low-gravity L6\,$\pm$\,1 dwarf based on the presence of strong VO absorption at 1.06 $\mu$m, 
the triangular shape of the $H$-band pseudo-continuum, and the red nature of its spectrum. EMIR is equipped 
with a 2048\,$\times$\,2048 pixel Teledyne HAWAII-2 HgCdTe detector with a pixel scale of 0.2 arcsec on the sky. 
We acquired the EMIR data of USco\,J155150.2$-$213457 with the $YJ$ spectroscopic filter, a 0.8 arcsec long slit, 
and the $YJ$ grism as the dispersive optical element. This instrumental configuration yields low-resolution 
spectra (R\,$\sim$\,740) covering the wavelength interval 0.92--1.33 $\mu$m. We obtained a total of 32 
individual spectra of 120 sec each (total on-source integration time was 3840 sec) following a nodding ABBA 
pattern with offsets of $\sim$10\arcsec. The observing conditions were photometric and the seeing was 1.2 arcsec
on the night of 8 April 2017\@. We also observed the field L6 dwarf 2MASS\,J10101480$-$0406499 
\citep{cruz03,jameson08a} with the same instrumental configuration and an exposure time of 4\,$\times$\,360 s. 
We used this field L6 dwarf as a spectral standard indicative of high-gravity features. We reduced the 
EMIR raw spectra in the same manner as the X-shooter data. We performed wavelength calibration using 
observations of an HgAr lamp and we corrected the instrumental response with the observations of a hot 
B3V-type star. We removed the telluric contribution from the targets data using the same hot star, since 
it was observed at related airmasses.

\begin{figure}
  \centering
  \includegraphics[width=\linewidth, angle=0]{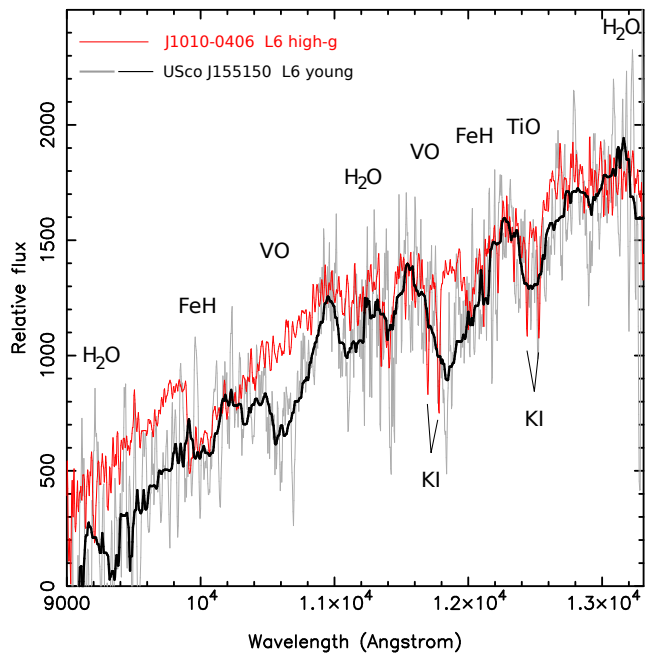}
  \caption{
GTC EMIR spectra of 2MASS\,J10101480$-$0406499 (red) and USco\,J155150.2$-$213457 (gray line depicts the 
original spectrum while the thick black line illustrates the median-filtered data). Both objects are classified 
as L6 high- and a low-gravity dwarfs, respectively. The size of the median filter is 41 pixels. The spectrum 
of the field dwarf was normalized to the observed flux of the UpSco object at 1.28--1.32 $\mu$m. 
}
  \label{fig_USco_GTC_XSH:plot_spec_UScoJ1551m2134_EMIR}
\end{figure}

{\bf{Figure \ref{fig_USco_GTC_XSH:plot_spec_UScoJ1551m2134_EMIR} shows the EMIR spectra. Although the 
optical-to-near-infrared rising slope of both USco\,J155150.2$-$213457 and the L6 standard is very similar 
to each other at these wavelengths, some individual features are clearly different. The molecular absorptions 
at $\sim$1.06 $\mu$m (VO), $\sim$1.18 $\mu$m (VO), and $\sim$1.25 $\mu$m (TiO) appear significantly stronger 
in the UpSco object, while the absorption at $\sim$0.99 $\mu$m (FeH) is weaker in relative terms probably 
because of the low gravity atmosphere of the UpSco source \citep[see also discussion in][]{mcgovern04,martin17a}.}}

%
%%%%%%%%%%%%%%%%%%%%%%%%%%%%%%%%%%%%%%%%%%%%%%%%%%%
%%%%% Figure: EMIR spectrum of UScoJ1550 %%%%%
%%%%%%%%%%%%%%%%%%%%%%%%%%%%%%%%%%%%%%%%%%%%%%%%%%%
%
%

%
%%%%%%%%%%%%%%%%%%%%%%%%%%%%%%%%%%%%%%%%%%
%%%%% Mid-IR WISE photometry %%%%%
%%%%%%%%%%%%%%%%%%%%%%%%%%%%%%%%%%%%%%%%%%
%
\section{AllWISE mid-infrared photometry}
\label{USco_GTC_XSH:WISE_midIR}

We cross-matched the list of 15 candidates with X-shooter spectra with the AllWISE catalogue
\citep{wright10} using a matching radius of five arcsec. Most targets were found within
0.3 arcsec of the position extracted from our VISTA catalogue. We found that 11 of them have 
reliable WISE $w1$ and $w2$ photometry (flags of A and B) with signal-to-noise ratios 
larger than six and a minimum of 19 flux measurements in each band 
(Table \ref{tab_USco_GTC_XSH:table_WISE_phot}).
The remaining four sources (abbreviated names are VISTA\,J1602$-$2057,
VISTA\,J1605$-$2403, VISTA\,J1609$-$2222, and VISTA\,J1614$-$2331) are detected in
the images at 3.6 and 4.5 $\mu$m but have no entry in the AllWISE catalogue mainly
because they are either very faint (2 cases) or close to another brighter star (3 cases, 
1 common to the very faint case).

\begin{figure*}
  \centering
  \includegraphics[width=0.49\linewidth, angle=0]{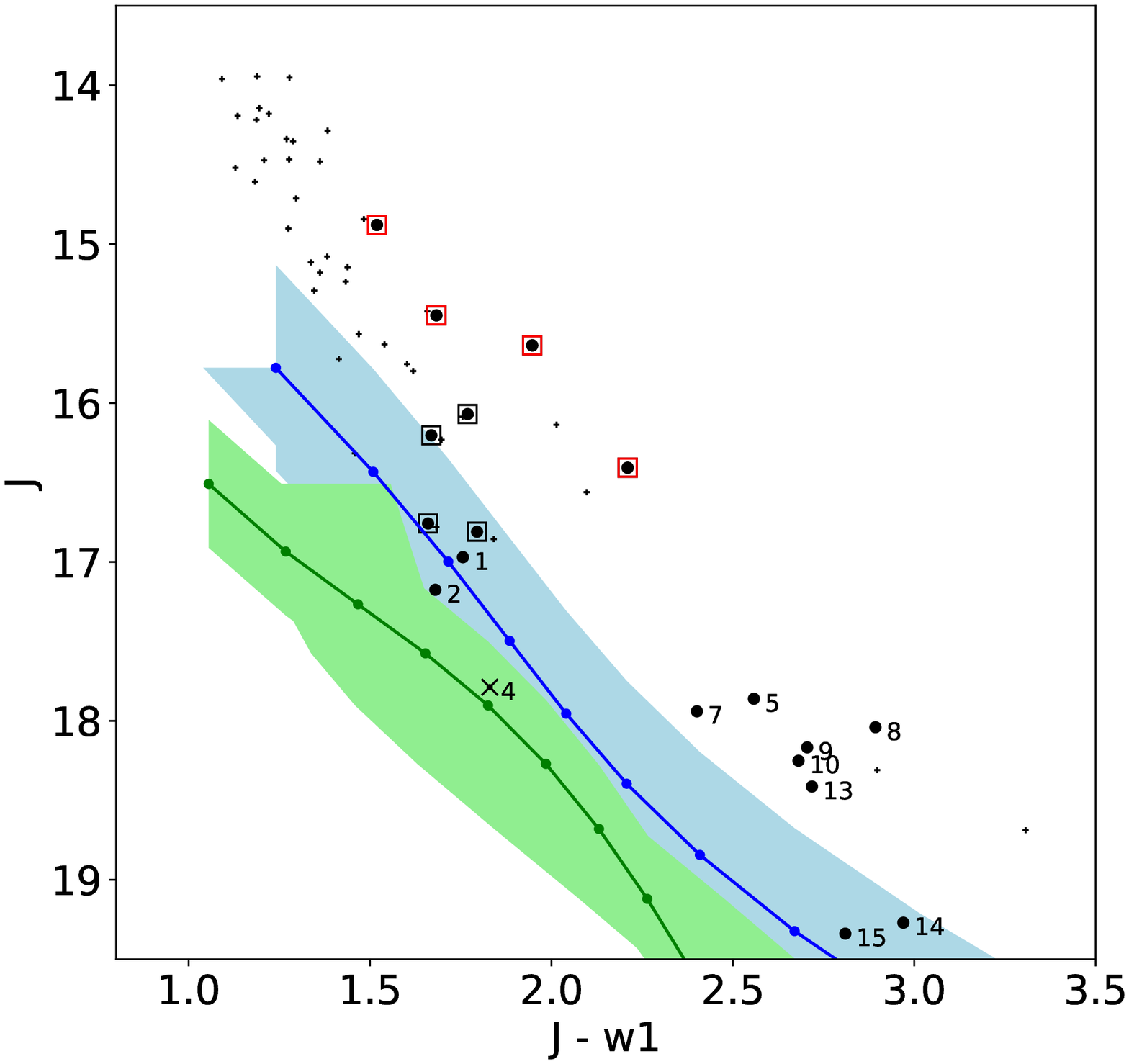}
  \includegraphics[width=0.49\linewidth, angle=0]{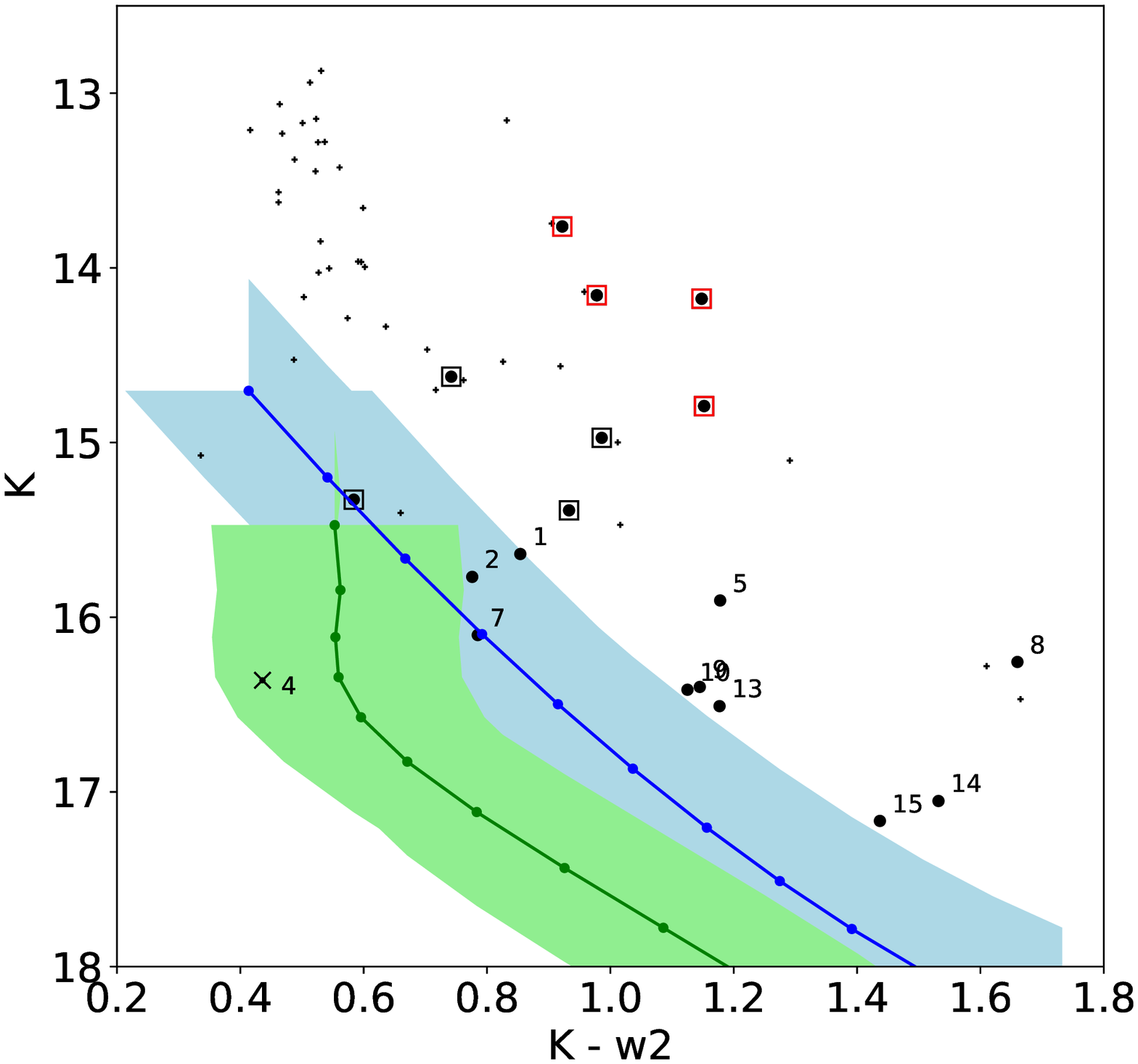}
  \caption{{\it{Left:}} ($J-w1$,$J$) colour-magnitude diagram for all VISTA $ZYJ$
candidates identified in L13 with WISE photometry (grey crossed).
Overplotted as black dots with their ID numbers are the UpSco candidates with X-shooter 
spectra (Table \ref{tab_USco_GTC_XSH:table_WISE_phot}) and the M9--L1
from \citet{lodieu08a} with GTC optical spectra as dots surrounded by squares
(those with red squares show mid-infrared excess). 
We added the sequence of young and field M7--L8 dwarfs from \citet{faherty16a} 
in blue and green with their dispersion, respectively.
}
  \label{fig_USco_GTC_XSH:CMD_VISTA_WISE}
\end{figure*}

\begin{figure}
  \centering
  \includegraphics[width=\linewidth, angle=0]{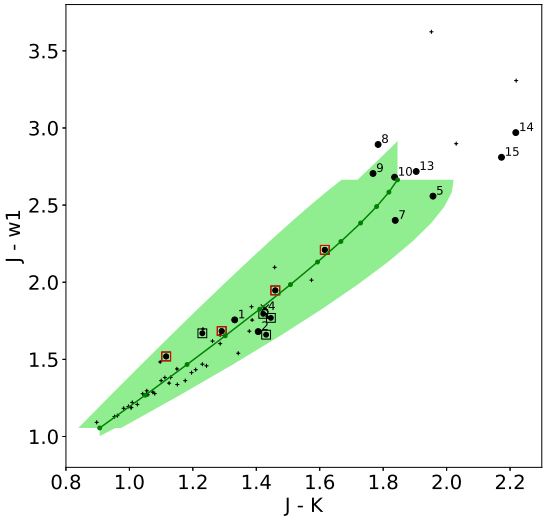}
  \caption{($J-K$,$J-w1$) colour-colour diagram for all VISTA $ZYJ$
candidates identified in L13 with WISE photometry (grey crosses).
Overplotted as black dots with their ID numbers are the UpSco candidates with X-shooter 
spectra (Table \ref{tab_USco_GTC_XSH:table_WISE_phot}) and the M9--L1
from \citet{lodieu08a} with GTC optical spectra as dots surrounded by squares
(those with red squares show mid-infrared excess).
We added the sequence of field M7--L8 dwarfs from \citet{faherty16a} 
in green with its dispersion.
}
  \label{fig_USco_GTC_XSH:CCD_VISTA_WISE}
\end{figure}

To define a clean infrared sequence of the lowest mass members in UpSco,
we also cross-correlated the list of 67 VISTA $ZYJ$ candidates
(Table 2 of L13) with AllWISE\@. This query retrieved 51 sources
with no deblending and reliable $w1$ and $w2$ photometry.
We show the resulting near-infrared vs mid-infrared colour-magnitude diagrams for
these two samples in Fig.\ \ref{fig_USco_GTC_XSH:CMD_VISTA_WISE}. We overplotted
the sequence of field M/L/T dwarfs derived from the polynomial fits given in
Table 14 of \citet*{dupuy12}, shifted to the distance of UpSco.
The field sequence lies well below the UpSco sequence because of the difference
in age between the association \citep[$\sim$5--10 Myr;][]{preibisch01,pecaut12,song12}
and field dwarfs (typically 1 Gyr or older).
We also added to this diagram the WISE photometry for the eight M9--L1 UpSco members
with near-infrared photometry from \citet{lodieu08a} listed in Table \ref{tab_USco_GTC_XSH:table_WISE_phot}.
In Fig.\ \ref{fig_USco_GTC_XSH:CCD_VISTA_WISE} we plot the ($J-K$,$J-w1$)
colour-colour diagram with the same symbology.

%
%%%%%%%%%%%%%%%%%%%%%%%%%%%%%%%%%%%%%
%%%%% Table: AllWISE photometry %%%%%
%%%%%%%%%%%%%%%%%%%%%%%%%%%%%%%%%%%%%
%
\begin{table}
 \centering
 \caption[]{{\it{Top panel:}} AllWISE photometry (MKO system) with errors for the 11 out the 
15 UpSco candidates with VLT/X-shooter (Table \ref{tab_USco_GTC_XSH:log_obs_XSH}). Numbers in 
brackets listed after the photometry depicts the signal-to-noise ratio (SNR).
The other targets do not have entries in the AllWISE catalogue but are detected on the
3.6 and 4.5 $\mu$m images.
{\it{Bottom panel:}} WISE photometry for M9--L1 dwarfs in \citet{lodieu08a} observed with
GTC/OSIRIS\@. Only VISTA\,J16083049$-$2335110 is detected in $w3$ (11.621+/-0.266 mag)
with a signal-to-noise ratio of 4.1\@.
 }
 \begin{tabular}{@{\hspace{0mm}}c @{\hspace{2mm}}c @{\hspace{2mm}}c@{\hspace{0mm}}}
 \hline
 \hline
Name      &     $w1$ & $w2$ \cr
 \hline
           & mag (SNR) &  mag (SNR) \cr
 \hline
15593638$-$2214159 & 15.214$\pm$0.045 (24) & 14.784$\pm$0.071 (15) \cr
16130231$-$2124285 & 15.495$\pm$0.055 (19) & 14.993$\pm$0.083 (13) \cr
16142256$-$2331178 &  --- $\pm$ ---  (---) &  --- $\pm$ ---  (---) \cr
16051705$-$2130449 & 15.958$\pm$0.060 (18) & 15.925$\pm$0.172 ( 6) \cr
16114437$-$2215446 & 15.303$\pm$0.045 (24) & 14.726$\pm$0.070 (15) \cr
16095636$-$2222457 &  --- $\pm$ ---  (---) &  --- $\pm$ ---  (---) \cr
16091868$-$2229239 & 15.539$\pm$0.051 (21) & 15.317$\pm$0.114 ( 9) \cr
16073161$-$2146544 & 15.147$\pm$0.040 (27) & 14.596$\pm$0.062 (17) \cr
16041304$-$2241034 & 15.462$\pm$0.049 (22) & 15.254$\pm$0.105 (10) \cr
16140756$-$2211522 & 15.570$\pm$0.051 (21) & 15.290$\pm$0.105 (10) \cr
16053909$-$2403328 &  --- $\pm$ ---  (---) &  --- $\pm$ ---  (---) \cr
16020000$-$2057341 &  --- $\pm$ ---  (---) &  --- $\pm$ ---  (---) \cr
16042042$-$2134530 & 15.695$\pm$0.051 (21) & 15.332$\pm$0.110 ( 9) \cr
16013692$-$2212027 & 16.300$\pm$0.076 (14) & 15.520$\pm$0.139 ( 7) \cr
16151270$-$2229492 & 16.529$\pm$0.100 (10) & 15.729$\pm$0.156 ( 6) \cr
 \hline
15472280$-$2139143 & 13.690$\pm$0.072 (15.0) & 13.029$\pm$0.033 (33.3) \cr
16060629$-$2335133 & 14.534$\pm$0.030 (35.8) & 13.987$\pm$0.042 (26.0) \cr
16072782$-$2239040 & 15.015$\pm$0.038 (28.3) & 14.455$\pm$0.066 (16.4) \cr
16073799$-$2242470 & 15.098$\pm$0.039 (27.7) & 14.743$\pm$0.075 (14.4) \cr
16082847$-$2315104 & 13.765$\pm$0.027 (40.2) & 13.179$\pm$0.030 (36.3) \cr
16083049$-$2335110 & 13.360$\pm$0.026 (41.4) & 12.841$\pm$0.029 (37.8) \cr % w3=11.621+/-0.266 (snr=4.1)
16122895$-$2159361 & 14.197$\pm$0.029 (37.2) & 13.639$\pm$0.039 (28.2) \cr
16144168$-$2351059 & 14.300$\pm$0.029 (37.4) & 13.881$\pm$0.041 (26.8) \cr
\hline
 \label{tab_USco_GTC_XSH:table_WISE_phot}
 \end{tabular}
\end{table}
%

%
%%%%%%%%%%%%%%%%%%%%%%%%%%%%%%%%%%%%%%%%%%%%%%%
%%%%% Figure: NIR vs midIR CMDs %%%%%
%%%%%%%%%%%%%%%%%%%%%%%%%%%%%%%%%%%%%%%%%%%%%%%
%
%

%
%%%%%%%%%%%%%%%%%%%%%%%%%%%%%%%%%%%%%%%%%%%%%%%
%%%%% Figure: NIR vs midIR col-col diag %%%%%
%%%%%%%%%%%%%%%%%%%%%%%%%%%%%%%%%%%%%%%%%%%%%%%
%
%

%
%%%%%%%%%%%%%%%%%%%%%%%%%%%%%%%%%%%%%%%%%%
%%%%% Membership analysis %%%%%
%%%%%%%%%%%%%%%%%%%%%%%%%%%%%%%%%%%%%%%%%%
%
\section{Characterisation of young L-type members of UpSco}
\label{USco_GTC_XSH:properties}

As illustrated in Figures \ref{fig_USco_GTC_XSH:CMD_iJJ}, \ref{fig_USco_GTC_XSH:spectra_XSH},
\ref{fig_USco_GTC_XSH:spectra_XSH_internal_compare}, and \ref{fig_USco_GTC_XSH:spectra_XSH_archive},
we have optical and near-infrared photometry and spectroscopy of a significant number of high-probability 
member candidates of UpSco ($\sim$5--10 Myr, d\,$\sim$\,145 pc, solar metallicity) that span almost 2.5 mag in the 
$J$-band (from 17.0 to 19.3 mag). We discuss the morphological changes observed in the spectra in subsequent 
sections. Because our targets likely share the same age and metallicity and are located at the same distance 
(the intra-cluster distances have a negligible impact in the following discussion), we attribute these changes 
to a sequence of spectral types and different effective temperatures ($T_{\rm eff}$), although additional 
effects, like the presence of dusty clouds or veiling due to strong accretion, may also imprint complex 
spectral signatures. Our goal is to establish a proper classification of these coeval objects, thus 
complementing the efforts made by other groups regarding the analysis of spectra of members of intermediate-age 
($>$10 Myr) stellar moving groups \citep[e.g.][]{allers13,gagne15a,filippazzo15,faherty16a,liu16a}.

\subsection{Photometric properties}
\label{USco_GTC_XSH:phot_properties}

We assessed further the membership of our VISTA candidates with the SDSS$i$ imaging obtained
with the GTC\@. Using the ($i-J$,$J$) colour-magnitude diagram of Fig.\ \ref{fig_USco_GTC_XSH:CMD_iJJ},
we already removed two likely contaminants from the sample of UpSco VISTA candidates
(see Section \ref{USco_GTC_XSH:GTC_phot}). The remaining sources, including $i$-band non-detections,
follow a sequence redder than field stars along the line of sight of UpSco.

We also check the mid-infrared photometry of our candidates. We find that UpSco members define 
a sequence with a high level of dispersion in the near-infrared vs mid-infrared diagrams presented in 
Fig.\ \ref{fig_USco_GTC_XSH:CMD_VISTA_WISE}, from $J$\,=\,14 mag down to $J$\,=\,19.5 mag 
(equivalent to $K$\,$\sim$\,13--17.5 mag). Late-M dwarfs show a trend towards
brighter magnitudes with younger ages. However, studies in young moving groups suggest that 
younger L dwarfs have similar magnitudes to older field L dwarfs in the near-infrared ($JHK$ bands)
with a tendancy to be brighter in the WISE passbands \citep{gagne15a,faherty16a,liu16a}.
We overplotted the sequence of field L dwarfs with its dispersion in the near-infrared to
mid-infrared diagrams to gauge the redness of our UpSco member candidates.
We observe that UpSco members lie above of the field sequence shifted to the distance of the association.
The candidates presented in this paper overlap and extend the sequence
of M9--L1 dwarfs confirmed spectroscopically in the near-infrared with Gemini/GNIRS
\citep[dots with open squares in Fig.\ \ref{fig_USco_GTC_XSH:CMD_VISTA_WISE}][]{lodieu08a,herczeg09}.
We note that VISTA\,J16083049$-$2335110 \citep[M8.5opt, M9ir;][]{lodieu08a} is clearly detected in 
the WISE $w3$ band with a magnitude of 11.621$\pm$0.266, advocating that this object most likely harbours 
a disk (Table \ref{tab_USco_GTC_XSH:table_WISE_phot}).
This source lies on top of the UpSco sequence in both near-infrared to mid-infrared colour-magnitude
diagrams (Fig.\ \ref{fig_USco_GTC_XSH:CMD_VISTA_WISE}), particularly in the ($K-w2$,$K$) diagram
(black dots with open red squares).
Three other sources appear above the sequence in the $K-w2$ colours (VISTA\,J16082847$-$2315104;
VISTA\,J15472280$-$2139143, and VISTA\,J16060629$-$2335133), which might suggest either the presence 
of a disk or a close-in companion although additional effects cannot be discarded at these ages 
(e.g.\ photometric errors, extinction; Table \ref{tab_USco_GTC_XSH:table_WISE_phot}).
The other possible explanation lies in the role of the accretion history of low-mass stars and
brown dwarfs during the early stages of their formation \citep{baraffe16}.
On the contrary, one of the 15 candidates with X-shooter spectra, USco\,16051705$-$2130449,
lies below the sequence, even below the sequence of field stars shifted to the distance of UpSco,
suggesting it might not be true member of UpSco but older.
One of the 15 targets might exhibit considerable flux excesses in the $W2$ passband: 
VISTA\,J1607$-$2146 (\#8). This object is discussed further in Section \ref{USco_GTC_XSH:spectral_classification_opt}, where we report the detection of H$\alpha$ emission 
in its GTC/OSIRIS optical spectrum.

Overall, the positions of 14 out of 15 of our candidates in several colour-colour magnitude over 
a wide wavelength range add credit to their membership to the UpSco association.

\subsection{Spectral classification}
\label{USco_GTC_XSH:spectral_classification}
\subsubsection{Word of caution}
\label{USco_GTC_XSH:word_caution}

First, we note that several authors have reported discrepancies between the optical and near-infrared spectral
classifications of young late-M and L dwarfs. This trend was mentioned first by \citet{martin01a}
and corroborated by the optical spectra of faint $\sigma$\,Ori members \citep{barrado01c}.
Later, \citet{luhman03b} found a systematic offset between optical
and $K$-band (2.0--2.4 $\mu$m) spectral types of young late-M dwarf members of IC\,348, the latter
being consistently 1--2 subclass later. This effect was already pointed out for members of the
$\rho$\,Ophiuchus region \citep*{luhman99b} and confirmed in Taurus \citep{luhman16c} and 
in other moving groups and associations
\citep{pecaut16a}. \citet{lucas01a} independently mentioned this effect comparing the observed
optical and $H$-band spectra with the physical parameters derived the synthetic models of \citet{allard01}.
\citet{allers13} also emphasise the fact that their qualitative classification of young L dwarfs
shows an offset of $\sim$1 subclass between infrared and optical spectral types.

In UpSco, \citet{bejar08} classified UScoCTIO\,108B as a M9.5 dwarf from its low-resolution optical 
(0.55--0.95 $\mu$m) spectrum by comparison with Ophiuchus members and old field stars. However, they
quote a spectral type of L3 from their $J$-band spectrum (1.15--1.3 $\mu$m) by comparison with field L dwarfs.
This discrepancy is comparable to the inconsistency between the infrared and optical 
spectral types of VISTA\,J160723.82$-$221102.0 (L1 vs M8.5) and VISTA\,J160603.75$-$221930.0  (L2 vs M8.75) 
reported by \citet{lodieu08a} and \citet{herczeg09}, respectively.

In this paper, we will classify our targets independently in the optical and in the near-infrared for
those followed up in both wavelength regions.
We would like to stress the importance of stellar clusters to define reliable photometric and
spectroscopic sequences of members that have distinct masses (therefore varying luminosities and
magnitudes) but same age, distance, and metallicity. The increasing numbers of bright, young L dwarf
member candidates (typically $<$L5) of stellar moving groups during the past years 
\citep[e.g.][]{gagne14a,malo14b,gagne15a,faherty16a,liu16a} contrasts with the limited numbers
of late-M and early-L type members of young clusters and star-forming regions with good-quality 
optical and near-infrared spectra \citep{barrado01c,martin03b,allers07,lodieu08a,herczeg09,cruz09,bayo11a},
some of them used as comparison in this work.
With this paper, we will alleviate this lack of data and will provide a sequence of young L-type
spectra for the age of 5--10 Myr and solar metallicity.

\subsubsection{Relative near-infrared classification}
\label{USco_GTC_XSH:spectral_classification_NIR_relative}

First, we compare the set of X-shooter spectra of our targets (Sect.\ \ref{USco_GTC_XSH:XSH_spec_obs};
Figs.\ \ref{fig_USco_GTC_XSH:spectra_XSH} and \ref{fig_USco_GTC_XSH:spectra_XSH_internal_compare}) 
against each other to reveal similarities and differences and define a relative classification
where later objects have steep pseudo-continuum slopes, the strongest water bands, and less
intense VO\@. We use abbreviated names below.

\begin{itemize}
\item [$\bullet$] VISTA\,J1559$-$2214 is the earliest target of all with $J$\,=\,16.970$\pm$0.009 mag
and $J-K$\,=\,1.33$\pm$0.02 mag.
\item [$\bullet$] VISTA\,J1613$-$2124 ($J$\,=\,17.18$\pm$0.01, $J-K$\,=\,1.41$\pm$0.02) 
is similar to VISTA\,J1559$-$2214 in the $J$-band while the blue part of the $H$-band appears 
in between the fluxes of VISTA\,J1559$-$2214 and VISTA\,J1611$-$2215\@. The $K$-band region is 
similar to VISTA\,J1605$-$2130 ($J$\,=\,17.79$\pm$0.02, $J-K$\,=\,1.43$\pm$0.03).
Overall, we classify this source as intermediate between the earliest source and 
VISTA\,J1605$-$2130\@.
\item [$\bullet$] VISTA\,J1605$-$2130 is relatively brighter than VISTA\,J1559$-$2214 in $J$ and $K$.
The blue part of the $H$-band exhibits more flux while the red part of the $H$-band is identical.
\item [$\bullet$] VISTA\,J1614$-$2331 ($J$\,=\,17.46$\pm$0.01, $J-K$\,=\,1.44$\pm$0.02)
is identical to VISTA\,J1605$-$2130 so we assign one subtype later than the earliest source to both objects.
\item [$\bullet$] VISTA\,J1611$-$2215 ($J$\,=\,17.86$\pm$0.02, $J-K$\,=\,1.96$\pm$0.03) 
is similar to VISTA\,J1613$-$2124 in $J$ but relatively much brighter in $H$ and $K$.
\item [$\bullet$] VISTA\,J1609$-$2222 ($J$\,=\,17.87$\pm$0.02, $J-K$\,=\,1.68$\pm$0.03)
shows almost no absorption in the VO band at 1.06\,$\mu$m,
displays a stronger CO absorption band at 2.3\,$\mu$m, and has a flatter $H$-band than any other source
in our sample, suggesting that it is a field L4.5 dwarf rejected as a member as of the association.
\item [$\bullet$] VISTA\,J1609$-$2229 ($J$\,=\,17.94$\pm$0.02, $J-K$\,=\,1.84$\pm$0.03)
is similar to VISTA\,J1613$-$2124 in $J$ but twice brighter than VISTA\,J1613$-$2124 and fainter than 
VISTA\,J1611$-$2215 in $H$ and $K$. Its overall SED is intermediate between VISTA J1613$-$2124 and
VISTA\,J1609$-$2222 and exhibits features of youth such as VO absorption at 1.06 micron and the peaked $H$-band.
\item [$\bullet$] VISTA\,J1602$-$2057 ($J$\,=\,18.50$\pm$0.03, $J-K$\,=\,2.12$\pm$0.04)
is difficult to classify: it shows much more straight
$J$-band spectrum with weak VO absorption than other objects in our sample. It also has a flat $H$-band
with weak water bands, and a more depressed red part of the $K$-band. However, it is almost as bright
as VISTA\,J1611$-$2215 in $H$ and $K$. Nonetheless, we cast doubts on its membership.
\item [$\bullet$] VISTA\,J1607$-$2146 ($J$\,=\,18.04$\pm$0.02, $J-K$\,=\,1.78$\pm$0.03)
is identical to VISTA\,J1614$-$2211 ($J$\,=\,18.25$\pm$0.02, $J-K$\,=\,1.84$\pm$0.04)
and VISTA\,J1605$-$2403 ($J$\,=\,18.32$\pm$0.03, $J-K$\,=\,1.94$\pm$0.04). 
They are similar to VISTA\,J1609$-$2229 in $J$ but appear
slightly brighter in $H$ and have similar flux level in $K$. We note that VISTA\,J1607$-$2146
shows an excess in the 1.27--1.32\,$\mu$m and is the reddest in our sample at these wavelengths.
We also note that VISTA\,J1605$-$2403 displays extra flux in the 2.08--2.015\,$\mu$m region
which might be due to a problem with the merging of the X-shooter orders.
\item [$\bullet$] VISTA\,J1604$-$2241 ($J$\,=\,18.17$\pm$0.03, $J-K$\,=\,1.77$\pm$0.04)
and VISTA\,J1604$-$2134 ($J$\,=\,18.41$\pm$0.03, $J-K$\,=\,1.90$\pm$0.04) are identical.
They are similar to VISTA\,J1607$-$2146 with shallower VO absorption band at 1.06\,$\mu$m,
similar flux in $H$ and slightly more flux in $K$. They are relatively brighter than VISTA\,J1611$-$2215
in $H$ and $K$, hence later than both VISTA\,J1607$-$2146 and VISTA\,J1611$-$2215\@.
\item [$\bullet$] VISTA\,J1615$-$2229 ($J$\,=\,19.34$\pm$0.06, $J-K$\,=\,2.17$\pm$0.08)
and VISTA\,J1601$-$2212 ($J$\,=\,19.27$\pm$0.07, $J-K$\,=\,2.22$\pm$0.08) are very similar.
They are the reddest objects in our sample and shows the strongest water absorption bands.
The VO absorption at 1.06$\mu$m is weaker than earlier sources, but the disappearance of 
this feature might be the result of lower temperature because both of them show clear features
characteristics of youth (peaked $H$-band, weak gravity-sensitive doublets).
\end{itemize}
%

%
%%%%%%%%%%%%%%%%%%%%%%%%%%%%%%%%%%%%%%%%%%%%%%%
%%%%% Figure: Magnitudes vs SpTypes %%%%%
%%%%%%%%%%%%%%%%%%%%%%%%%%%%%%%%%%%%%%%%%%%%%%%
%
% This plot was created with the IDL program spectra_Xshooter_USco_Deep_VISTA.pro
%
\begin{figure*}
  \centering
  \includegraphics[width=0.49\linewidth, angle=0]{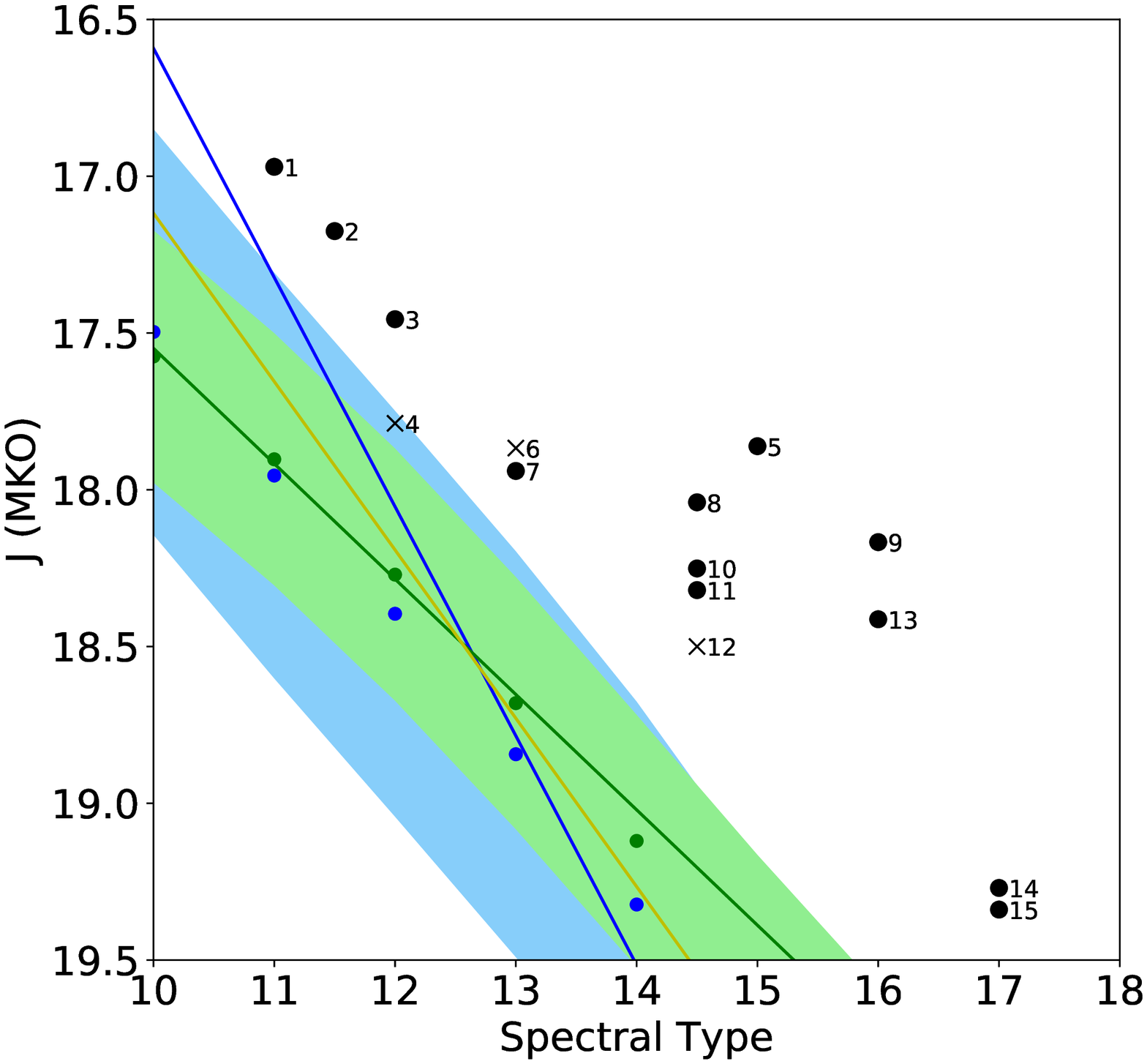}
  \includegraphics[width=0.49\linewidth, angle=0]{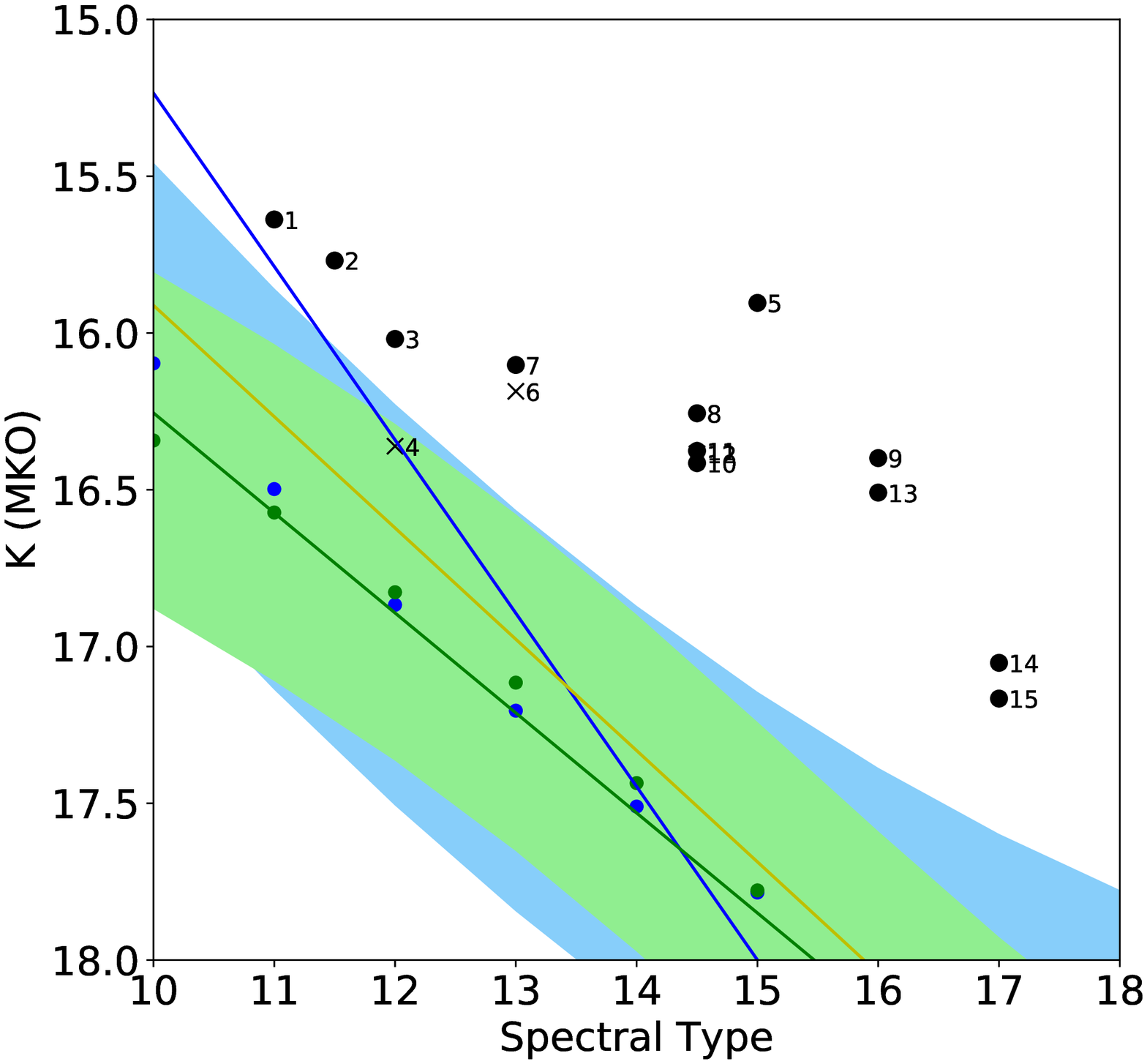}
  \includegraphics[width=0.49\linewidth, angle=0]{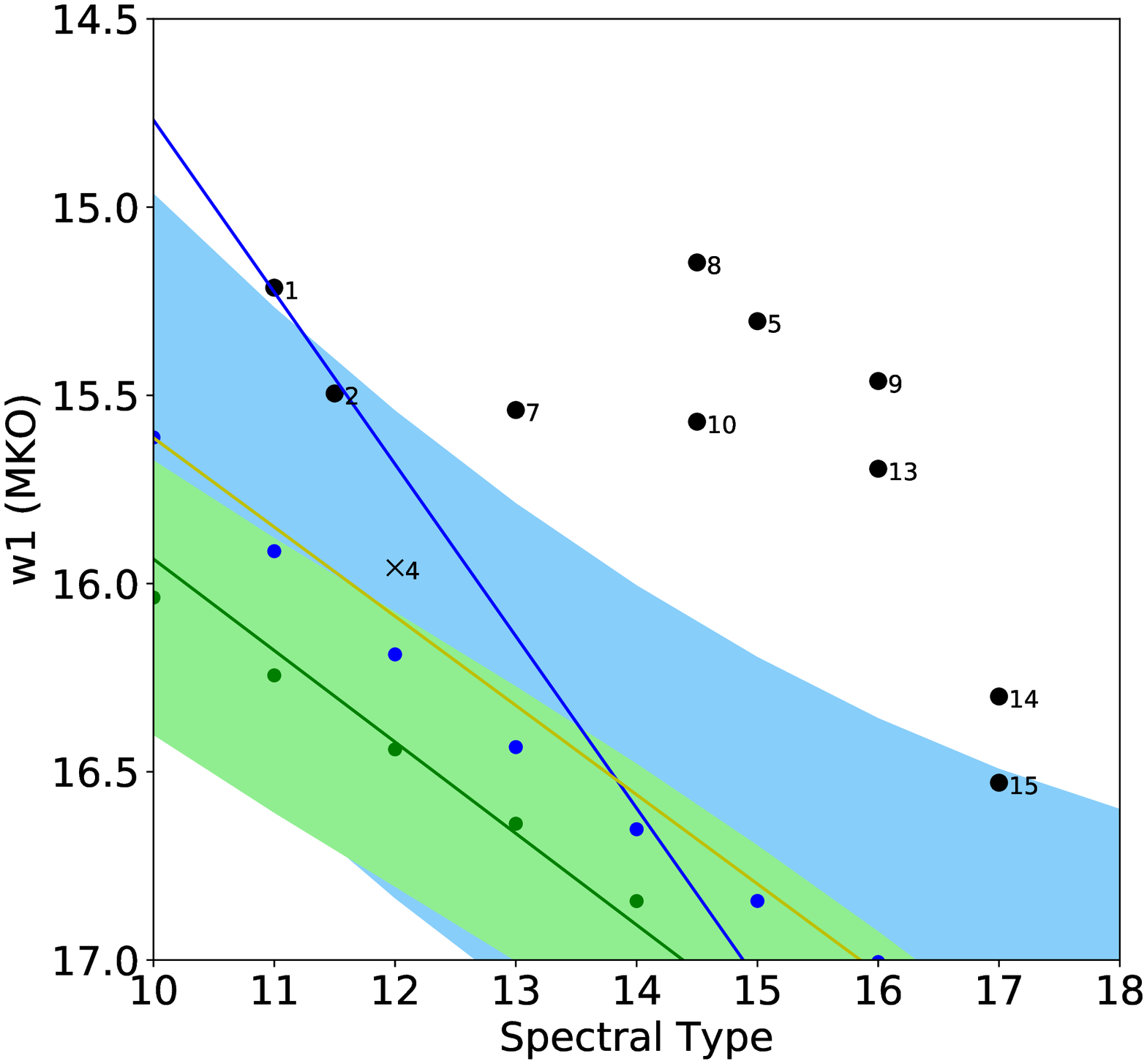}
  \includegraphics[width=0.49\linewidth, angle=0]{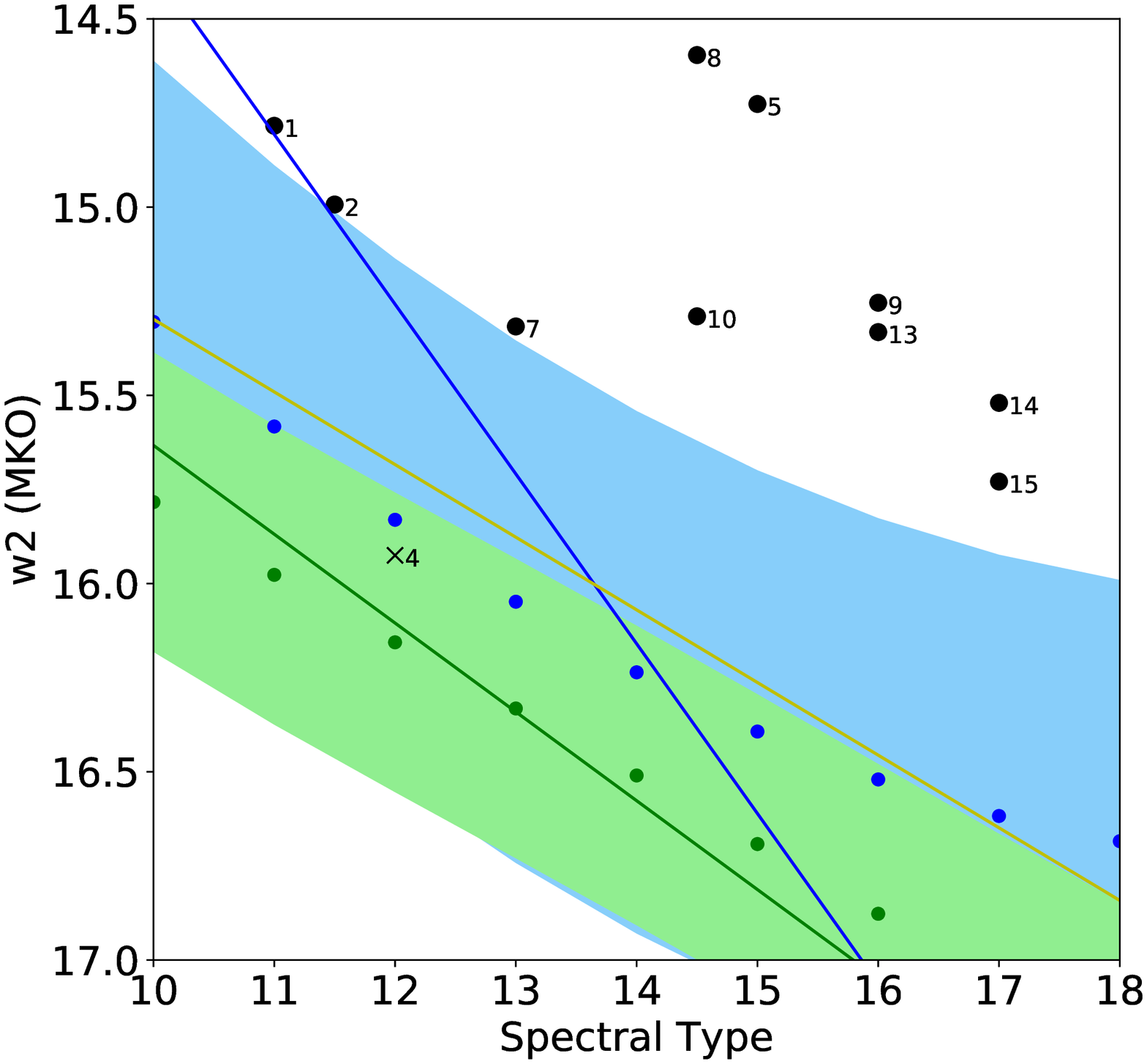}
  \caption{Observed $J$, $K$, $w1$, $w2$ magnitudes vs NIR spectral type for all UpSco candidates
with X-shooter spectra black dots). Our targets are labelled with numbers following the order listed
in Table \ref{tab_USco_GTC_XSH:table_WISE_phot}.
Photometric error bars are plotted but usually smaller than the symbol size for $J$ and $K$ magnitudes.
We overplotted the polynomial of the sequences of young and field L dwarfs from \citet{faherty16a} with
blue and green dots with their dispersion (light blue and light green). We also added the polynomial fits
for the field, INT-G and VLG samples of \citet{liu16a} in green, yellow, and blue, respectively.
The axis represents spectral types, where 10$\equiv$L0, 12$\equiv$L2, etc\ldots{}.
}
  \label{fig_USco_GTC_XSH:diagram_mag_SpT}
\end{figure*}
\subsubsection{Tentative absolute near-infrared classification}
\label{USco_GTC_XSH:spectral_classification_NIR_absolute}

In this section, we attempt to define a tentative spectral classification by comparison with the known
sequence of field L dwarf using templates downloaded from the SpeX 
archive\footnote{http://pono.ucsd.edu/$\sim$adam/browndwarfs/spexprism/index\_old.html}.

We also compared our spectra to the two known mid-L dwarf reported as photometric and
spectroscopic members in USco: 1RXS\,J1609291$-$210524 \citep[L4$^{+1}_{-2}$;][]{lafreniere08,lafreniere10a}
and USco\,J1551502$-$213457 \citep[L6$\pm$1;][]{penya16a} as well as a sample of young L dwarfs whose
spectra are available from J.\ Gagn\'e's webpage \citep{gagne15c}\footnote{https://docs.google.com/spreadsheets/d/1136rRdcjHZJoe00mt4kPeWU8ZNORcPe-w4cipkDKGRc/edit?pref=2\&pli=1\#gid=0} 
and in the Pleiades \citep{zapatero14b}. We should point out that 1RXS\,J1609291$-$210524 might 
harbour a disk \citep{wu15a} so we should be cautious using it as a spectral template. We detail 
our comparisons below going from the brightest to the faintest sources. The results are given in 
Table \ref{tab_USco_GTC_XSH:physical_parameters} and shown in Fig.\ \ref{fig_USco_GTC_XSH:spectra_XSH_compare}.

\begin{itemize}
\item [$\bullet$] VISTA\,J1559$-$2214 is the earliest member in our sample. Its overall shape is 
best reproduced by an early-L dwarf; the $J$-band is best fit by a L1 while the $H+K$ resemble
a L0\@. We assign a near-infrared spectral type of L1.0$\pm$0.5 to this source
(Fig.\ \ref{fig_USco_GTC_XSH:spectra_XSH_compare}), 
consistent with the overall SED of 2MASS\,J23255604$-$0259508 \citep[L1$\gamma$;][]{gagne15c}.
\item [$\bullet$] The spectra of VISTA\,J1613$-$2124 and VISTA\,J1614$-$2331 
look similar and are well reproduced by the SpeX spectrum of Kelu\,1
\citep[L2opt; L3ir\_pec][]{ruiz97,kirkpatrick99,stumpf08}. Therefore, we classify VISTA\,J1613$-$2124
as a L1.5 and VISTA\,J1614$-$2331 as L2 member of UpSco with an uncertainty of half a subclass.
\item [$\bullet$] The spectrum of VISTA\,J1605$-$2130 looks similar to the SpeX spectrum of Kelu\,1\@.
The absorption at 1.06 micron is almost absent and the $H$-band spectrum is not as peaky as for
the other candidates, casting doubt on its spectroscopic membership. We classify this object
as a L2 dwarf.
\item [$\bullet$] VISTA\,J1611$-$2215 is best fit by the SpeX spectrum of a field L6 dwarf
(Fig.\ \ref{fig_USco_GTC_XSH:spectra_XSH_compare}) although the water bands in $H$ and $K$ suggest
a slightly earlier spectral type. It is brighter in $H$ and $K$ than 1RXS\,J1609291$-$210524
\citep[L4$^{+1}_{-2}$;][]{lafreniere08,lafreniere10a} and shows an excess in the colour-magnitude
diagrams using WISE passbands (Section \ref{USco_GTC_XSH:WISE_midIR}; Fig.\ \ref{fig_USco_GTC_XSH:CMD_VISTA_WISE}) 
so we assign it a spectral type of L5 throughout the paper but keep in mind that its spectral
type is most likely L2exc$\pm$1 with an infrared excess.  
\item [$\bullet$] VISTA\,J1609$-$2222 is well fit by a field L4.5 dwarf from the SpeX archive.
as described earlier, its spectrum does not show features of youth, fact corroborated by the
comparison with a young L3$\gamma$ \citep{gagne15c}. Hence, we discard this source as a member
of UpSco and assign it a spectral type of dL3.0$\pm$0.5\@.
\item [$\bullet$] The spectrum of VISTA\,J1609$-$2229 is perfectly fit by the spectrum of
2MASS\,J01531463$-$6744181 \citep[L3$\gamma$;][]{reid08a,gagne15c}. The object shows clear
features of youth such as the VO absorption at 1.06\,$\mu$m and the peaked $H$-band so
we classify it as a L3.0$\pm$0.5 member of UpSco.
\item [$\bullet$] VISTA\,J1607$-$2146, VISTA\,J1614$-$2211, and VISTA\,J1605$-$2403 show a SED
similar to the SpeX L4.5 template and are comparable to the near-infrared spectrum of
1RXS\,J1609291$-$210524. Hence, we classify them as L4.5$\pm$0.5 members of UpSco.
We note that VISTA\,J1607$-$2146 exhibits a possible excess in $w2$ as discussed in
Section \ref{USco_GTC_XSH:phot_properties}.
\item [$\bullet$] VISTA\,J1602$-$2057 is difficult to classify and could not assign it a spectral
type because none of the field or young L dwarf template display a good fit. We do not detect
strong water bands in its infrared spectrum. We leave its classification open at this stage.
\item [$\bullet$] VISTA\,J1604$-$2134 and VISTA\,J1604$-$2241 are well fit by the spectrum of
USco\,J1551502$-$213457 \citep[L6$\pm$1;][]{penya16a}. They are redder than VISTA\,J1611$-$2215
that we classify as L5 and earlier than the two coolest sources (see next bullet). 
Their infrared spectra compare well with the SpeX L7 template, except for a stronger VO absorption
and peaked $H$-band feature. Therefore, we classify them as L6$\pm$1 member of UpSco.
\item [$\bullet$] VISTA\,J1601$-$2212 and VISTA\,J1615$-$2229 are the coolest members and appear
later than any other of our target. In the classification of field L dwarfs, the reddest sources
in $H$ and $K$ have spectral types of L7 and L7.5\@. Moreover, the 1.5--2.4\,$\mu$m region is
very similar to the SED of Calar 21 \citep[L7$\pm$1;][]{zapatero14b}. Both of them are fairly 
well fit by a field L7 spectrum, except for a stronger VO absorption, a peaked $H$-band feature, 
and more flux in $K$. Therefore, we assign to them the latest spectral type, L7$\pm$1\@.
\end{itemize}
%

%
%%%%%%%%%%%%%%%%%%%%%%%%%%%%%%%%%%%%%%%%%%%%%
%%%%% Table: SpT, EWs, RVs, Teff, grav %%%%%
%%%%%%%%%%%%%%%%%%%%%%%%%%%%%%%%%%%%%%%%%%%%%
%
\begin{table*}
 \centering
 \caption[]{Coordinates (J2000) of VISTA candidates for which we derive near-infrared spectral types 
(SpT$_{\rm NIR}$),
measured pseudo-equivalent widths (in \AA{}) of the Na{\small{I}} line at 1.138\,$\mu$m and two potassium 
doublets in the $J$-band (Fig.\ \ref{fig_USco_GTC_XSH:plot_EW_doublets}), radial velocities from the positions 
of the potassium doublets and via the cross-correlation technique (RV\_CC).
To infer the true radial velocity from the lines, we should subtract $-$9.05$\pm$4.83 km/s which is the 
radial velocity of the reference star  VISTA\,J1559$-$2214\@.
We note that 1611$-$2215 might be of earlier type with an infrared excess (L2.0exc).
 }
{\small
 \begin{tabular}{@{\hspace{0mm}}c @{\hspace{2mm}}c @{\hspace{2mm}}c @{\hspace{2mm}}c @{\hspace{2mm}}c @{\hspace{2mm}}c @{\hspace{2mm}}c @{\hspace{2mm}}c @{\hspace{2mm}}c @{\hspace{2mm}}c@{\hspace{0mm}}}
 \hline
 \hline
ID & Name        &     $J$ & SpT$_{\rm NIR}$ & EW$_{Na{\small{I}}}$ &  $\lambda_{K{\small{I}}}$ EW$_{K{\small{I}}}$ & $\lambda_{K{\small{I}}}$ EW$_{K{\small{I}}}$ & RV\_index & RV\_CC \cr
 \hline
   & VISTA\,J     &  mag &               &  \AA{} & (\AA{},\AA{}) (\AA{},\AA{})   & (\AA{},\AA{}) (\AA{},\AA{})     &  km/s & km/s \cr
\hline
 1 & 15593638$-$2214159 & 16.970 & L1.0$\pm$0.5 & 0.37 & (1243.11,1252.10) (0.18,0.23) & (1168.90,1177.05) (0.25,0.35) & $-$9.11$\pm$4.83 &     Reference     \cr % cand2
 2 & 16130231$-$2124285 & 17.175 & L1.5$\pm$0.5 & 0.33 & (1243.05,1252.03) (0.21,0.23) & (1168.93,1177.12) (0.27,0.48) &  $-$8.9$\pm$14.5 &  $-$3.7$\pm$1.3   \cr % cand55
 3 & 16142256$-$2331178 & 17.456 & L2.0$\pm$0.5 & 0.41 & (1243.12,1252.12) (0.16,0.22) & (1169.02,1177.13) (0.44,0.43) & $-$10.4$\pm$38.8 &    14.5$\pm$1.3   \cr % cand58
 4 & 16051705$-$2130449 & 17.788 & dL2.0$\pm$0.5 & 0.36 & (1243.17,1252.09) (0.27,0.26) & (1168.90,1177.17) (0.67,0.79) &  $-$0.2$\pm$17.7 &    17.2$\pm$2.6   \cr % cand19
 5 & 16114437$-$2215446 & 17.861 & L5.0$\pm$1.0 & 0.43 & (1243.03,1252.06) (0.14,0.16) & (1168.76,1177.05) (0.19,0.32) & $-$23.9$\pm$15.9 &     0.6$\pm$11.6  \cr % cand50
 6 & 16095636$-$2222457 & 17.867 & dL3$\pm$0.5  & 0.35 & (1243.15,1252.10) (0.67,0.66) & (1169.02,1177.16) (0.68,0.66) &     9.4$\pm$19.1 &    13.7$\pm$6.7   \cr % cand44
 7 & 16091868$-$2229239 & 17.940 & L3.0$\pm$0.5 & 0.25 & (1243.11,1252.08) (0.25,0.27) & (1168.94,1177.13) (0.30,0.36) &  $-$1.2$\pm$9.2  &     6.8$\pm$7.0   \cr % cand41
 8 & 16073161$-$2146544 & 18.040 & L4.5$\pm$0.5 & 0.24 & (1243.11,1252.02) (0.16,0.21) & (1168.88,1176.92) (0.30,0.37) &  $-$8.1$\pm$9.4  &     2.4$\pm$6.1   \cr % cand32
 9 & 16041304$-$2241034 & 18.167 & L6.0$\pm$1.0 & 0.31 & (1243.10,1252.17) (0.21,0.26) & (1168.89,1177.06) (0.14,0.40) &  $-$4.1$\pm$24.1 &     9.0$\pm$7.8   \cr % cand70
10 & 16140756$-$2211522 & 18.251 & L4.5$\pm$0.5 & 0.38 & (1243.05,1252.05) (0.14,0.20) & (1168.95,1177.00) (0.21,0.37) &  $-$4.5$\pm$15.1 &     1.4$\pm$5.8   \cr % cand71
11 & 16053909$-$2403328 & 18.320 & L4.5$\pm$0.5 & 0.43 & (1243.14,1252.08) (0.13,0.28) & (1169.03,1177.23) (0.27,0.41) &     6.6$\pm$24.5 &     3.8$\pm$7.9   \cr % cand73
12 & 16020000$-$2057341 & 18.500 &     ????     & 0.50 & (1242.97,1252.24) (0.32,0.37) & (1169.02,1176.88) (0.26,0.48) & $-$10.9$\pm$63.1 &     1.5$\pm$7.5   \cr % cand74
13 & 16042042$-$2134530 & 18.413 & L6.0$\pm$1.0 & 0.26 & (1243.06,1252.12) (0.19,0.21) & (1168.88,1177.19) (0.22,0.34) &  $-$1.9$\pm$9.6  &     5.5$\pm$6.9   \cr % cand75
14 & 16013692$-$2212027 & 19.270 & L7.0$\pm$1.0 & 0.15 & (1243.12,1252.13) (0.13,0.20) & (1168.89,1177.00) (0.33,0.35) &     2.3$\pm$7.2  &    13.4$\pm$10.4  \cr % cand96
15 & 16151270$-$2229492 & 19.339 & L7.0$\pm$1.0 &  --- & (1243.07,1252.10) (0.15,0.30) & (1168.75,1177.13) (0.40,0.40) &  $-$5.9$\pm$7.5  &    18.3$\pm$13.1  \cr % cand97
 \hline
 \label{tab_USco_GTC_XSH:physical_parameters}
 \end{tabular}
}
\end{table*}
%

%
%%%%%%%%%%%%%%%%%%%%%%%%%%%%%%%%%%%%%%%%%
%%%%% Table: Membership: summary %%%%%
%%%%%%%%%%%%%%%%%%%%%%%%%%%%%%%%%%%%%%%%%
%
\begin{table}
 \centering
 \caption[]{Summary of the membership of all candidates with near-infrared spectroscopy presented
in this paper. We give their ID plotted in several figures and names. We list several criteria: 
(1) photometric sequence, (2) near-infrared spectral types,
(3) strength of the VO absorption at 1.06\,$\mu$m, (4) presence of the peaked $H$-band, 
(5) pseudo-equivalent widths of gravity-sensitive features, (6) radial velocities. 
``Y'' stands for ``Yes, it is a member'', ``N''
for ``non-member'', and ``?'' for ``maybe''. We consider an object as a member if it satisfies
five or six of the criteria.
We note that 1611$-$2215 might be of earlier type with an infrared excess (L2.0exc).
}
 \begin{tabular}{@{\hspace{0mm}}c @{\hspace{2mm}}c @{\hspace{2mm}}c @{\hspace{2mm}}c @{\hspace{2mm}}c @{\hspace{1mm}}c @{\hspace{1mm}}c @{\hspace{1mm}}c @{\hspace{1mm}}c @{\hspace{1mm}}c@{\hspace{0mm}}}
 \hline
 \hline
ID & Name               & SpT$_{\rm NIR}$ & SpT$_{\rm OPT}$ & phot & SpT  & VO  & $H$-peak & EW  &  RV  \cr
 \hline
   & VISTA\,J            &      &    &     &     &     &     &     &     \cr
 \hline
 1 & 1559$-$2214 & L1.0$\pm$0.5 & M8.5$\pm$0.5 &  Y  &  Y  &  Y  &  Y  &  Y  &  Y  \cr % USco_cand2
 2 & 1613$-$2124 & L2.0$\pm$0.5 & M9.0$\pm$0.5 &  Y  &  Y  &  Y  &  Y  &  Y  &  Y  \cr % USco_cand55
 3 & 1614$-$2331 & L1.5$\pm$0.5 & M9.5$\pm$0.5 & Y  &  Y  &  Y  &  Y  &  Y  &  Y  \cr % USco_cand58
 4 & 1605$-$2130 & dL2.0$\pm$0.5 & L1--L2       & ?  &  Y  &  ?  &  N  &  ?  &  ?  \cr % USco_cand19
 5 & 1611$-$2215 & L5.0$\pm$1.0 & L2--L3       & Y  &  Y  &  Y  &  Y  &  Y  &  Y  \cr % USco_cand50
 6 & 1609$-$2222 & dL4.5$\pm$0.5 & L3          & Y  &  N  &  N  &  N  &  N  &  Y  \cr % USco_cand44
 7 & 1609$-$2229 & L3.0$\pm$0.5 & L2--L3       & Y  &  Y  &  Y  &  Y  &  Y  &  Y  \cr % USco_cand41
 8 & 1607$-$2146 & L4.5$\pm$0.5 & L2--L3       & Y  &  Y  &  Y  &  Y  &  Y  &  Y  \cr % USco_cand32
 9 & 1604$-$2241 & L6.0$\pm$1.0 & L1--L2       & Y  &  Y  &  Y  &  Y  &  Y  &  Y  \cr % USco_cand70
10 & 1614$-$2211 & L4.5$\pm$0.5 & L1--L2       & Y  &  Y  &  Y  &  Y  &  Y  &  Y  \cr % USco_cand71
11 & 1605$-$2403 & L4.5$\pm$0.5 & L2.0$\pm$0.5 & Y  &  Y  &  Y  &  Y  &  Y  &  Y  \cr % USco_cand73
12 & 1602$-$2057 & ????         & ????         & ?  &  ?  &  N  &  N  &  ?  &  Y  \cr % USco_cand74
13 & 1604$-$2134 & L6.0$\pm$1.0 & L1--L2       & Y  &  Y  &  Y  &  Y  &  Y  &  Y  \cr % USco_cand75
14 & 1601$-$2212 & L7.0$\pm$1.0 & L4.0$\pm$1.0 & Y  &  Y  &  N  &  Y  &  Y  &  Y  \cr % USco_cand96
15 & 1615$-$2229 & L7.0$\pm$1.0 & L4.0$\pm$1.0 & Y  &  Y  &  N  &  Y  &  Y  &  Y  \cr % USco_cand97
\hline
 \label{tab_USco_GTC_XSH:table_membership_summary}
 \end{tabular}
\end{table}

Overall, this absolute spectral classification agrees well with the relative classification
described in the previous section.
The spectral sequence progresses from L1--L2 for the brightest VISTA candidates, to mid-L 
for the group of sources around $J$\,$\sim$\,18 mag, to late-L for the two faintest objects 
with $J$\,$\sim$\,19.3 mag. We define the first sequence for young L dwarf members
of the UpSco association with an age of $\sim$5--10 Myr that should serve as ``benchmark'' for
comparison with other young systems discussed in the literature or to be discovered
in the near future.

\subsubsection{Optical spectral classification}
\label{USco_GTC_XSH:spectral_classification_opt}

We proceed in a similar manner as for the near-infrared classification to assign optical spectral
types to the targets with GTC/OSIRIS low-resolution spectra (Sect.\ \ref{USco_GTC_XSH:optical_LowRes}).
In Fig.\ \ref{fig_USco_GTC_XSH:plot_spec_optical_LowRes} we plot the optical spectra of all sources 
classified as L dwarfs in the near-infrared and observed in June 2016 (left panel) and May 2017
(right panel). 

From Figs.\ \ref{fig_USco_GTC_XSH:plot_spec_optical_LowRes} and \ref{fig_USco_GTC_XSH:plot_spec_optical_LowRes_Ha}, we see H$\alpha$ in emission in four L-type objects: VISTA\,J1615$-$2229, VISTA\,J1607$-$2146, VISTA\,J1602$-$2057, and VISTA\,J1611$-$2215
(Table \ref{tab_USco_GTC_XSH:logs_spectro_GTC}). The line strength is moderate in 
VISTA\,J1615$-$2229, and this object does not show evidence of mid-infrared flux excesses up to 
4.5\,$\mu$m. Therefore, the origin of the H$\alpha$ emission could be chromospheric. 
Chromospheric emission is expected in young M and L dwarfs \citep[see review by][]{luhman12b}. 
VISTA\,J1607$-$2146 also has a moderate H$\alpha$ emission and significant mid-infrared flux excess 
at 4.5\,$\mu$m (see object \#8 in Figs. \ref{fig_USco_GTC_XSH:CMD_VISTA_WISE} and \ref{fig_USco_GTC_XSH:CCD_VISTA_WISE}), which implies the presence of a warm disk and 
possible accretion. VISTA\,J1602$-$2057 and VISTA\,J1611$-$2215 show, however, considerably 
strong H$\alpha$ emission although it is not  resolved at the resolution of our data. Both objects 
could be accreting material and their optical spectra may be affected by some veiling
(Fig.\ \ref{fig_USco_GTC_XSH:plot_spec_optical_LowRes_Ha}), thus contributing to unreliable determinations of spectral types at optical wavelengths. This scenario is reinforced by the presence 
of possible mid-infrared flux excesses in VISTA\,J1611$-$2215 (object \#5 in 
Fig.\ \ref{fig_USco_GTC_XSH:diagram_mag_SpT}), which suggests the existence of a disk from which 
the central source is still accreting even at the age of UpSco (5--10 Myr). As discussed in the review by \citet{luhman12b}, it seems that disks manage to survive at ages of $\sim$10 Myr around very low 
mass sources, see also \citet{luhman12c}. VISTA\,J1607$-$2146 and VISTA\,J1611$-$2215, whose 
masses are likely below 15 M$_{\rm Jup}$, prove that both disks and accretion phenomena can 
persist up to $\sim$10 Myr, unless these sources are much younger than the cluster (which is possible 
only for VISTA\,J1611$-$2215 given the fact that it is brighter than all other related objects at all magnitudes). VISTA\,J1607$-$2146 and VISTA\,J1611$-$2215 should be added to the limited number 
of planetary-mass objects hoarbouring disks, e.g.\ SOri\,60 \citep{zapatero07b,luhman08c} or 
2MASS\,J04414489$+$2301513 \citep{adame11}.

In Fig.\ \ref{fig_USco_GTC_XSH:plot_spec_optical_LowRes} we overplotted in green Sloan optical
spectra of old field L dwarfs that best match the spectra of targets to assign tentative spectral types.
We classify our targets into several groups, best fit by L1--L2, L2--L3, and L4 in the optical 
by direct comparison with Sloan spectra. We classify the two faintest targets 
in our sample as L4opt$\pm$1 (compared L7ir). Consequently, we find a trend towards earlier 
spectral types in the optical than in the near-infrared marked by the presence of TiO and VO
at optical wavelengths for sources classified as early-L dwarfs in the near-infrared.
We find an optical sequence going from late-M dwarfs to early-L, and mid-L dwarfs that matches
the photometric sequence and near-infrared spectral sequence in UpSco where brightest sources 
exhibit earlier spectral types.
We overplotted in red two known young members of Taurus: KPNO6 (M8.5) and KPNO\,4 (M9.5) 
\citep{briceno98,luhman03a} which reproduce well the optical spectra of the late-M UpSco members,
thus, reinforcing our classification scheme and the young nature of these UpSco candidates.

Puzzled by this systematic trend, we increased the sample of optical spectra of UpSco members classified 
as M9--L1 in the near-infrared \citep[Gemini/GNIRS spectra in Table 1 and Fig.\ 1 of][]{lodieu08a}. 
We plot these optical spectra
in Fig.\ \ref{fig_USco_GTC_XSH:plot_spec_optical_backup}, where the earliest is at the bottom and
quote both optical and near-infrared spectral types next to their names. We compare them to the two 
young sources discussed above and UScoCTIO\,108B to assign optical spectral types between M8.5 and M9.5\@.
We observe a small spectral type interval at optical than infrared wavelengths (M8.5--M9.5 vs M9--L1),
which most likely reflects the fact that oxides remain stronger at low gravities due to the low 
pressures although VO and TiO have been severely depleted in field early-L dwarfs.
We clearly detect H$_{\alpha}$ in emission in three of these UpSco members 
(Table \ref{tab_USco_GTC_XSH:logs_spectro_GTC}).

\subsubsection{Adopted spectral classification}
\label{USco_GTC_XSH:spectral_classification_adopted}

We have decided to assign both optical and near-infrared spectral types to our targets. Nonetheless,
we have adopted the near-infrared spectral types based on three arguments:

\begin{itemize}
\item A significant fraction of the flux of late-M and L dwarfs is emitted at near-infrared wavelengths. 
The optical flux up to 1.0 $\mu$m barely represents $<$10\,\% of the total flux. Both optical and near-infrared 
wavelengths have spectral features that depend on temperature, gravity, and metallicity. 
Figure \ref{fig_USco_GTC_XSH:spectra_XSH_compare} shows the comparison of the combined optical and near-infrared 
spectral energy distributions (SEDs) 
of UpSco dwarfs against SEDs of spectral types based on the optical and near-infrared classifications. 
The overall SED of UpSco dwarfs is better matched by the SEDs corresponding to the near-infrared classification.

\item The TiO and VO bands, which are the basis of optical spectral scheme, depend strongly on gravity, 
as we can see from the presence of VO at 1.06 microns 
(Fig.\ \ref{fig_USco_GTC_XSH:plot_spectra_GTCplusXSH}).
Nonetheless, the optical-to-infrared slope of the spectra is best reproduced by the slope of the near-infrared 
spectral types.
\item The UpSco spectral sequence spans $\sim$2.4 mag in $J$ (17.0--19.3 mag), which would correspond to 
field L dwarfs with spectral types in the L0--L6 range, in agreement with our near-infrared spectral
range but larger than our optical spectral range.
\end{itemize}

We do not share the opinion of constructing a spectral classification for young L dwarfs based on 
optical spectral types ignoring the differences between visible and infrared spectra as described in 
\citet{luhman16c}. This is a valid proposal for field L dwarfs but we argue that the low gravity 
affecting the sodium and potassium doublets
in the infrared should have an even stronger effect on the optical spectral energy distribution of young
L dwarfs than on the infrared since the optical region is mainly shaped by alkalis.
The redness of some L dwarf members of young moving groups with intermediate ages might originate 
from the presence of dust and/or clouds in their atmospheres or debris disk \citep{zakhozhay17a}.

\subsubsection{Photometric vs spectroscopic sequence}
\label{USco_GTC_XSH:spectral_classification_phot_vs_spec}

We have shown that our targets define a photometric sequence in colour-magnitude and colour-colour 
diagrams and span a $\sim$2.5 magnitude range, from $J$\,$\sim$\,17.0 mag down to $J$\,=\,19.3 mag. 
They also define a spectroscopic sequence, from L1 to L7 in the near-infrared and from M8.5 to L4 in 
the optical. We naturally expect such an agreement if our candidates are members of the same region
although additional parameters can enter into the game such as binarity and presence of disks at the 
age of UpSco ($\sim$5--10 Myr).

In Table \ref{tab_USco_GTC_XSH:table_membership_summary} we summarise the various criteria used
in this work to assign membership to the 15 UpSco candidates with near-infrared spectral types.
In  Fig.\ \ref{fig_USco_GTC_XSH:diagram_mag_SpT} we plot observed magnitudes ($J$, $K$, $w1$, $w2$) 
as a function of near-infrared spectral types to put together the photometric and spectroscopic sequences.
We see that the two objects (VISTA\,J1609$-$2222, \#6; VISTA\,J1602$-$2057, \#12) classified as 
non-members do not stand out in these diagrams.
However, VISTA\,J1605$-$2130 (\#4) lies well below the sequences in all diagrams and its near-infrared
spectrum did not show clear presence of VO absorption and presence of a peaked $H$-band
(Section \ref{USco_GTC_XSH:spectral_classification_NIR_absolute}) suggesting that it is a contaminant.
Object number 5 (VISTA\,J1611$-$2215) tends to lie above the sequence due to its late type compared 
to the photometric sequence but we cannot discard it as earlier type member with an infrared 
excess (L2exc vs normal L5) at this stage because all other features typical of young objects are present.

\subsection{Spectral indices}
\label{USco_GTC_XSH:spectral_indices}

We have compiled a large number of spectral indices defined in the literature to classify M, L, and
T dwarfs in the field and at young ages by independent teams 
\citep{tokunaga99,reid01a,testi01,burgasser02,geballe02,mclean03,slesnick04,burgasser06a,allers07,allers13}.
We refer the reader to these papers for the definitions of the indices.
We computed these spectral indices for a variety of L dwarfs, including the SDSS sequence spanning
optical spectral types between L0 and L8 \citep{schmidt10b}, L dwarfs downloaded from Sandy Leggett's
webpage\footnote{http://staff.gemini.edu/~sleggett/LTdata.html}, those in \citet{reid01a},
and young L0--L5 dwarfs classified as $\beta$ and $\gamma$ following \citet{cruz09}, \citet{bonnefoy14a}, 
\citet{allers13}, and 
\citet{gagne15c}\footnote{https://docs.google.com/spreadsheets/d/ \newline 1136rRdcjHZJoe00mt4kPeWU8ZNORcPe-w4cipkDKGRc/}
plus the recent additions of PSO\,J318.5338$-$22.8603 \citep[L7pec;][]{liu13a} and 
VHS\,J125601.92$-$125723.9 \citep[L7pec;][]{gauza15a}.
We report the values of the spectral indices for all sources in Table \ref{tab_USco_GTC_XSH:table_spec_indices} 
and plot them as a function of spectral type in 
Figs \ref{fig_USco_GTC_XSH:plot_spectral_indices_1}--\ref{fig_USco_GTC_XSH:plot_spectral_indices_4}
in Appendix \ref{app_USco_GTC_XSH:spectral_indices}.
In those figures, we plot the infrared spectral types of our UpSco sources and the spectral types
published in the literature for the comparison samples of old field, beta, and gamma L dwarfs.
We note that the spectral types come from various sources in the literature, which does not
necesarily represent a one-to-one correlation with effective temperature.
We distinguish field L dwarfs from their younger counterparts by the colours of the symbols: green for
old fied dwarfs, blue triangles for $\beta$ L dwarfs, red pentagons for $\gamma$ L dwarfs, and black dots
for our UpSco sample.
We plot all the spectral indices along with the 2$^{nd}$-order polynomial fits in the figures in
Appendix \ref{app_USco_GTC_XSH:spectral_indices}. We plot a sub-sample of three indices discussed
below in Fig.\ \ref{fig_USco_GTC_XSH:plot_spectral_indices_SpT}.

We carried out a Kolmogorov-Smirnov statistical test to identify the spectral indices most sensitive
to gravity. We compared the distributions of our UpSco objects with old field L dwarfs for each index
individually. After ordering them, we identified the following indices as most sensitive to age (in this
order): H-cont, CH4-H, FeHH, VOz, and H20-K\@. We repeated this procedure comparing the $\gamma$ L dwarfs
and the old field L dwarfs (both samples from the literature) and identified four of the indices
aforementionned among the five best ones (H-cont, CH4-H, VOz, and H20-K). We inspected visually these
indices and observed indeed strong dependency on gravity for those indices. We conclude that the
H-cont, VOz, CH4-H, H2O-K, and FeH-H can be primarily used to the youth of members of moving groups
\citep[e.g.][]{gagne14a,malo14b,gagne15c,faherty16a}.
On the contrary, we identified some spectral indices that appear to be nearly independent of the gravity 
(age) of L dwarfs despite the scatter observed in the measurements. The K1 index, which measures
the rise of the spectra between 2.0 and 2.14 $\mu$m \citep*{tokunaga99}, is convenient for spectral
classification independently of gravity.

We observe that most of the water vapour indices defined in the literature are fairly insensitive to gravity,
with a few exceptions like H2O-K
and tend to either increase or decrease with later spectral types, demonstrating that the community has
used water bands as the main features for L dwarf classification. Moreover, we observe significant
differences between our sample and objects in the literature which we attribute to the lower 
signal-to-noise ratios in the NIR spectra of our UpSco targets (especially close to the main telluric
bands) compared to other studies using high quality spectra to derive those indices. 
Summarizing, the H-cont, VOz, CH4-H, and FeH-H spectral indices can be used to discriminate
surface gravity for late-M and L spectral types, while water-based indices are proxies for spectral
typing, with the only exception of H2O-K\@.

%
%%%%%%%%%%%%%%%%%%%%%%%%%%%%%%%%%%%%%%%%%%%%%%%%%%
%%%%% Figure: EW gravity-sensitive doublets %%%%%
%%%%%%%%%%%%%%%%%%%%%%%%%%%%%%%%%%%%%%%%%%%%%%%%%%
%
\begin{figure*}
  \centering
  \includegraphics[width=0.90\linewidth, angle=0]{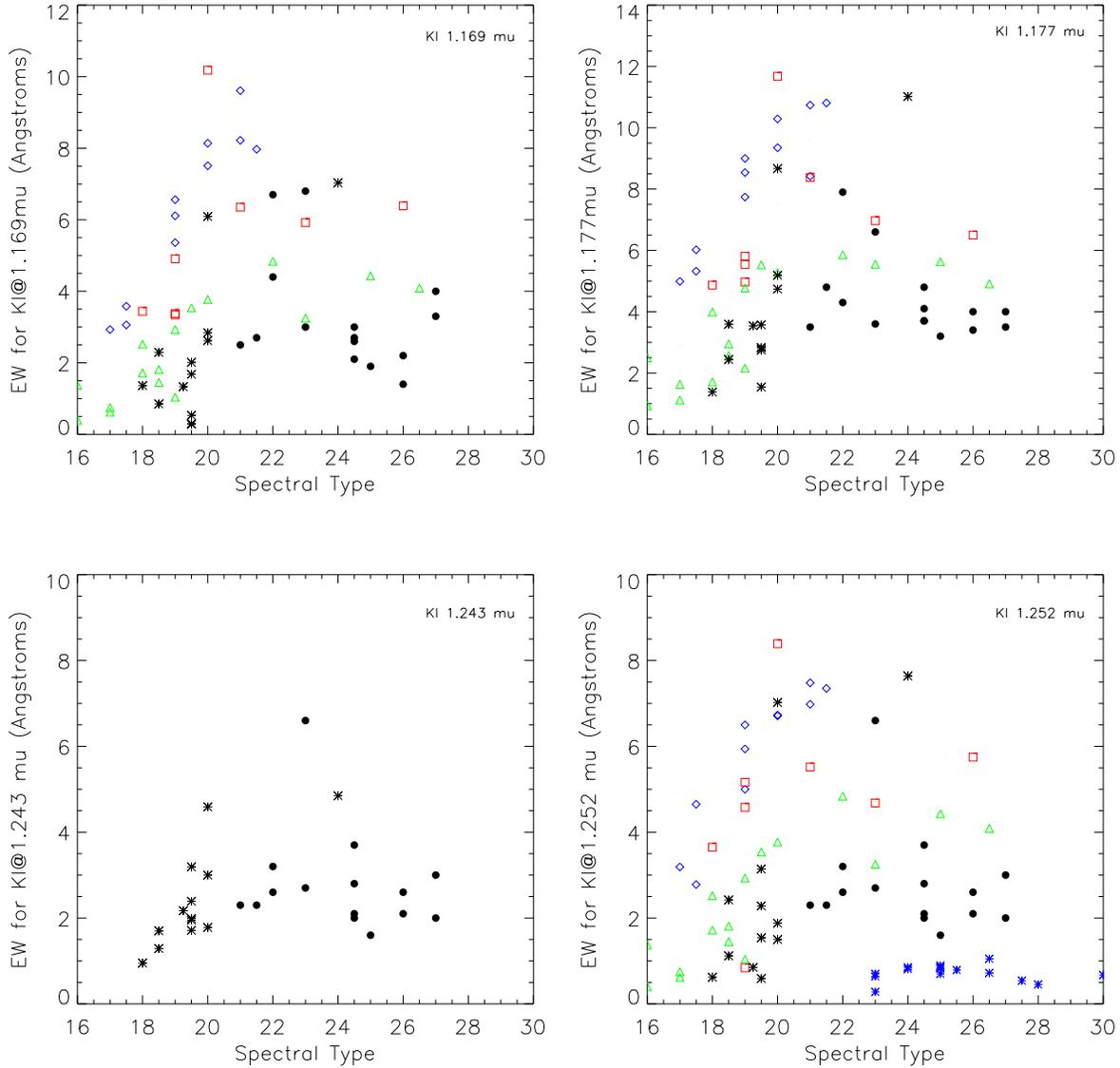}
  \caption{Pseudo-equivalent widths of four gravity-sensitive lines/doublets as a
function of NIR spectral types (20$\equiv$L0, 25$\equiv$L5, etc\ldots{}) for our USco
targets (filled dots) compared with field dwarfs (blue diamonds), low gravity dwarfs
(red squares), very low gravity sources \citep[green triangles;][]{allers13} and other
young objects from the literature \citep[black asterisks;][]{bonnefoy14a,manjavacas14}.
}
  \label{fig_USco_GTC_XSH:plot_EW_doublets}
\end{figure*}
%

%
%%%%%%%%%%%%%%%%%%%%%%%%%%%%%%%%%%%%%%%%%%%%%%%
%%%%% Figure: Spectral Indices vs SpType %%%%%
%%%%%%%%%%%%%%%%%%%%%%%%%%%%%%%%%%%%%%%%%%%%%%%
%
\begin{figure*}
  \centering
  \includegraphics[width=0.31\linewidth, angle=0]{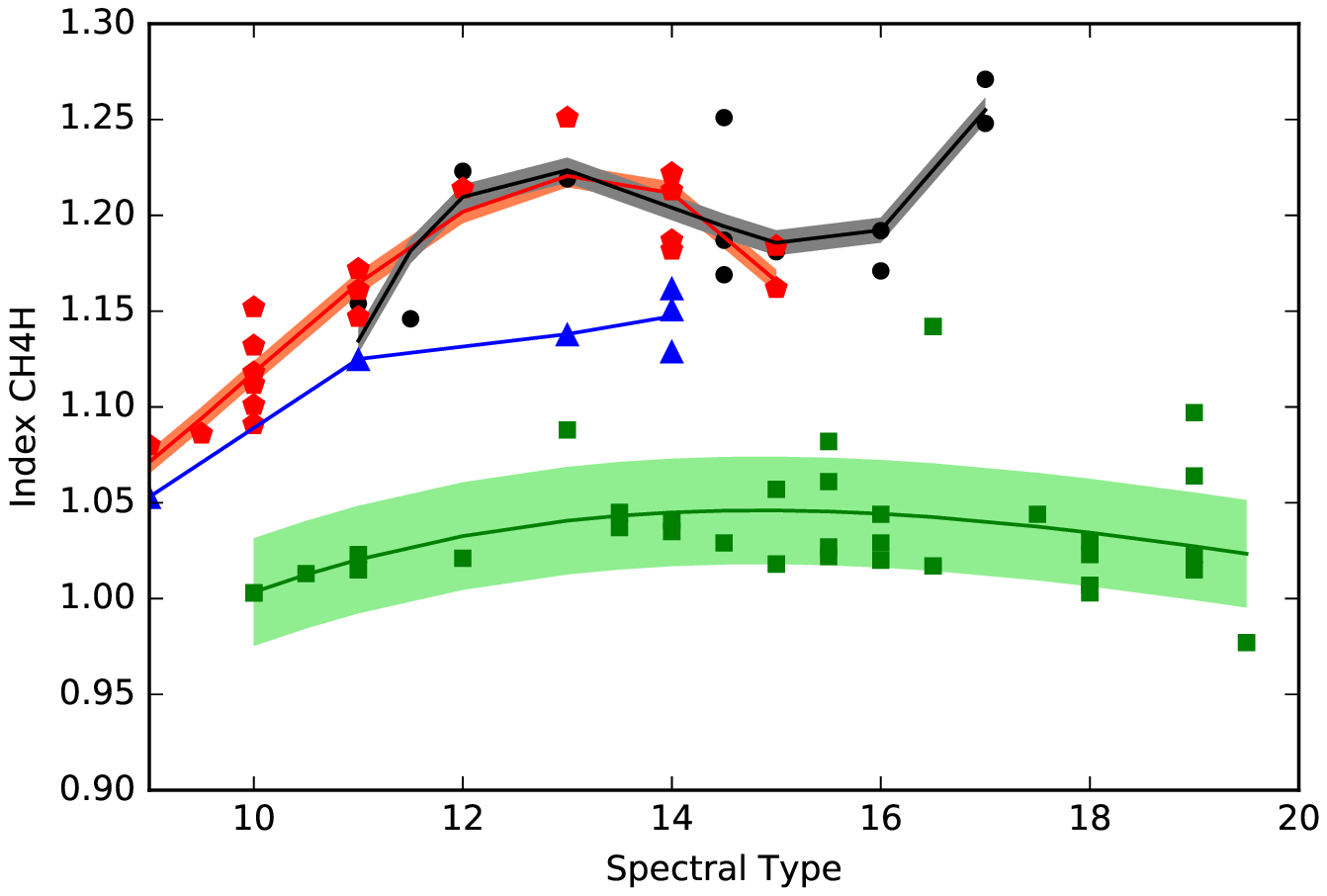}
  \includegraphics[width=0.31\linewidth, angle=0]{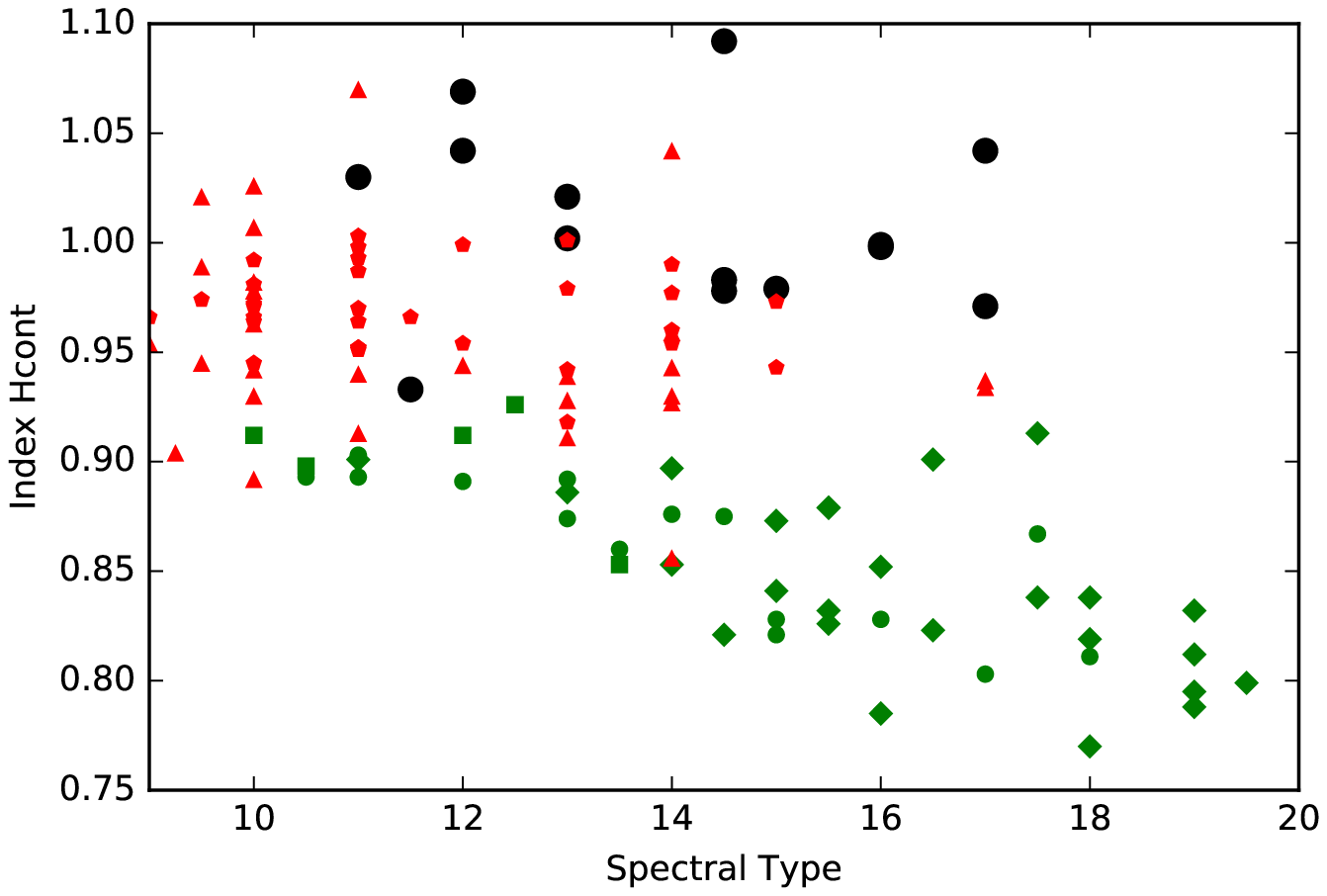}
  \includegraphics[width=0.31\linewidth, angle=0]{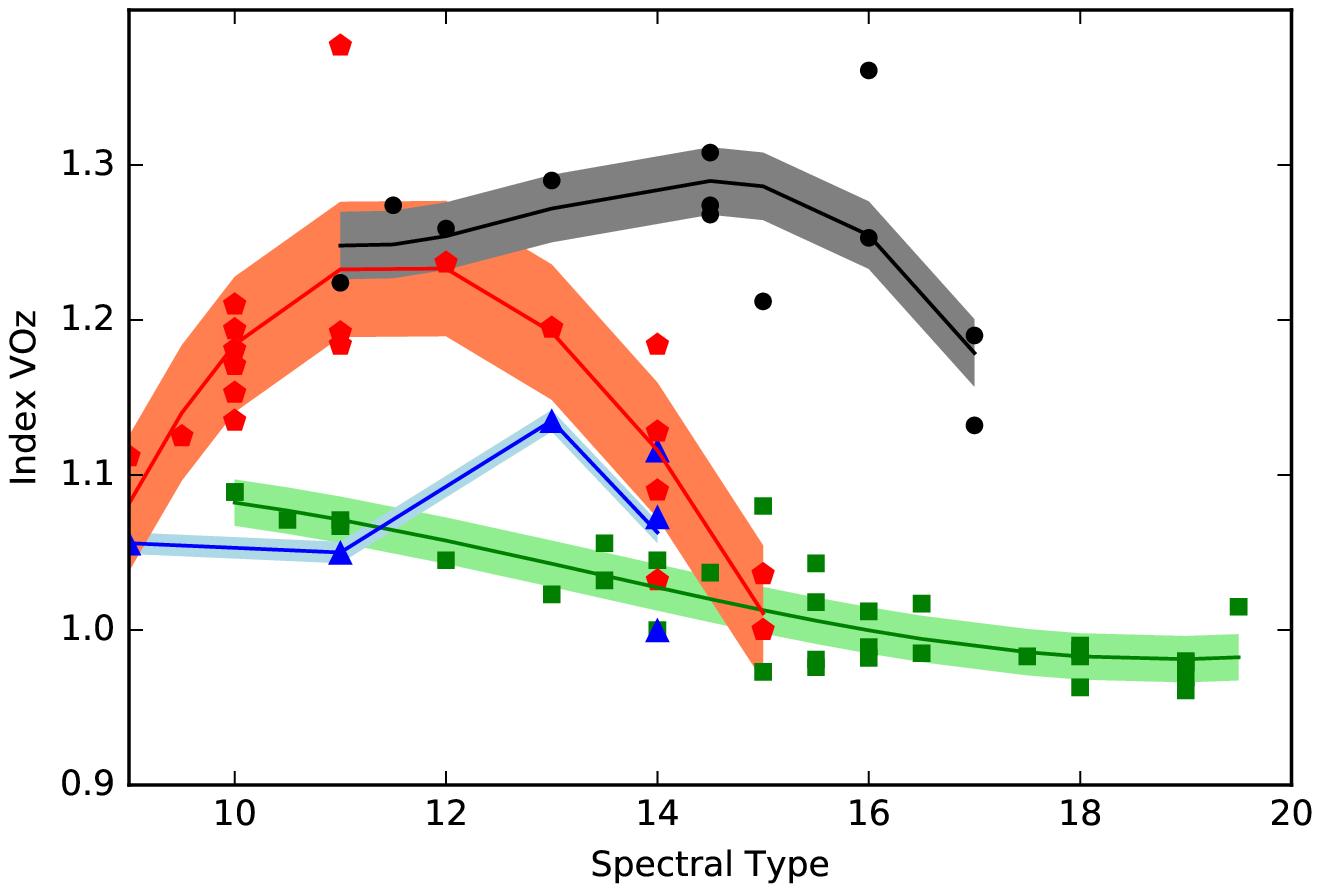}
  \caption{Near-infrared spectral indices defined in the literature vs.\ NIR
spectral types for our UpSco candidates (black dots), young $\beta$ (blue triangles) and $\gamma$
(red pentagons) L dwarfs \citep{allers13,bonnefoy14a,gagne15c}, and field L dwarfs with spectral types
taken from the literature \citep[green squares;][]{reid01a,geballe02,mclean03,knapp04,golimowski04a,chiu06}.
The CH4-H, H-cont, and VO indices are among the best indices to discriminate young and old field
L dwarfs based on our statistical test.
}
  \label{fig_USco_GTC_XSH:plot_spectral_indices_SpT}
\end{figure*}
%

%
%%%%%%%%%%%%%%%%%%%%%%%%%%%%%%%%%%%%%%%%%%%%%%%
%%%%% Figure: GTC+XSH spectra %%%%%
%%%%%%%%%%%%%%%%%%%%%%%%%%%%%%%%%%%%%%%%%%%%%%%
%
\begin{figure*}
  \centering
  \includegraphics[width=0.97\linewidth, angle=0]{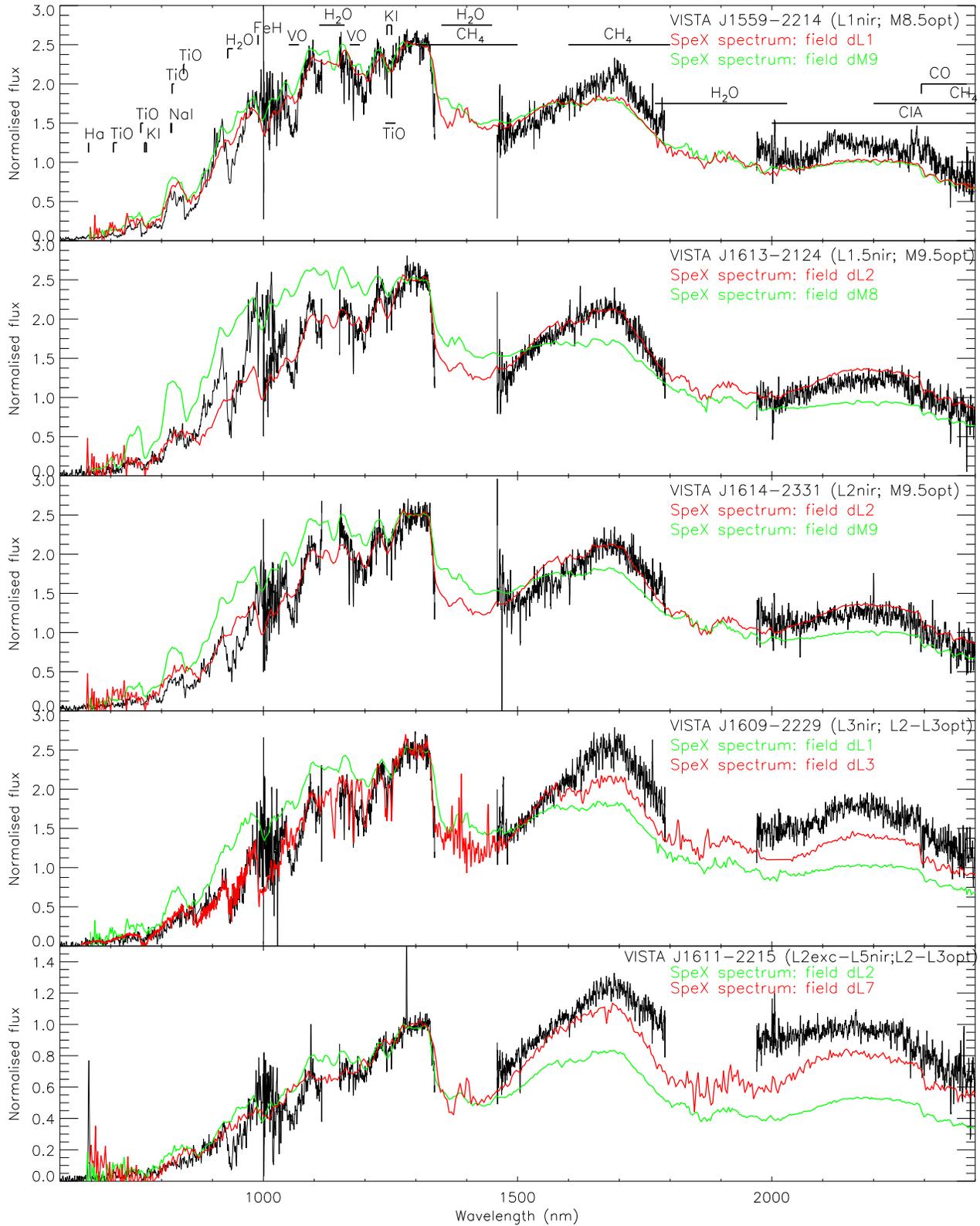}
  \caption{Optical (black) plus near-infrared (red; smoothed by a factor of 21) spectral energy
distribution for UpSco targets with spectroscopic data from GTC/OSIRIS and VLT/X-shooter, respectively:
From top to bottom: VISTA\,J1559$-$2214 (M8.5opt, L1ir); VISTA\,J1613$-$2124 (M9opt, L1.5ir); 
VISTA\,J1614$-$2331 (M9.5opt, L2ir); VISTA\,J1609$-$2229 (L2--L3opt, L3ir), and
VISTA\,J1611$-$2215 (L2--L3opt, L2exc--L5ir).
Overplotted are SpeX spectra for spectral types corresponding to the infrared and optical
classifications of our targets, demonstrating that the overall spectral energy distribution
matches best the infrared classification.
All spectra are normalised at 1.30--1.32 microns.
}
  \label{fig_USco_GTC_XSH:plot_spectra_GTCplusXSH}
\end{figure*}

\begin{figure*}
  \centering
  \includegraphics[width=\linewidth, angle=0]{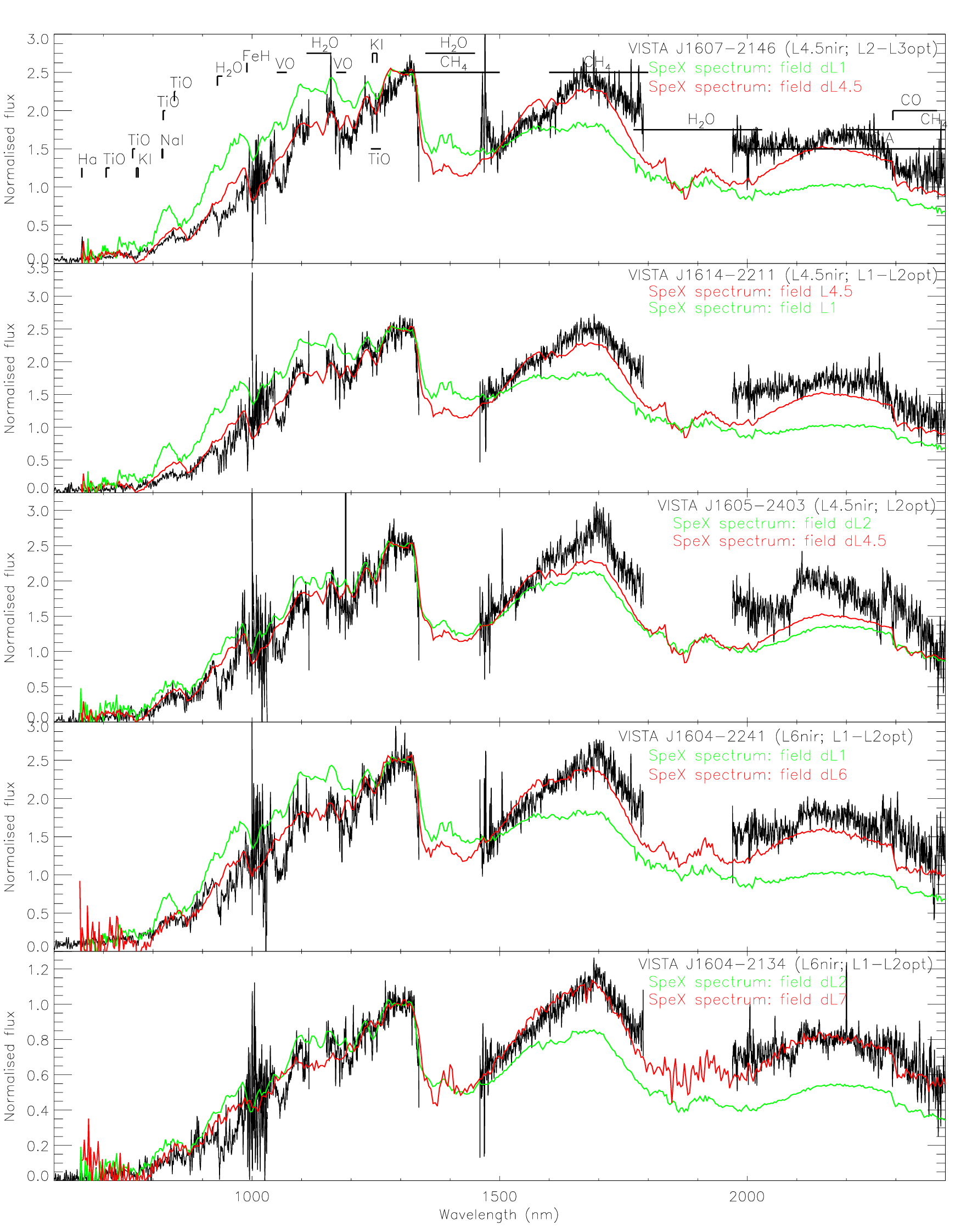}
  \contcaption{
From top to bottom: VISTA\,J1607$-$2146 (L2--L3opt, L4.5ir); VISTA\,J1614$-$2211 (L1--L2opt, L4.5ir);
VISTA\,J1605$-$2403 (L2opt, L4.5ir); VISTA\,J1604$-$2241 (L1--L2opt, L6ir); and
VISTA\,J1604$-$2134 (L1--L2opt, L6ir)}
\end{figure*}

\begin{figure*}
  \centering
  \includegraphics[width=\linewidth, angle=0]{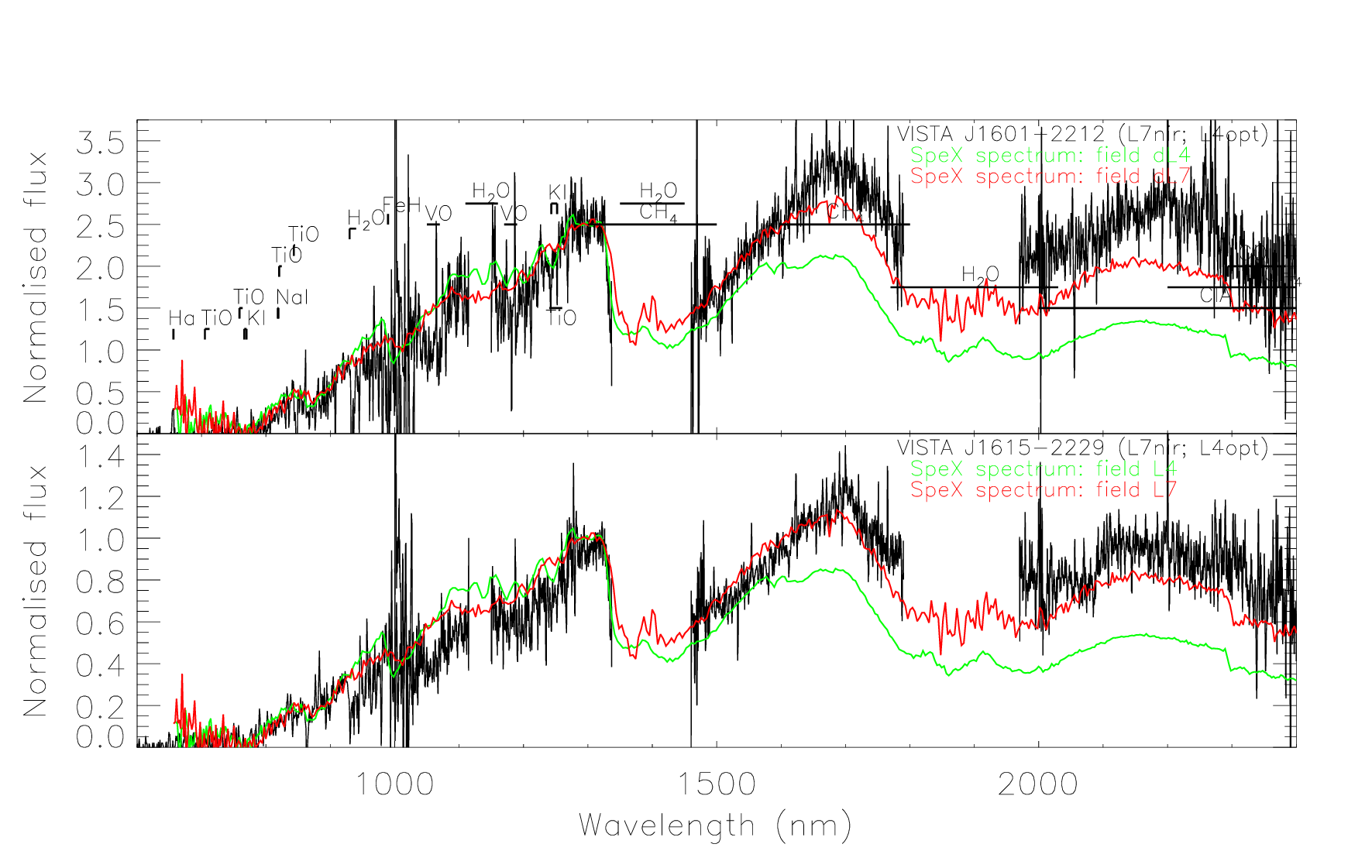}
  \contcaption{
From top to bottom, the two latest member candidates: VISTA\,J601$-$2212 (L4opt, L7ir) and 
VISTA\,J615$-$2229 (L4opt, L7ir).
}
\end{figure*}
\subsection{Gravity sensitive features}
\label{USco_GTC_XSH:spectral_features}

Brown dwarf members of open clusters, star-forming regions, and young moving groups are known 
to exhibit features characteristics of youth, including pseudo-equivalent widths of potassium 
and sodium doublets weaker than for older field dwarfs as well as the presence of a peaked $H$-band 
\citep{martin96,luhman98,bejar99,lucas00,zapatero00,cushing00,gorlova03,mcgovern04,slesnick04,lodieu08a,slesnick08,scholz12b,alves12,alves13a,bonnefoy14a,manjavacas14,muzic14,muzic15}. 
In Fig.\ \ref{fig_USco_GTC_XSH:plot_EW_doublets}, we plot the pseudo-equivalent widths of the 
potassium (K{\small{I}}) doublets at 1.169/1.177 $\mu$m and 1.243/1.252 $\mu$m
as a function of spectral type. We can clearly see that most of our measurements lie below the sequence 
of field M and L dwarfs (blue symbols) and are consistent with very low gravity dwarfs (green symbols) 
defined by \citet{allers13}. These objects would correspond to $\gamma$ L dwarfs in the scale of
\citet{cruz09}. Our sample of UpSco members is comparable in size to the sample of very low gravity 
sources in \citet{allers13} with the main difference that ours belong to the same region.
Only two objects have stronger gravity-sensitive features: 
VISTA\,J1609$-$2222 (\#6; dL4.5) shows stronger potassium features consistent with young L dwarfs or field
dwarfs. USco\,1605$-$2130 (\#4) exhibits stronger absorptions at 1.169/1.177 $\mu$m but not at
1.243/1.252 $\mu$m, suggesting that the doublet with the bluest wavelength might be more gravity-dependent
than the other which more often used in the literature.

We find that all X-shooter spectra exhibit weaker potassium (1.169/1.177\,$\mu$m, 1.243/1.252\,$\mu$m)
doublets, except two objects (VISTA\,J1605$-$2130, \#4; VISTA\,J1609$-$2222, \#6). 
We also find that all but three (VISTA\,J1605$-$2130, \#4; VISTA\,J1609$-$2222, \#6; 
VISTA\,J1602$-$2057, \#12) show a peaked $H$-band, adding strong 
credit to the membership of 12 of our VISTA candidates to the UpSco association 
(Fig.\ \ref{fig_USco_GTC_XSH:spectra_XSH}; Table \ref{tab_USco_GTC_XSH:physical_parameters}).
We also observe a weak sodium doublet at around 0.82\,$\mu$m in the GTC OSIRIS optical spectra
of the two brightest sources, conclusion further confirmed by the direct comparison with 
spectral template of similar spectral types from the Taurus and Chamaeleon star-forming regions 
kindly provided by Kevin Luhman (Fig.\ \ref{fig_USco_GTC_XSH:plot_spec_optical_LowRes}).
Despite the low resolution and poor signal-to-noise of the optical spectra of the other targets, 
the sodium doublets of the L2--L3opt dwarfs appear weaker than those of the SDSS template spectra 
whereas we cannot draw any definite conclusion for the two faintest sources.
We find the same features in the X-shooter spectra of known UpSco members \citep{luhman12c}
(Fig.\ \ref{fig_USco_GTC_XSH:spectra_XSH_archive}) and M9--L1 dwarfs from \citet{lodieu08a} 
observed with GTC/OSIRIS (Fig.\ \ref{fig_USco_GTC_XSH:plot_spec_optical_backup}).
These findings are corroborated by the spectral indices displayed in the bottom panels 
of Fig.\ \ref{fig_USco_GTC_XSH:plot_spectral_indices_SpT}.

The other striking feature present in the near-infrared spectrum of young dwarfs is the VO absorption
at 1.05--1.08\,$\mu$m \citep{cushing05}. \citet{allers13} defined a specific index centered on that
absorption to assign gravity classes of L0--L4 dwarfs (see Appendix \ref{app_USco_GTC_XSH:spectral_indices}
for the comparison with our sources). Indeed, at these spectral types, the lower the 
gravity, the stronger is the depression around 1.06\,$\mu$m due to condensation effects and hydride 
opacities. This feature is visible in the spectrum of USco\,J1551502$-$213457 \citep[L6$\pm$1;][]{penya16a}.
Two of the 15 spectra (VISTA\,J1609$-$2222 (\#6); VISTA\,J1602$-$2057 (\#12)) lack this feature
(Fig.\ \ref{fig_USco_GTC_XSH:spectra_XSH}) while a tiny absorption is seen in the spectrum of
USco\,1605$-$2130 (\#4), which together with other criteria, cast doubt on their youth, hence, 
membership to the UpSco association. The two faintest sources with the
latest spectral types show a weaker absorption at 1.06 microns but the comparison with field L7 dwarfs
shown in Fig.\ \ref{fig_USco_GTC_XSH:plot_spectra_GTCplusXSH} demonstrate that VO is still present in 
$\sim$5--10 Myr-old late-L dwarfs.

In Fig.\ \ref{fig_USco_GTC_XSH:plot_spectra_GTCplusXSH} we show the full spectral energy distribution 
combining GTC/OSIRIS optical and VLT/X-shooter infrared spectra of our L dwarf members of UpSco.
This figure shows the impact of gravity on the full 600--2500 nm region for L dwarfs, in particular
the remnants of alkali lines and oxyde bands. For this reason we have decided to assign both optical
and near-infrared spectral types to our targets.

\subsection{Radial velocities}
\label{USco_GTC_XSH:RVs}

We computed the radial velocities of our 15 VISTA candidates in two ways. On the one hand,
we used cross-correlation function using the spectrum of highest signal-to-noise as reference,
and, on the other hand, the positions of the individual lines of the gravity-sensitive doublets.

\begin{figure}
  \centering
  \includegraphics[width=\linewidth, angle=0]{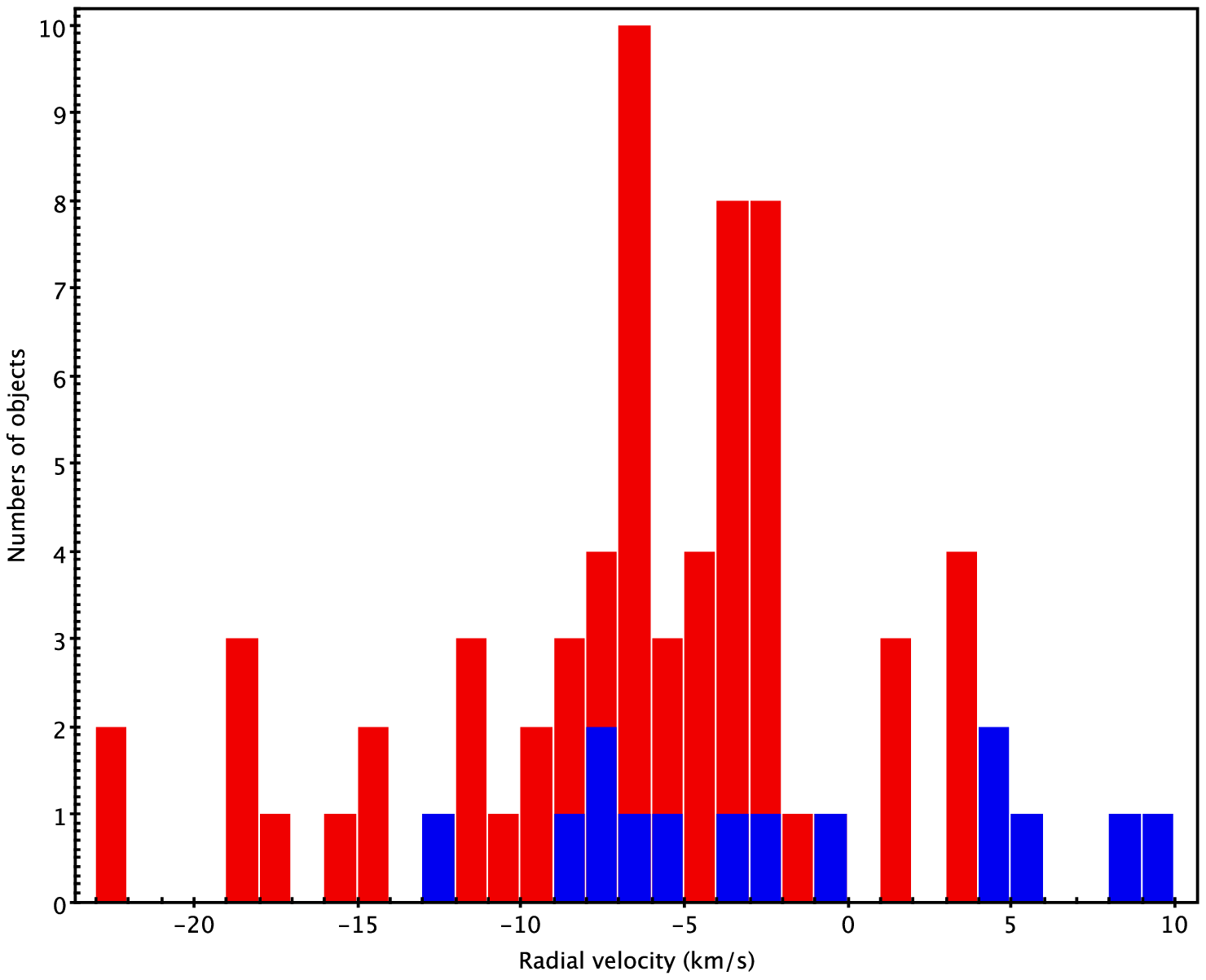}
  \caption{Histogram of our radial velocity measurements (blue) compared to observed 
radial velocities of high mass members of UpSco \citep[red; Table 12 in][]{pecaut12}.
}
  \label{fig_USco_GTC_XSH:plot_RVs}
\end{figure}

We used the {\tt{fxcor}} task under IRAF to measure the offsets between each candidate and
the reference chosen as VISTA\,J1559$-$2214, the brightest candidate ($J$\,=\,16.97 mag)
with the highest signal-to-noise X-shooter spectrum. We cleaned the spectra for the strongest
telluric bands and smoothed them by a factor of seven. We included the keywords necessary to correct
the relative Doppler shifts for the Earth barycentric motion. We considered four regions free of
telluric bands for the correlation, located between 1.02--1.10\,$\mu$m, 1.17--1.33\,$\mu$m, 1.49--1.77\,$\mu$m,
and 2.05--2.36\,$\mu$m. The relative radial velocities and their associated error bars (between 1.3 and
13.1 km/s typically increasing with lower signal-to-noise spectra) are quoted in 
Table \ref{tab_USco_GTC_XSH:physical_parameters}. The final radial velocities 
(Fig.\ \ref{fig_USco_GTC_XSH:plot_RVs}) are obtained by subtracting the value of $-$9.05$\pm$4.83 km/s 
obtained for VISTA\,J1559$-$2214 from the spectral lines (see next paragraph).
The true uncertainty on these measurements is set by the slit of 1.2 arcsec used in the NIR arm
of X-shooter, giving a resolution of $\sim$4000 (i.e.\ a spectral point every 0.06 nm) corresponding 
to an uncertainty in radial velocity of $\sim$15 km/s.

We also measured the observed positions of the potassium (K{\small{I}}) doublets at 1.169/1.177\,$\mu$m and
1.243/1.252\,$\mu$m. We compared these values to the nominal values taken from the NIST atomic spectra 
database\footnote{http://physics.nist.gov/PhysRefData/ASD/lines\_form.html}. We multiplied the
difference by the speed of light to derive a Doppler shift in km/s after correcting for the barycentric
velocities. We calculated the error bars
for each object based on the dispersion in the four different measurements (or three only when one
was clearly discrepant). The final error bars represent the quadratic sum of the individual measurements
and the uncertainty on the reference object (4.83 km/s). These uncertainties are much larger than
the ones from the cross-correlation technique but more in line with the spectral resolution of the
X-shooter spectra and the faintness of our objects. We list the final radial velocities in 
Table \ref{tab_USco_GTC_XSH:physical_parameters}. 

We find that both methods give compatible values within 1$\sigma$ of the error bars. We plot the Doppler 
shifts from the cross-correlation technique corrected for the radial velocity of our reference
(VISTA\,J1559$-$2214; $-$9.05$\pm$4.83 km/s) in Fig.\ \ref{fig_USco_GTC_XSH:plot_RVs}. 
We compare the distribution of our candidates with the distribution of radial velocities of higher mass
member \citep[values in Table 12 of][and references therein \citep{duflot95a,barbier_brossart00,gontcharov06,chen11a,dahm12}]{pecaut12} whose values range from $-$10 to 0 km/s. The two faintest members exhibit large
radial velocities but with larger uncertainties also consistent with the mean velocities of
UpSco members. The only source that lies 2$\sigma$ away from the higher mass members with the
largest Doppler shifts is VISTA\,J1605$-$2130 (dL2) so we set its radial velocity membership to ``?''\@.
It is worth noting that this source was also a doubtful member because it displays blue infrared
colour, its near-infrared spectrum did not show obvious youth features, and its discrepant radial velocity.
All the other candidates with X-shooter spectra can not be discarded as members based on the 
low accuracy of our radial velocity measurements (Table \ref{tab_USco_GTC_XSH:table_membership_summary}).
Overall, we have identified two clear non members based on their $I$-band photometry, two
non member from their near-infrared spectroscopy and/or radial velocity, and one doubtful 
currently unclassified. We infer a minimum contamination level of 4/28--5/28\,=\,14--18\% among 
photometric candidates and 2/15--3/15\,=\,13.3--20\% among candidates with spectroscopic follow-up.

%
%%%%%%%%%%%%%%%%%%%%%%%%%%%%%%%%%%%%%%%%%%%%%%%
%%%%% Figure: Histogram of RVs %%%%%
%%%%%%%%%%%%%%%%%%%%%%%%%%%%%%%%%%%%%%%%%%%%%%%
%
%

%
%%%%%%%%%%%%%%%%%%%%%%%%%%%%%%%%%%%%%%%%%%%%%%%
%%%%% Figure: SEDs from SVO (I) %%%%%
%%%%%%%%%%%%%%%%%%%%%%%%%%%%%%%%%%%%%%%%%%%%%%%
%
\begin{figure*}
  \centering
  \includegraphics[width=0.49\linewidth, angle=0]{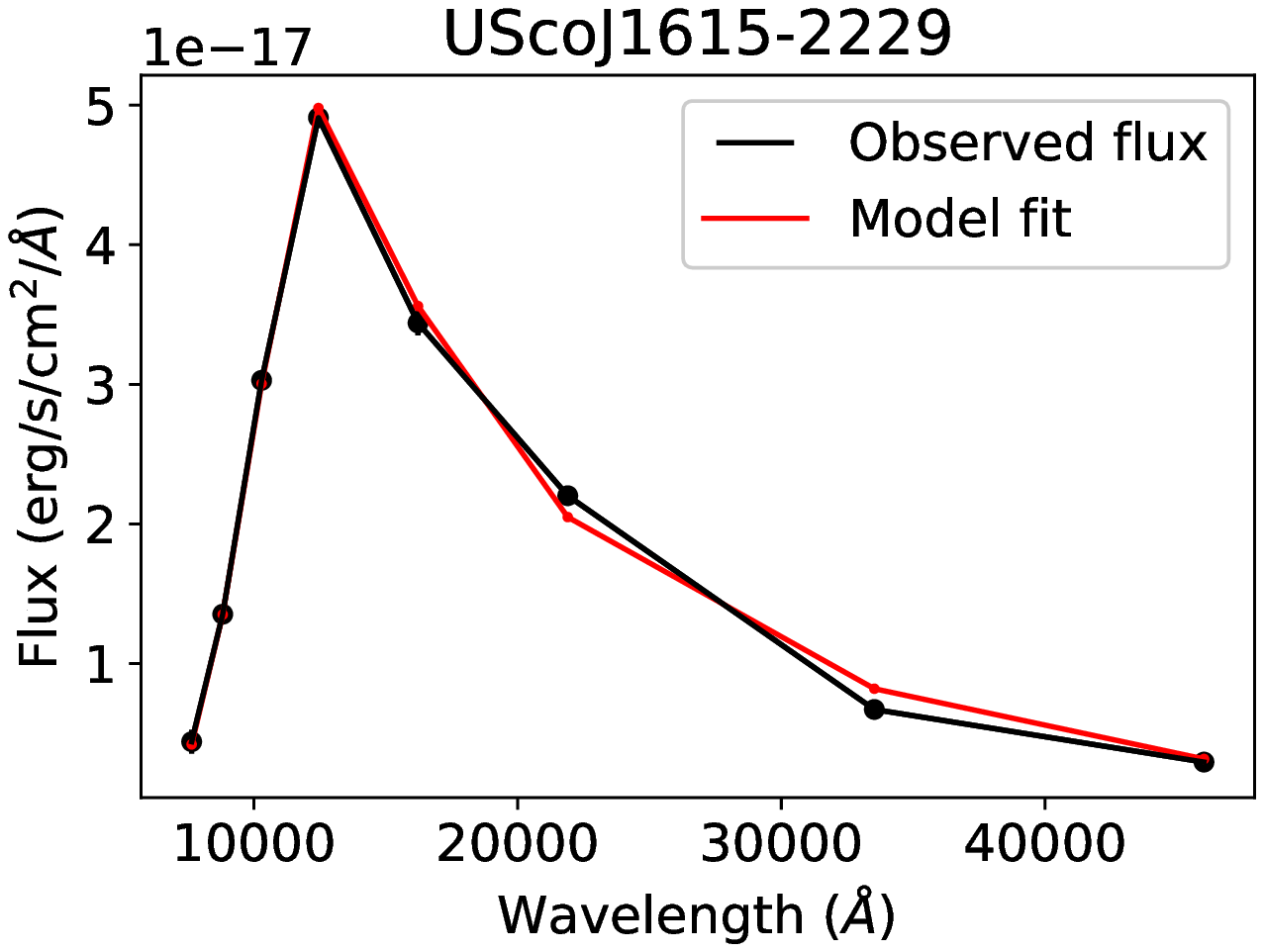}
  \includegraphics[width=0.49\linewidth, angle=0]{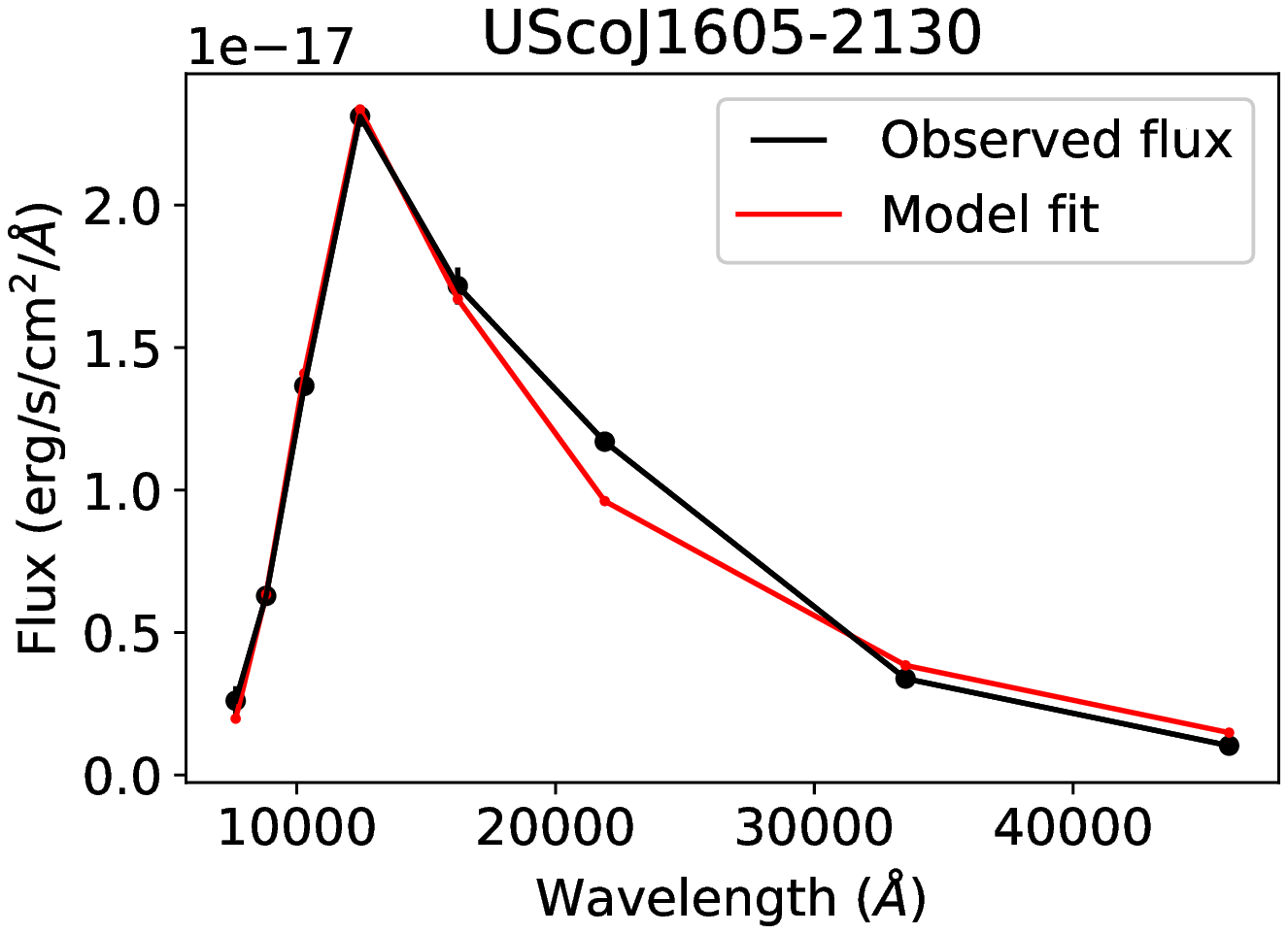}
  \includegraphics[width=0.49\linewidth, angle=0]{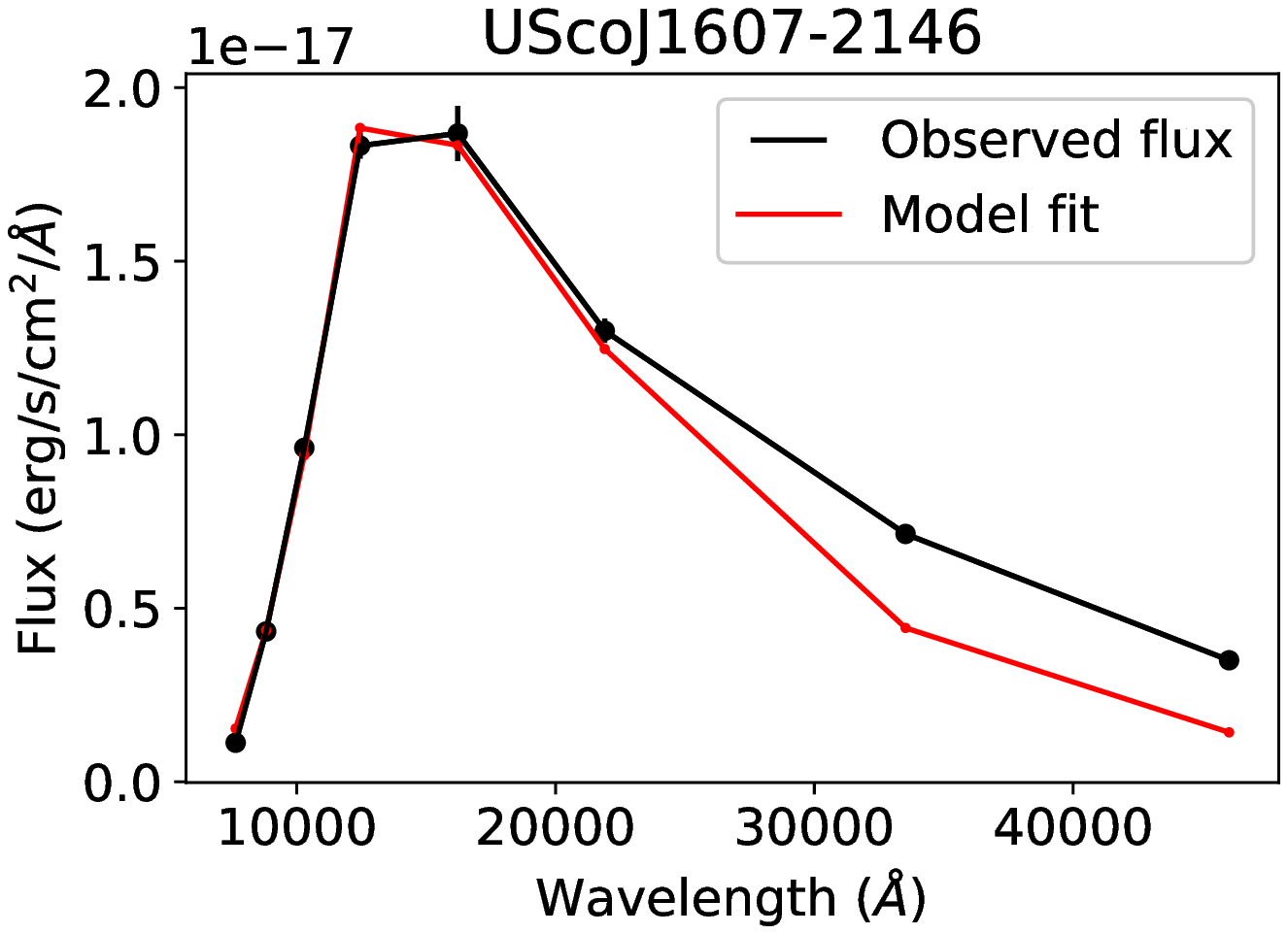}
  \includegraphics[width=0.49\linewidth, angle=0]{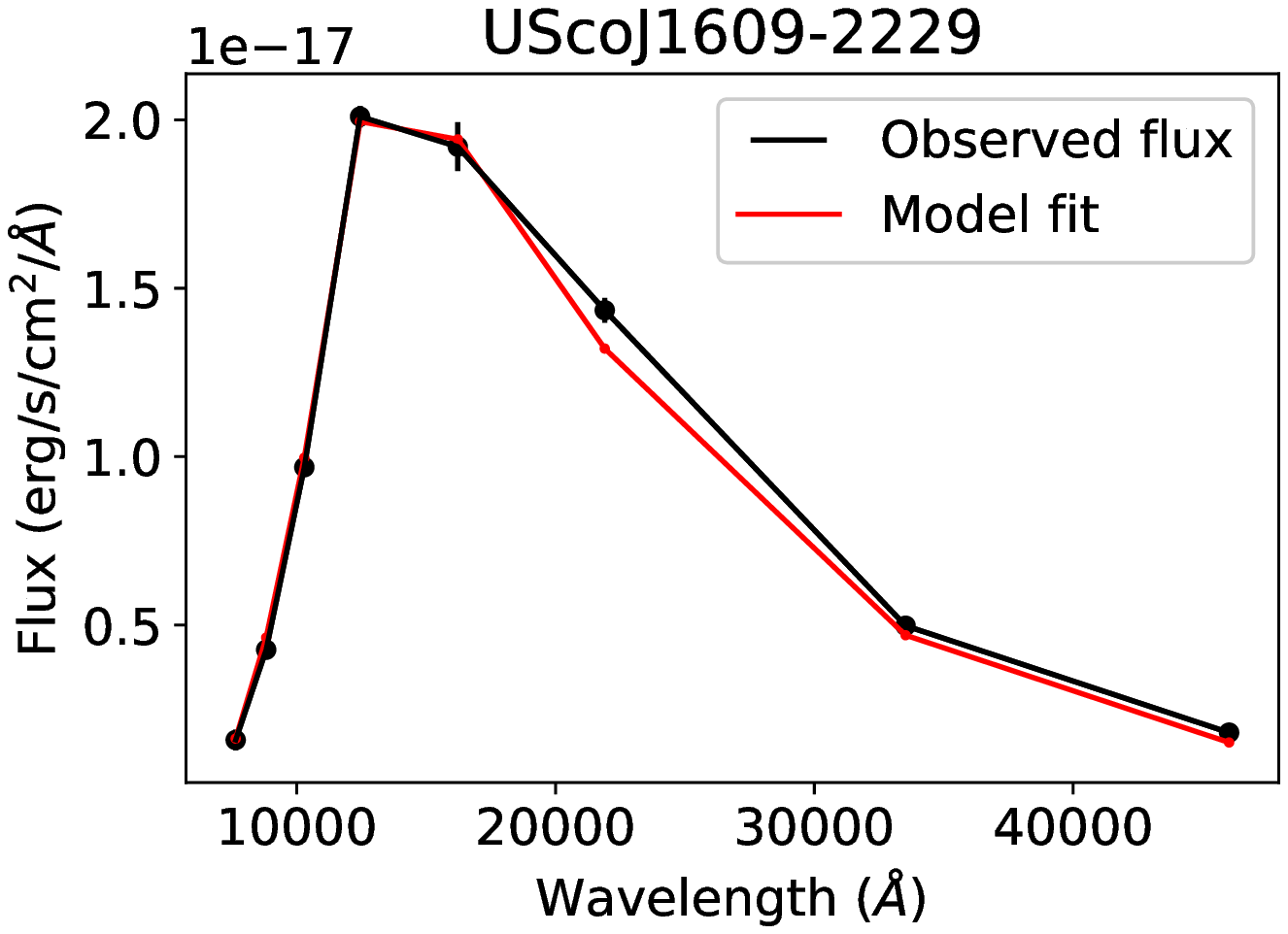}
  \includegraphics[width=0.49\linewidth, angle=0]{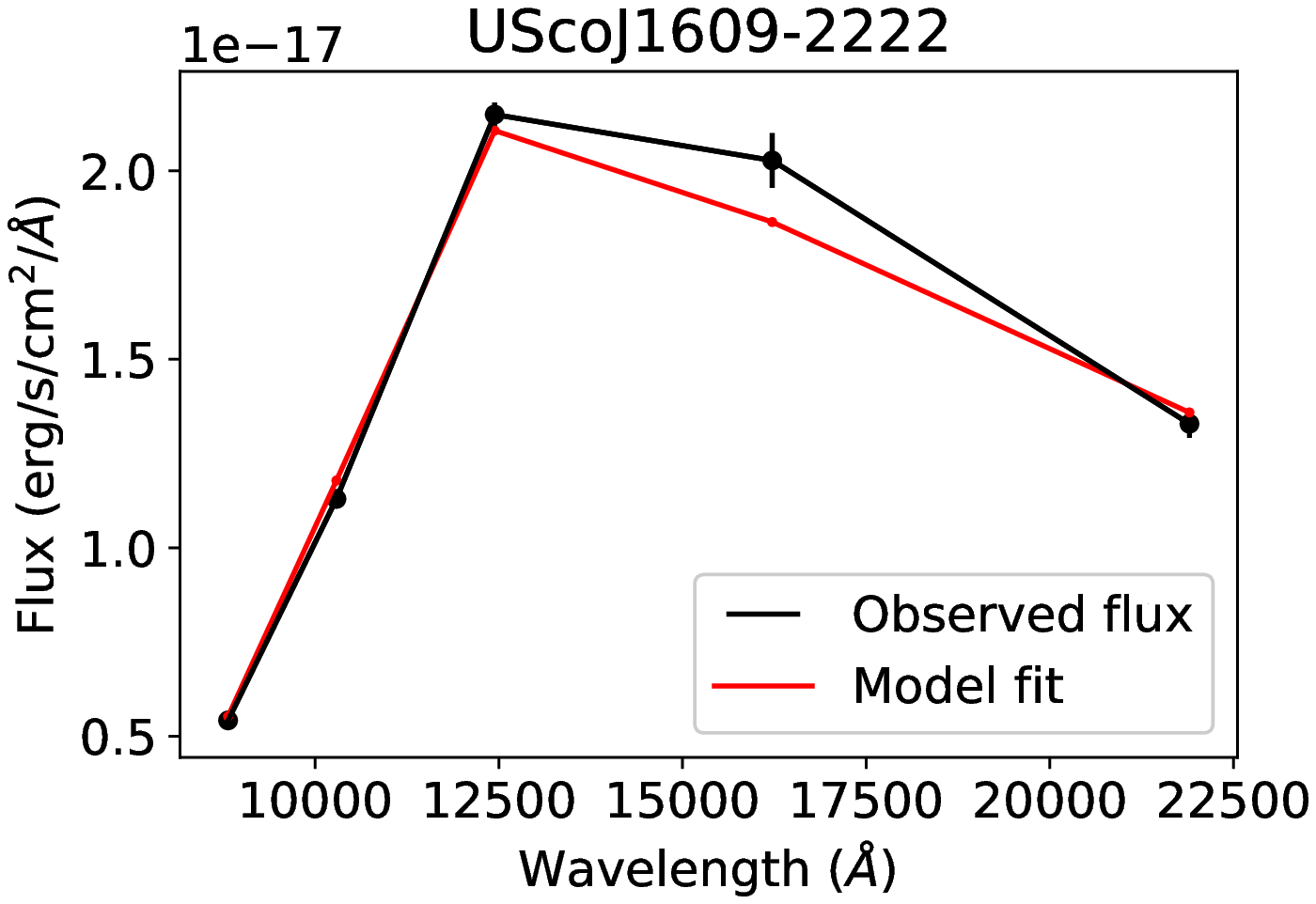}
  \includegraphics[width=0.49\linewidth, angle=0]{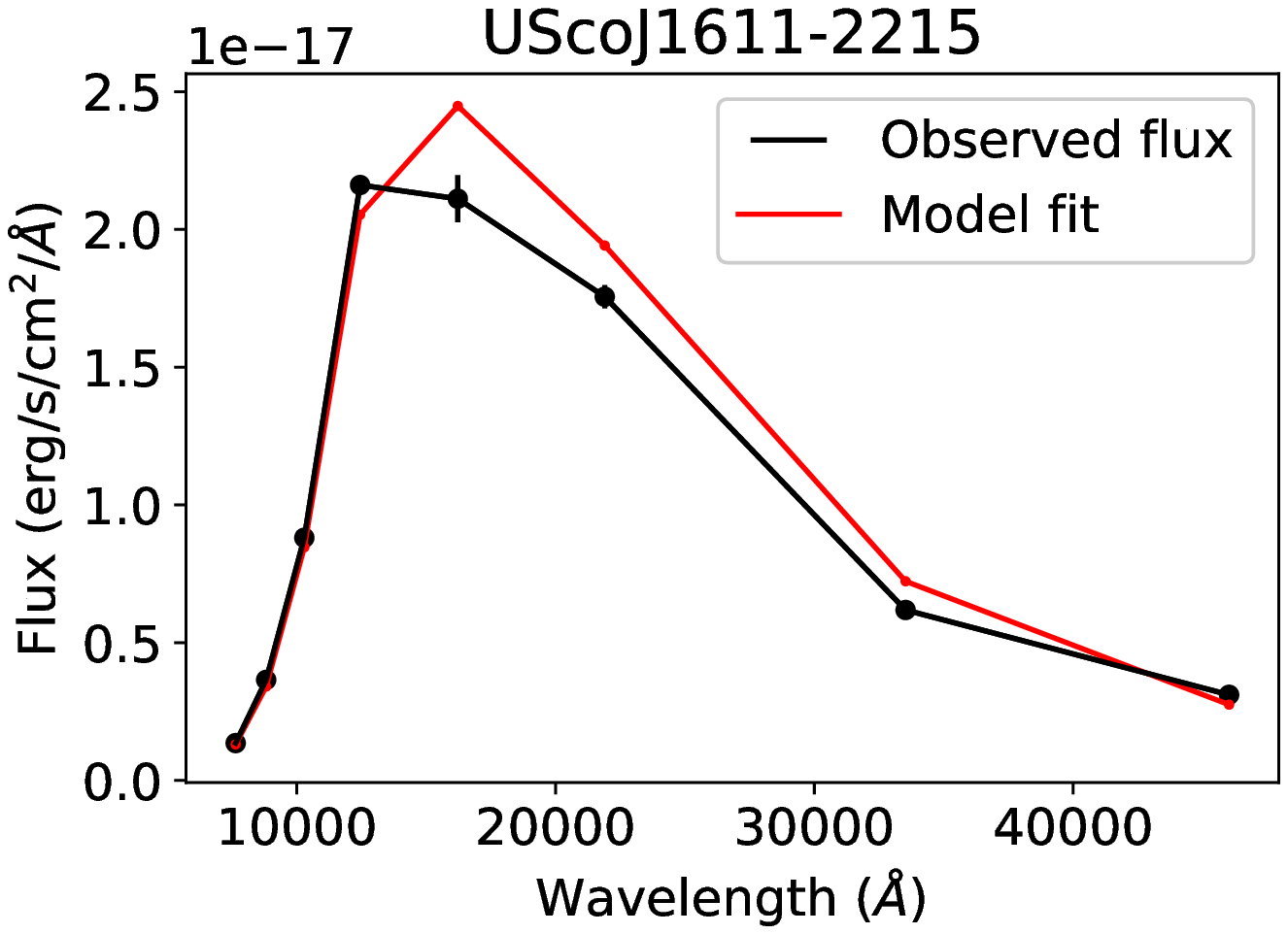}
  \includegraphics[width=0.49\linewidth, angle=0]{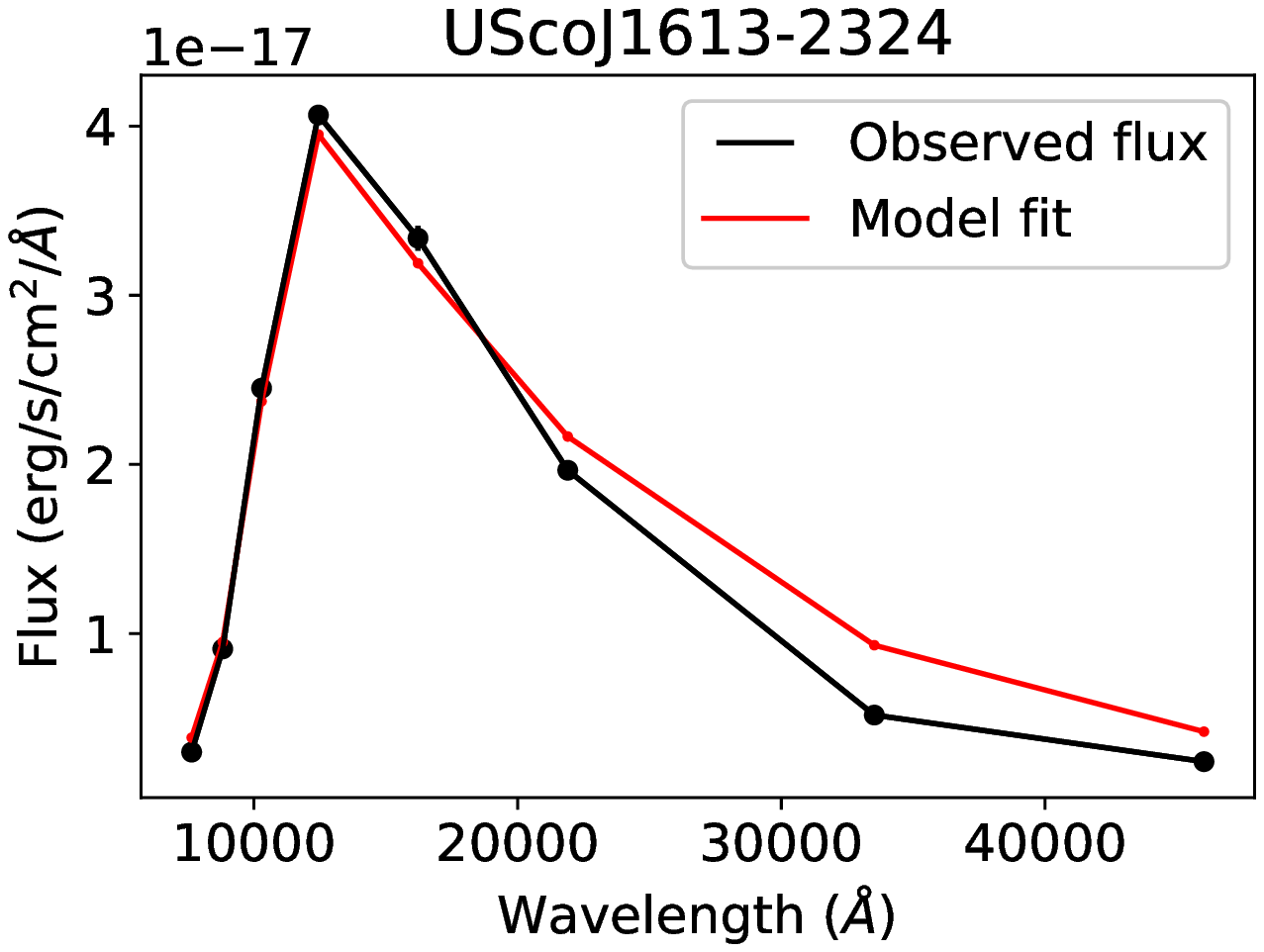}
  \includegraphics[width=0.49\linewidth, angle=0]{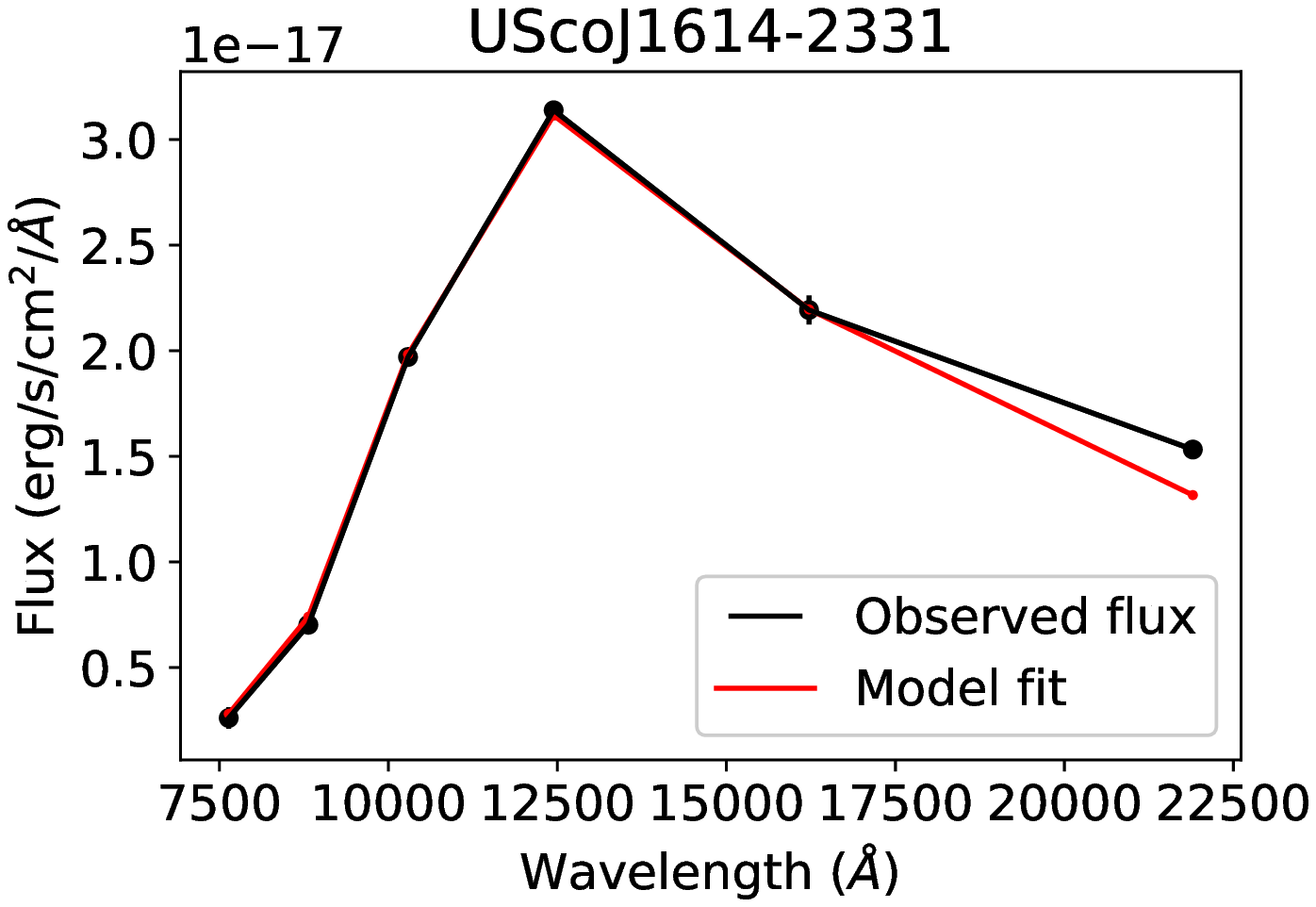}
  \caption{Spectral energy distributions (red dots) of UpSco candidates compared to
theoretical models (blue spectra) to derive temperatures, gravities, and metallicities.
}
  \label{fig_USco_GTC_XSH:SED_VO_1}
\end{figure*}
%

%
%%%%%%%%%%%%%%%%%%%%%%%%%%%%%%%%%%%%%%%%%%%%%%%
%%%%% Figure: SEDs from SVO (II) %%%%%
%%%%%%%%%%%%%%%%%%%%%%%%%%%%%%%%%%%%%%%%%%%%%%%
%
\begin{figure*}
  \centering
  \includegraphics[width=0.49\linewidth, angle=0]{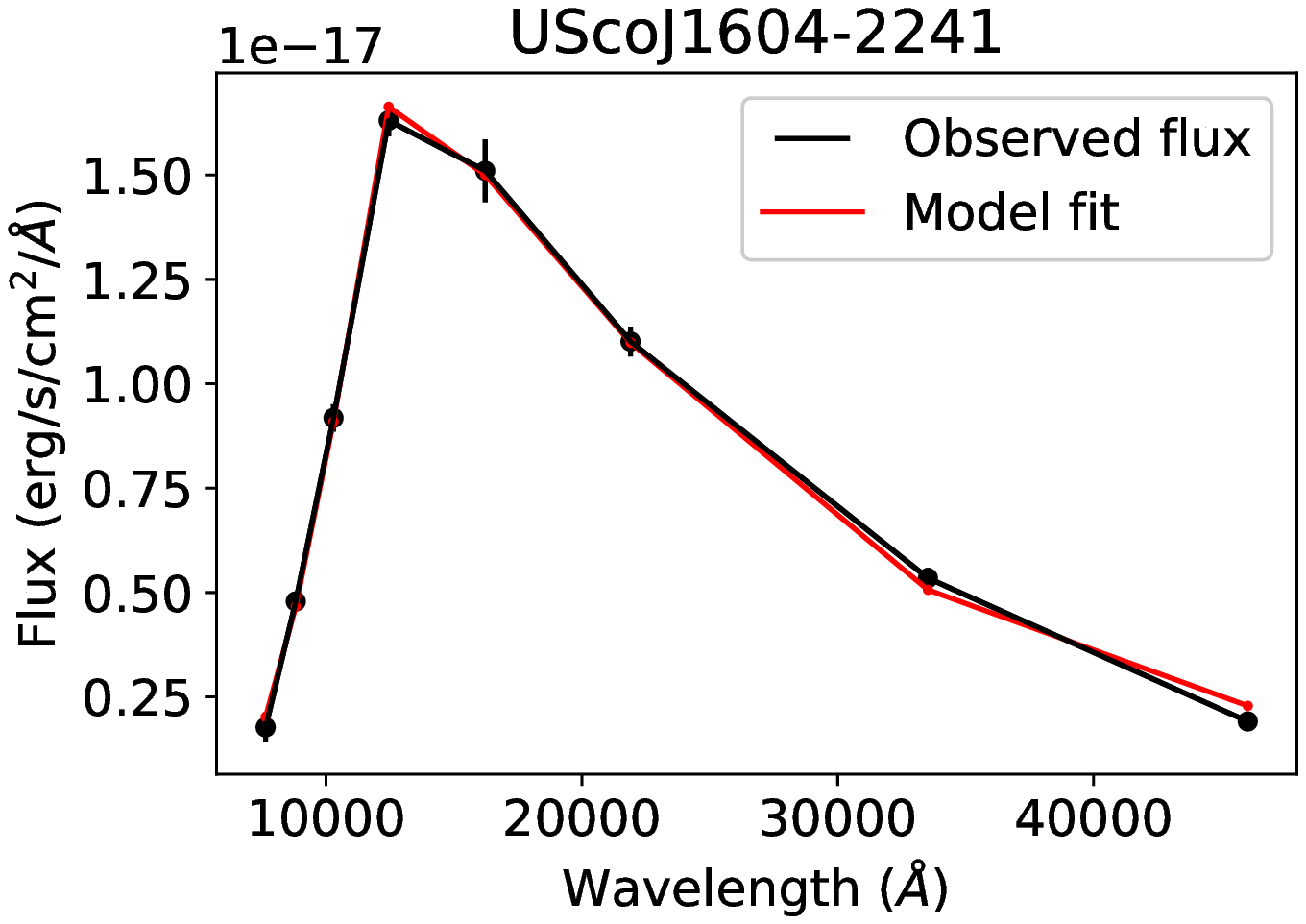}
  \includegraphics[width=0.49\linewidth, angle=0]{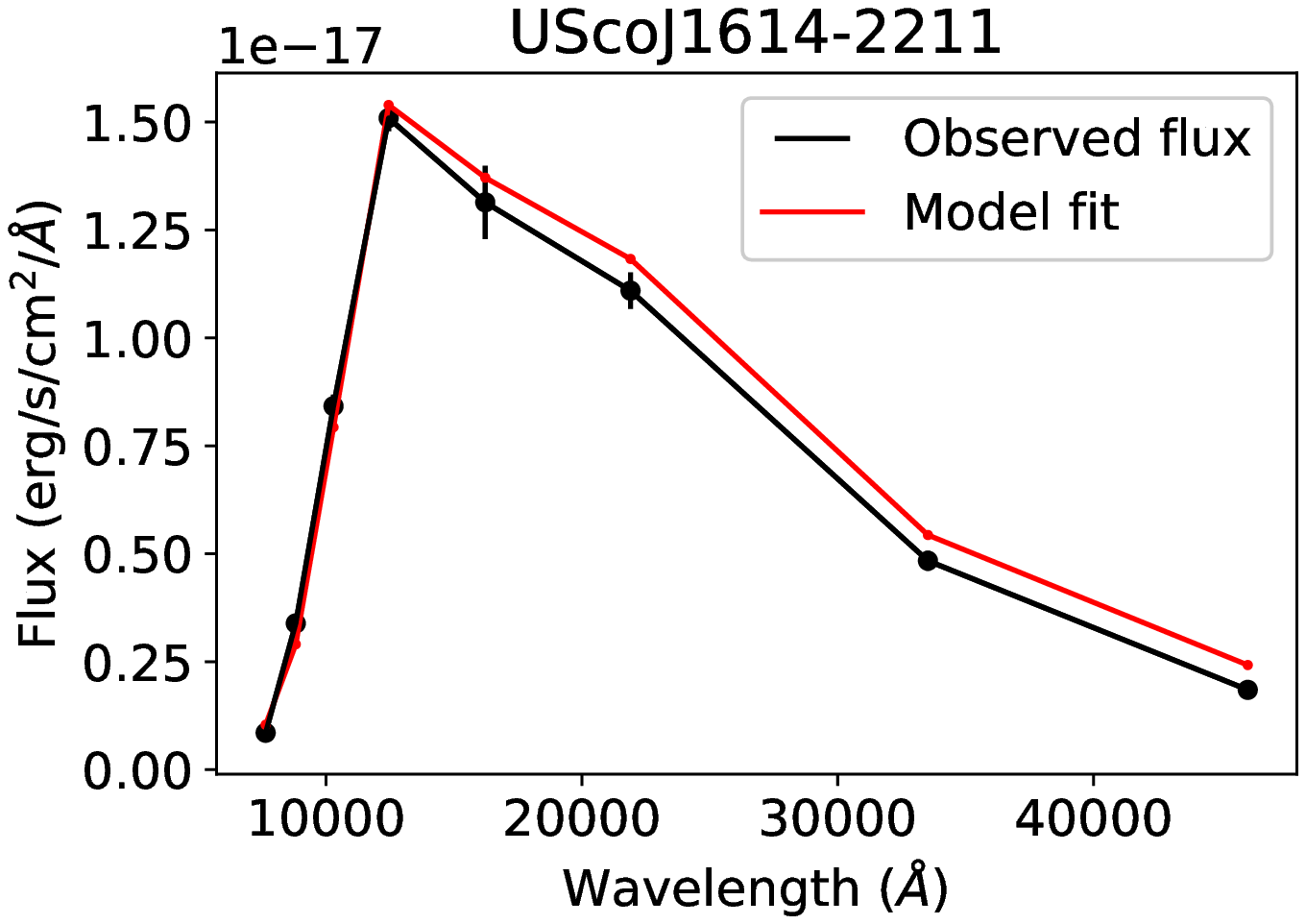}
  \includegraphics[width=0.49\linewidth, angle=0]{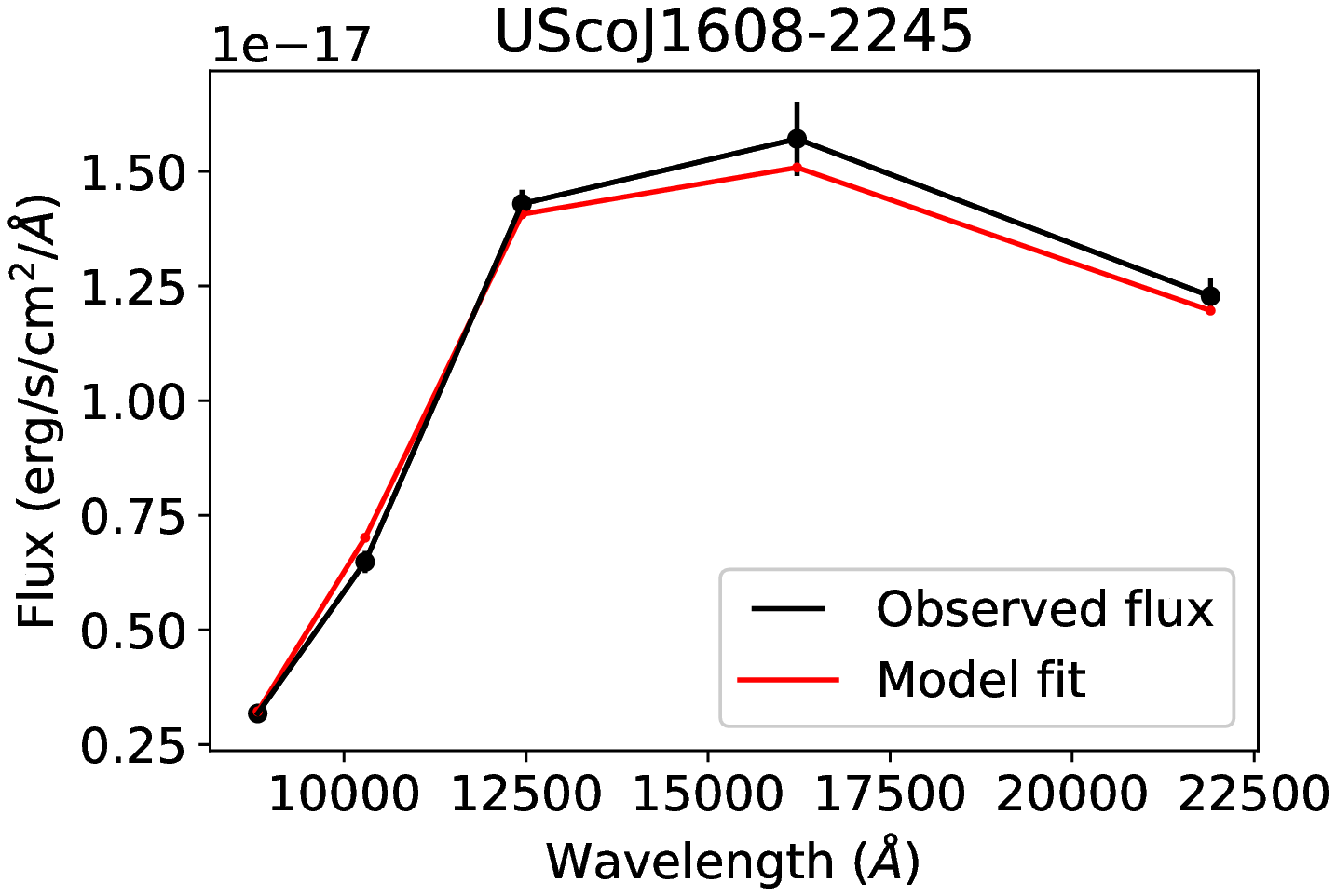}
  \includegraphics[width=0.49\linewidth, angle=0]{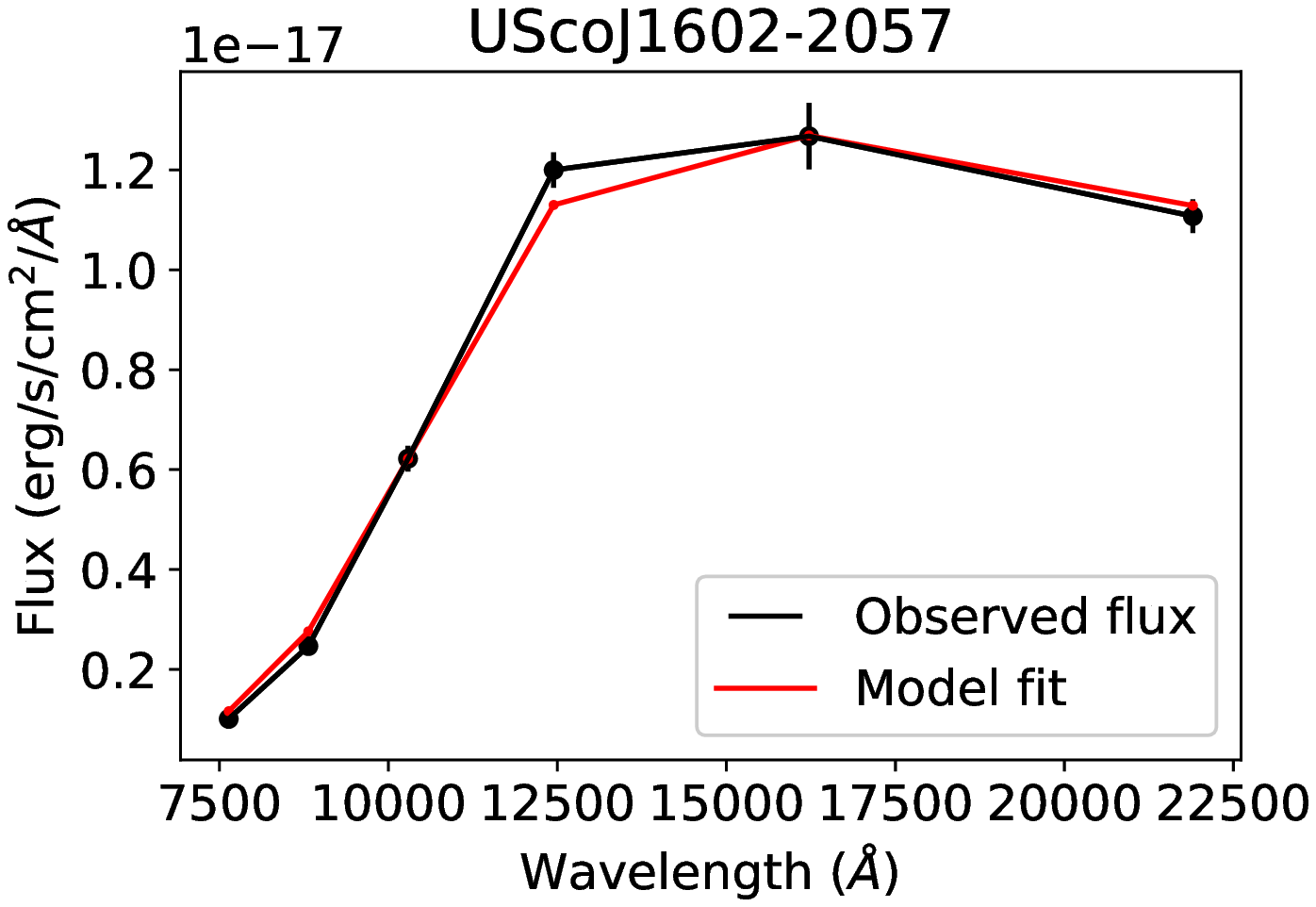}
  \includegraphics[width=0.49\linewidth, angle=0]{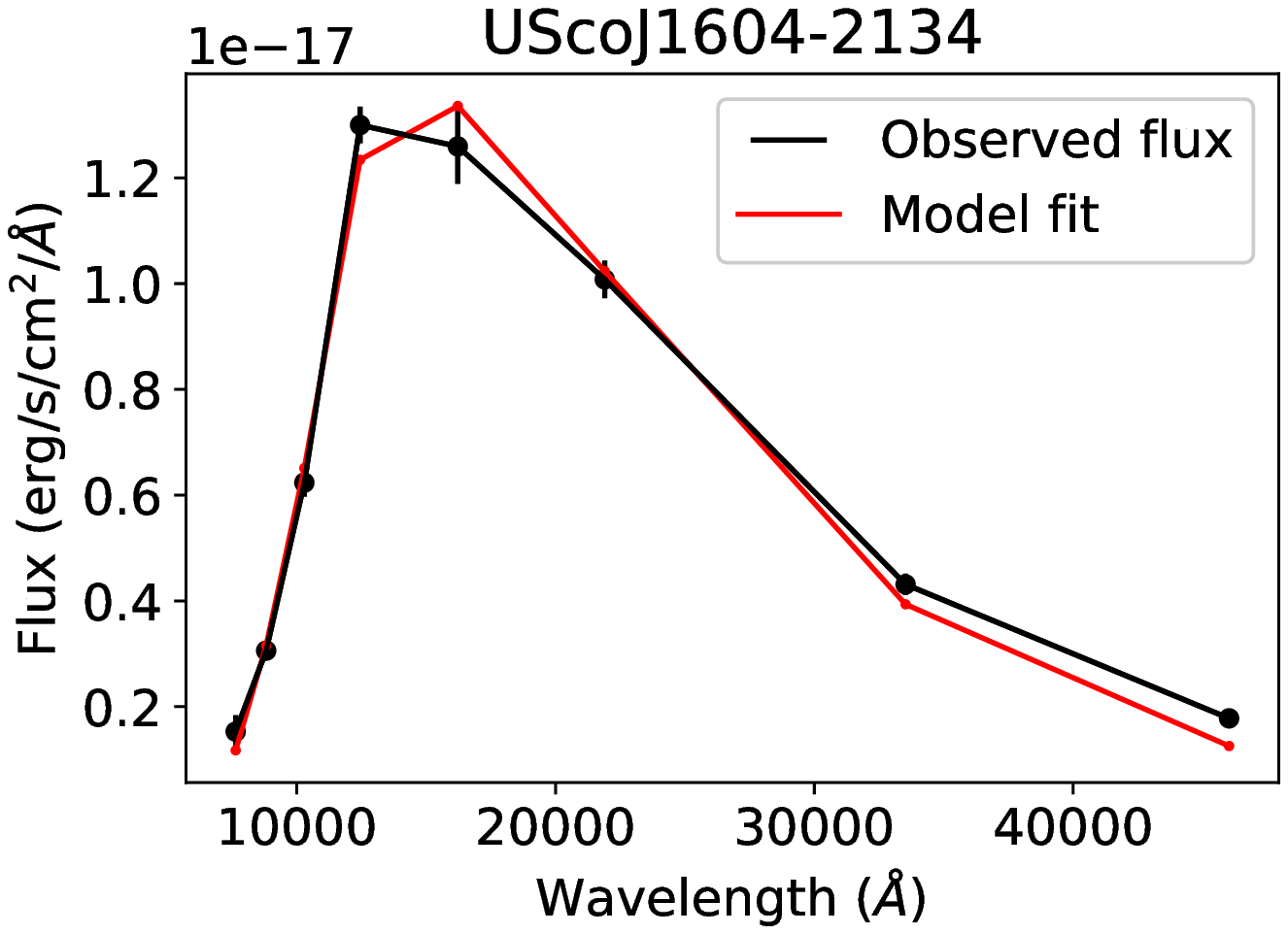}
  \includegraphics[width=0.49\linewidth, angle=0]{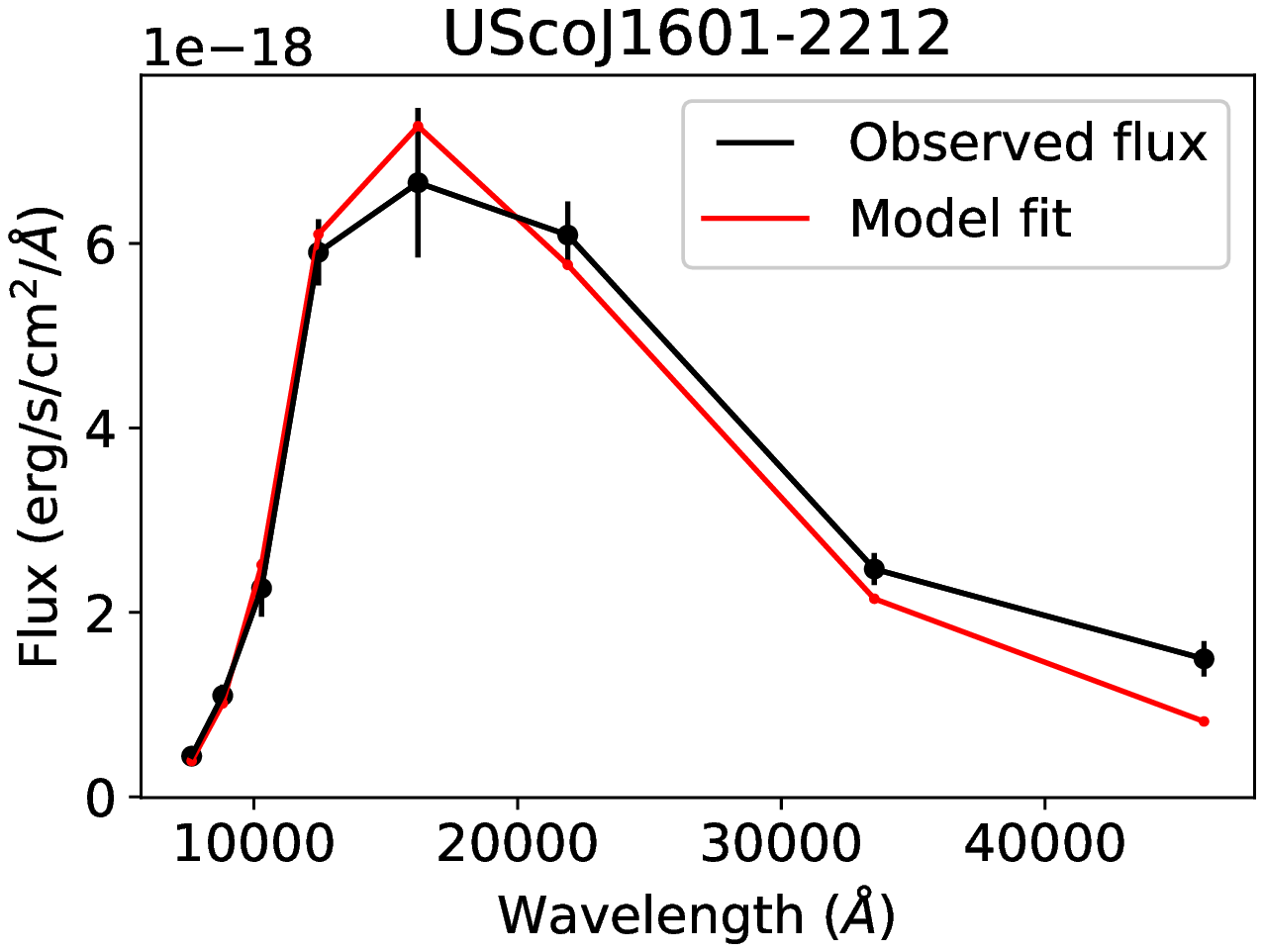}
  \includegraphics[width=0.49\linewidth, angle=0]{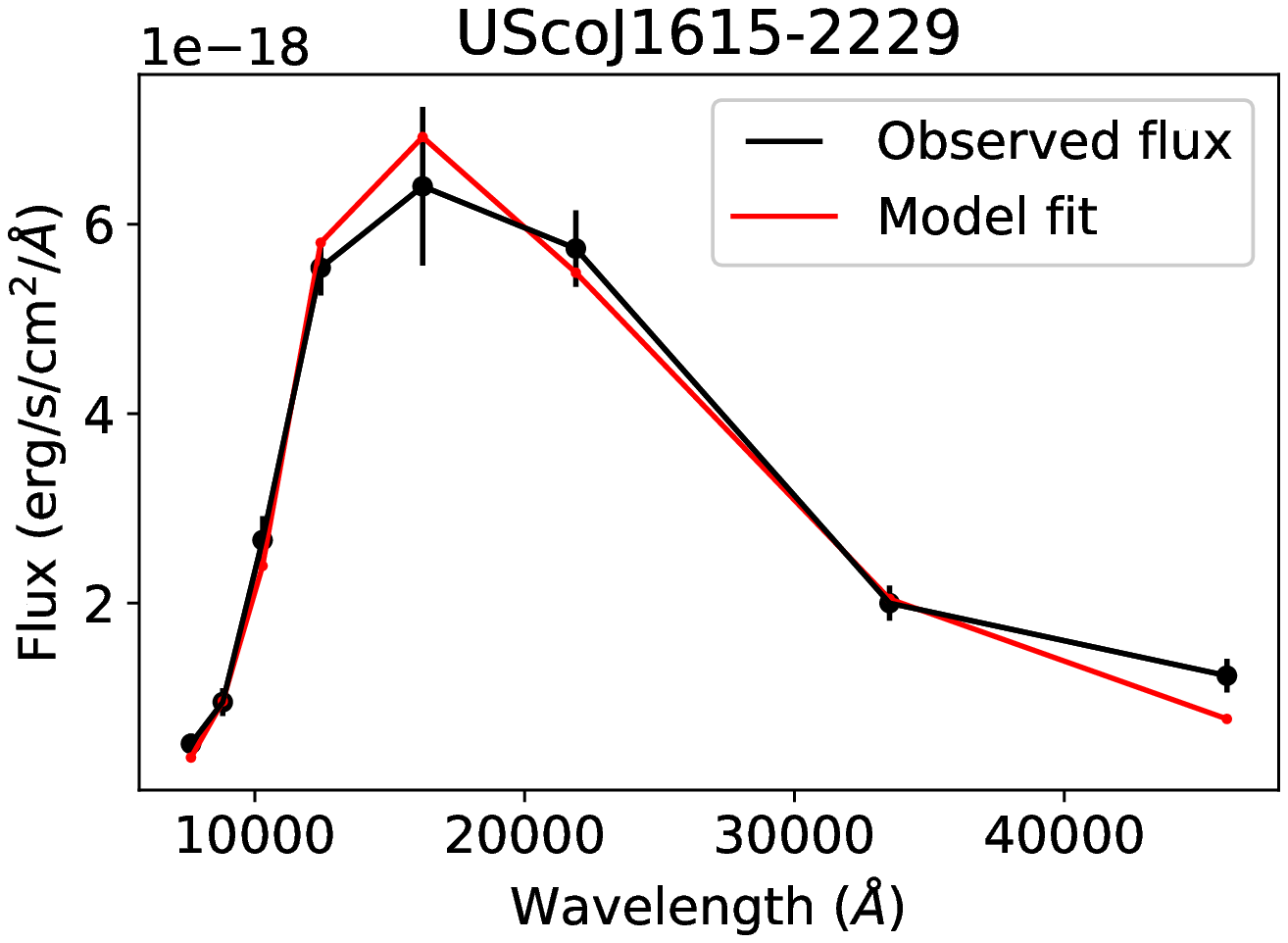}
  \contcaption{
}
  \label{fig_USco_GTC_XSH:SED_VO_2}
\end{figure*}
%

%
%%%%%%%%%%%%%%%%%%%%%%%%%%%%%%%%%%%%%%%%%%%%%%%
%%%%% Figure: VOSA Teff %%%%%
%%%%%%%%%%%%%%%%%%%%%%%%%%%%%%%%%%%%%%%%%%%%%%%
%
\begin{figure}
  \centering
  \includegraphics[width=\linewidth, angle=0]{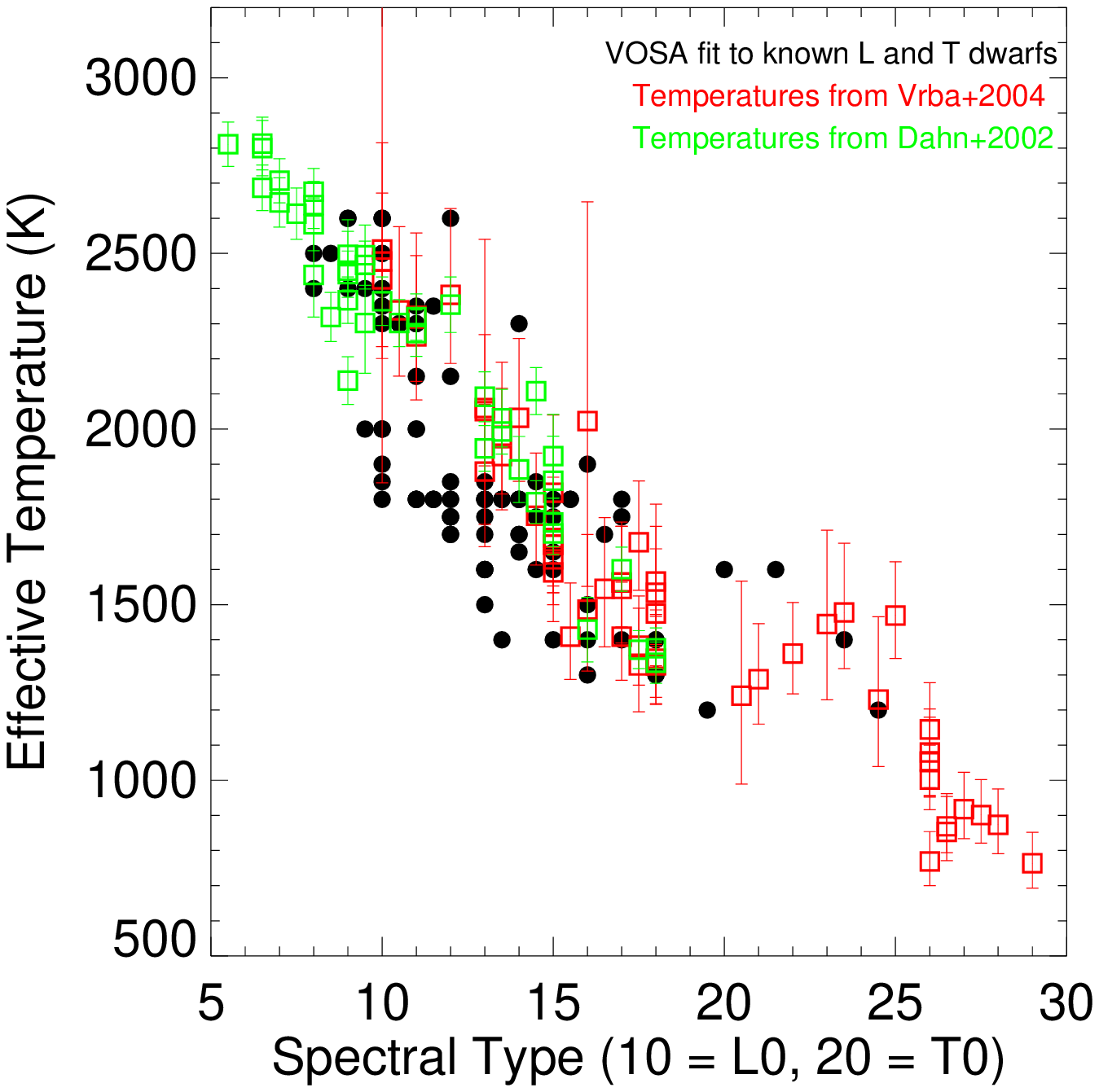}
  \caption{Effective temperatures of known late-M, L, and T dwarfs with spectral types 
published by \citet[green symbols;][]{dahn02} and \citet[red symbols;][]{vrba04} 
compared to the model-dependent T$_{\rm eff}$ derived from the VOSA fitting tool 
(black dots). The typical error bar of VOSA temperatures is around 50--100\,K\@.
}
  \label{fig_USco_GTC_XSH:VOSA_Teff}
\end{figure}
\subsection{Effective temperatures and gravities}
\label{USco_GTC_XSH:Teff_models}

We have used the spectral energy distribution (SED) analyser tool implemented in the Spanish virtual 
observatory \citep[VOSA;][]{bayo08} to infer the effective temperatures (T$_{\rm eff}$), surface gravities, 
and metallicities of our UpSco candidates by comparison with a set of theoretical models. 
We investigated the validity of the VOSA fitting tool on known field late-M, L, and T dwarfs 
before applying it to our sample. We input in VOSA the samples of ultracool dwarfs with effective
temperature derived by \citet{dahn02} and \citet{vrba04}. We used the VOSA photometry tool to extract
optical and infrared photometry from various large-scale surveys, including Sloan, 2MASS, UKIDSS,
VISTA, and WISE to construct their spectral energy distributions. We kept in the final sample only
sources with at least five photometric points in the near- and mid-infrared. We fit their distribution with
three different types of models incorporated in VOSA: BT-Settl \citep{allard12}, BT-DUSTY \citep{allard09}, 
and BT-CIFIST \citep{baraffe15} for the L dwarfs and only BT-Settl for the T dwarfs because of the
validity range of the other two models. We plot the T$_{\rm eff}$ from VOSA with black dots in
Fig.\ \ref{fig_USco_GTC_XSH:VOSA_Teff} and compare them with those of \citet{dahn02} and \citet{vrba04} 
shown in green and red, respectively. We observe that the models reproduce well the overall trend in
temperature for the full high-gravity late-M, L, and T dwarf sequence \citep{dahn02,vrba04}, but
with a very large dispersion of about 500\,K\@. We find that model fits tend to 
under-estimate the T$_{\rm eff}$ for L dwarfs by a few hundreds of Kelvins but this trend is only
valid for old field dwarfs (the effect is unknown at young ages).

For our purposes, we compared the photometric SEDs including the GTC OSIRIS $i$-band photometry, the VISTA 
$ZYJ$ magnitude (L13), the UKIDSS $HK$ photometry 
\citep[$\sim$0.88--2.5\,$\mu$m;][]{hewett06}, and the WISE $w1$\,$+$\,$w2$ passbands (3.4 and 4.6 $\mu$m) 
to three sets of models with a limited range of parameters: BT-Settl \citep{allard12}, BT-DUSTY \citep{allard09}, 
and BT-CIFIST \citep{baraffe15} for T$_{\rm eff}$\,=\,1200--3500\,K, surface gravities between 2.5 and 5.5 dex, 
and metallicities ranging from $-$0.5 to 0.5 dex. In other words, we assumed solar 
metallicity for members of UpSco, which is expected for such a young nearby OB association. We also 
assumed a mean distance of $\sim$145 pc for UpSco \citep{deBruijne97}.

We list the physical parameters derived from the photometric fits in Table \ref{tab_USco_GTC_XSH:table_SED_fits}
and show the SEDs in Figs.\ \ref{fig_USco_GTC_XSH:SED_VO_1}--\ref{fig_USco_GTC_XSH:SED_VO_2}. We note
that the best fit is obtained for different models depending on the targets. All our
targets have solar metallicity with a typical error of $\pm$0.2 dex. All objects exhibit gravity
in the $\log$(g)\,=\,3.5--5.0 range with uncertainties of 0.5 dex (the steps of the models).
We expect $\log$(g)\,=\,4.0 dex for brown dwarfs still contracting at the age of UpSco, while
$\log$(g)\,=\,2.5 derived for VISTA\,J1614$-$2211 whose gravity is too low according to model predictions.
The effective temperatures lie between 1500\,K and 1900\,K with uncertainties less than 60\,K\@.
One source, VISTA\,J1604$-$2241, is best fit with a $\log$(g) of 5.5 dex, which is quite high
for young brown dwarfs, highlighting current uncertainties in fitting schemes and atmospheric models.
Keeping in mind current flaws in theoretical models reproducing spectra of brown dwarfs,
the large dispersion of temperatures observed for L dwarfs, and the fact that young L dwarfs
might be $\sim$100--300\,K cooler than their old counterparts \citep{filippazzo15}, our T$_{\rm eff}$
represent first-guess estimates. More accurate temperatures could be inferred if we knew the range of 
radii for UpSco L-type brown dwarfs but such a measurement is currently available only for one (isolated)
eclipsing brown dwarf binary in Orion \citep[0.48--0.68 R$_{\odot}$;][]{stassun06,stassun07b}.

We have also taken advantage of the template fitting procedure in VOSA, which uses
a sequence of spectral libraries: the NIRSPEC Brown Dwarf spectroscopic survey \citep{mclean03,mclean07},
L and T dwarf data archive \citep{knapp04,golimowski04a,chiu06}, the SpeX prism 
library\footnote{http://pono.ucsd.edu/$\sim$adam/browndwarfs/spexprism/index\_old.html},
the Keck LRIS spectra of late-M, L and T dwarfs \citep{kirkpatrick91,kirkpatrick99},
and the database of Collinder\,69 \citep{bayo11a}. We observe that the best fits were returned
from the first three archives, which use mainly near-infrared spectra instead of the full
optical/near-infrared/mid-infrared range. We see that a few targets are well fit by known
templates although the majority tend to return earlier spectral types consistent within 2--3 
sub-classes with those assigned in this paper (Sect.\ \ref{USco_GTC_XSH:spectral_classification_NIR_absolute};
Table \ref{tab_USco_GTC_XSH:table_membership_summary}), except for VISTA\,J1613$-$2124 
classified as a T0 dwarf from direct comparison with a 1.0--2.5 micron spectrum.
On the light of the discrepancies observed between photometric and spectroscopic spectral
types, we caution the use of the template fitting procedure to assign temperatures especially
at young ages. We list the assigned spectral types from template fitting in the last column of 
Table \ref{tab_USco_GTC_XSH:table_SED_fits}.

\subsection{Bolometric corrections}
\label{USco_GTC_XSH:BCs}

\begin{figure}
  \centering
  \includegraphics[width=\linewidth, angle=0]{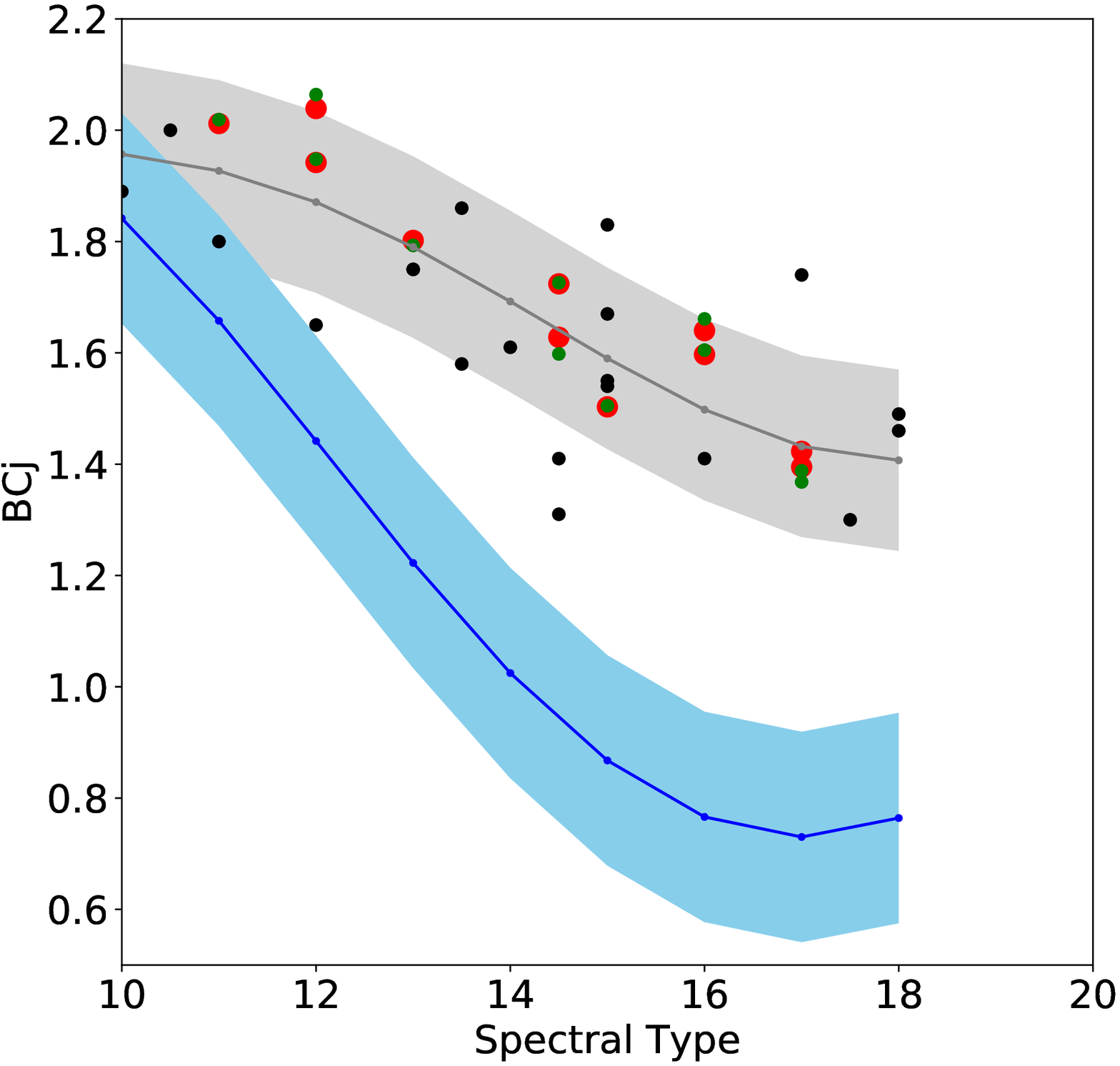}
  \includegraphics[width=\linewidth, angle=0]{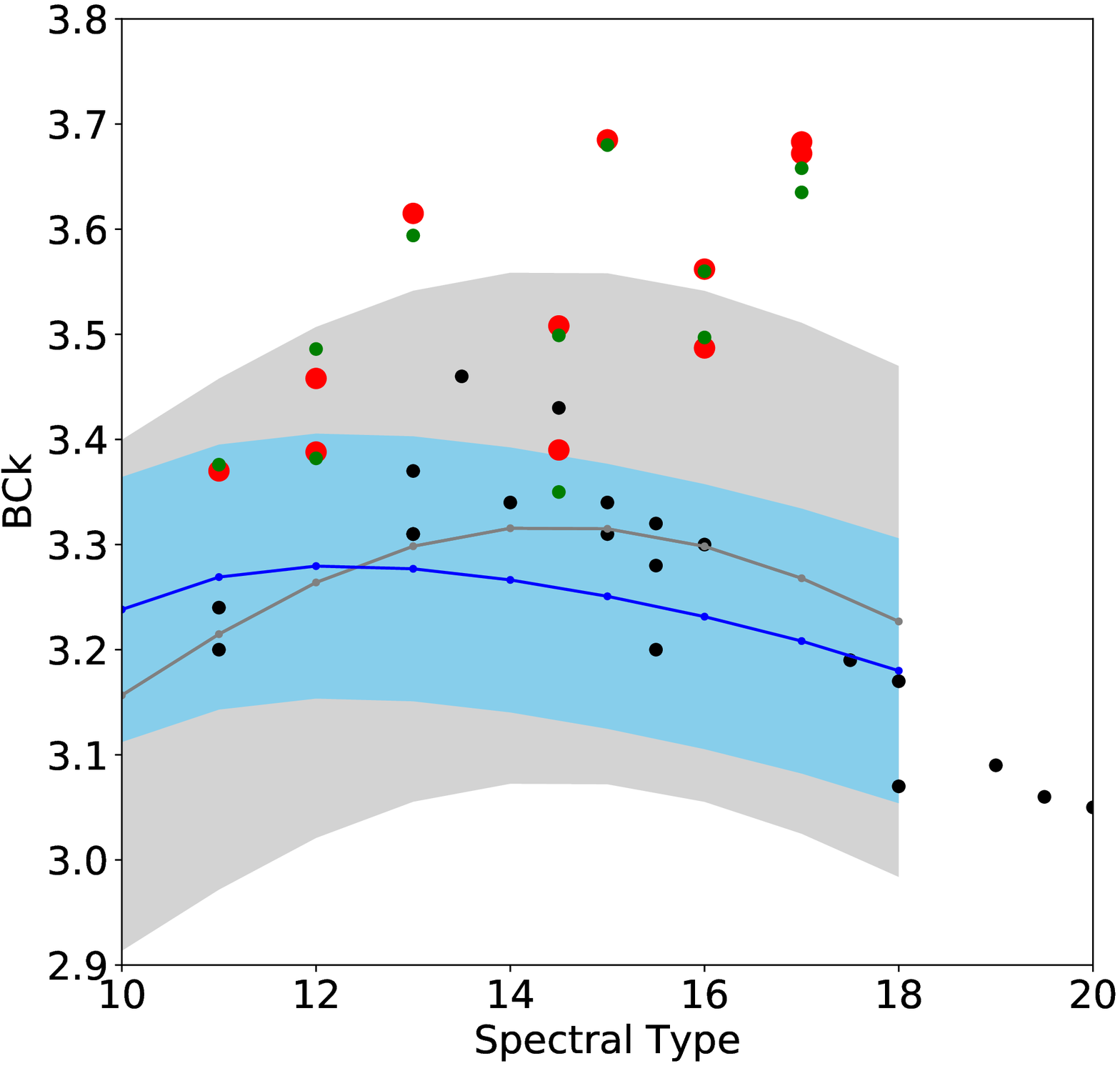}
  \caption{$J$-band (top) and $K$-band (bottom) bolometric correction vs.\ spectral type for 
field L dwarfs \citep[black dots;][]{dahn02,golimowski04a} along with our UpSco candidates with
optical and infrared spectroscopy (red dots) and infrared spectroscopy only (green dots).
We overplotted the polynomial fits and dispersions  of the BC$_{J}$ and BC$_{K}$ relations
presented in \citet{filippazzo15} for field (grey) and young (blue) L dwarfs.
Note for x-axis: 10$\equiv$L0, 12$\equiv$L2, etc\ldots{}.
}
  \label{fig_USco_GTC_XSH:BC_vs_SpT}
\end{figure}

We computed bolometric corrections (BCs) for our UpSco candidates, assuming that the observed
spectroscopic and photometric fluxes come from a single source (i.e.\ the target). First, we computed the
BCs including the optical and near-infrared spectra from the GTC and VLT, respectively. We integrated 
the spectra from $\sim$563 nm to 2480 nm and added the flux from the WISE passbands (up to 5.34\,$\mu$m). 
We did not include the objects lacking WISE photometry (Table \ref{tab_USco_GTC_XSH:table_WISE_phot}).
We integrated the solar metallicity BT-Settl models for a $\log$(g) or 4.5 dex, and temperatures
of 2000,K, 1800\,K, and 1600\,K for L1--L2, L4$\pm$1, and L6--L7 dwarfs, respectively.
We integrated the BT-Settl models for wavelenths shortwards of 563 nm and longer than 5.56 $\mu$m
(upper limit of the WISE $w2$ filter)
The contribution from the 2000\,K and 1600\,K models represents only $\sim$2.6\% and $\sim$5.4\%, 
respectively.
We assumed an absolute solar bolometric magnitude of 4.74 for the Sun, following the IAU recommendation.
We report their bolometric magnitudes and corrections in the top panel of Table \ref{tab_USco_GTC_XSH:table_BC}
and plot their $J$- and $K$-band BCs as red dots in Fig.\ \ref{fig_USco_GTC_XSH:BC_vs_SpT} compared
to the BCs of field L dwarfs (black dots) taken from \citet{dahn02} and \citet{golimowski04a}.
In Fig.\ \ref{fig_USco_GTC_XSH:BC_vs_SpT}, we added the polynomial fits and their associated dispersions
reported for young and old field L dwarfs in blue and grey, respectively \citep{filippazzo15}.

In a second step, we calculate the BCs for our candidates integrating their near-infrared spectra only
and adding the WISE photometry to gauge the contribution of the optical spectrum. The flux 
concentrated in the aforementioned models shortwards of 1000 nm and longwards of the WISE $w2$ passband 
represent 10\% at most in all cases. We obtained similar BCs within $\pm$0.04 mag for all sources with
both optical$+$infrared and infrared-only contributions, suggesting that the contribution from the optical 
regions is very small in our targets.
We list the BCs for the candidates with WISE photometry and infrared spectra in the bottom panel of 
Table \ref{tab_USco_GTC_XSH:table_BC} and plot their BCs as green dots in Fig.\ \ref{fig_USco_GTC_XSH:BC_vs_SpT}.
We note that our $K$-band BC for our L1 member of UpSco is comparable to the value obtained by
\citet{todorov10a} for M9.5--L0 members of Taurus (BC$_{K}$\,$\sim$\,3.40 mag).

In the top panel of Fig.\ \ref{fig_USco_GTC_XSH:BC_vs_SpT} depicting the BC$_{J}$ as a function of
spectral types we observe a decreasing trend of the BC with later spectral type for young UpSco members
and for field L dwarfs. Our BC$_{J}$ lie within the dispersion of field L dwarfs derived by
\citet{filippazzo15}. 
In the case of BC$_{K}$ (bottom panel in Fig.\ \ref{fig_USco_GTC_XSH:BC_vs_SpT}), we see a slight
increase towards later spectral types, a different trend to the one observed for the samples of
field and young L dwarfs presented in \citet{filippazzo15}. This has a clear impact on the luminosity 
and, more importantly, on the mass of young L dwarfs because higher values of BC$_{K}$ indicate that 
objects become less luminous, therefore, less massive.
Fig.\ \ref{fig_USco_GTC_XSH:BC_vs_SpT} provides an additional illustration of what is seen in 
Fig.\ \ref{fig_USco_GTC_XSH:plot_spectra_GTCplusXSH}, where the spectra of UpSco objects appear 
systematically redder in the $K$-band than the corresponding high-gravity sources of related 
near-infrared spectral type.

%
%%%%%%%%%%%%%%%%%%%%%%%%%%%%%%%%%%%%%
%%%%% Table: SED fits from SVO %%%%%
%%%%%%%%%%%%%%%%%%%%%%%%%%%%%%%%%%%%%
%
\begin{table}
 \centering
 \caption[]{Results of the best fits to the photometric spectral energy distribution
of UpSco members with spectral types (SpT) from this study through direct comparison with 
theoretical models using the VOSA tool.
Uncertainties on the temperatures, gravities, and metallicities are $\pm$50\,K, 
$\pm$0.25--0.50 dex, and $\pm$0.2 dex, respectively. These errors do not include
neither the systematics nor the accuracy of the procedure ($\pm$500\,K in T$_{\rm eff}$).
The last column gives the spectral types from template fitting using the spectral library 
from NIRSPEC (1), the L and T dwarf archive (2), and the IRTF/SpeX archive (3).
}
 \begin{tabular}{@{\hspace{0mm}}c @{\hspace{1mm}}c @{\hspace{1mm}}c @{\hspace{2mm}}c @{\hspace{2mm}}c @{\hspace{2mm}}c @{\hspace{2mm}}c @{\hspace{2mm}}c@{\hspace{0mm}}}
 \hline
 \hline
ID & Name      &  SpT$_{\rm NIR}$ &   Model &  T$_{\rm eff}$ &  $\log$(g) & [M/H] & Template \cr
 \hline
   & VISTA\,J   &      &         &   K            &    cgs     &       &          \cr
 \hline
 1 & 1559-2214 & L1.0 & bt-settl        & 1800 & 5.0 &  0.0 & L0 (1) \cr % UScoJ1559-2214
 2 & 1613-2124 & L1.5 & bt-dusty        & 1800 & 4.5 &  0.0 & T0 (2) \cr % UScoJ1613-2124
 3 & 1614-2331 & L2.0 & bt-dusty        & 1900 & 4.5 &  0.0 & L0 (1) \cr % UScoJ1614-2331
 4 & 1605-2130 & dL2.0 & bt-settl        & 1800 & 5.0 &  0.0 & L3 (3) \cr % UScoJ1605-2130
 5 & 1611-2215 & L5.0 & bt-settl-cifist & 1700 & 4.0 &  0.0 & L6 (2) \cr % UScoJ1611-2215
 6 & 1609-2222 & dL4.5  & bt-dusty        & 1600 & 5.0 &  0.0 & L2 (1) \cr % UScoJ1609-2222
 7 & 1609-2229 & L3.0 & bt-settl-cifist & 1800 & 5.0 &  0.0 & L2 (1) \cr % UScoJ1609-2229
 8 & 1607-2146 & L4.5 & bt-settl-cifist & 1800 & 5.0 &  0.0 & L2 (1) \cr % UScoJ1607-2146
 9 & 1604-2241 & L6.0 & bt-dusty        & 1600 & 5.5 &  0.0 & L4 (2) \cr % UScoJ1604-2241
10 & 1614-2211 & L4.5 & bt-settl-cifist & 1500 & 2.5 &  0.0 & L4.5 (2) \cr % UScoJ1614-2211
11 & 1605-2403 & L4.5 & bt-settl        & 1600 & 3.5 &  0.0 & L2 (1) \cr % UScoJ1605-2403
12 & 1602-2057 & ???? & bt-dusty        & 1600 & 4.5 &  0.0 & L3 (2) \cr % UScoJ1602-2057
13 & 1604-2134 & L6.0 & bt-settl-cifist & 1750 & 5.0 &  0.0 & L5 (2) \cr % UScoJ1604-2134
14 & 1601-2212 & L7.0 & bt-settl-cifist & 1700 & 4.0 &  0.0 & L6 (2) \cr % UScoJ1601-2212
15 & 1615-2229 & L7.0 & bt-settl-cifist & 1700 & 4.0 &  0.0 & L6 (2) \cr % UScoJ1615-2229
\hline
 \label{tab_USco_GTC_XSH:table_SED_fits}
 \end{tabular}
\end{table}
%

%
%%%%%%%%%%%%%%%%%%%%%%%%%%%%%%%%%%%%%%%%%%%%%%%
%%%%% Table: Bolometric corrections %%%%%
%%%%%%%%%%%%%%%%%%%%%%%%%%%%%%%%%%%%%%%%%%%%%%%
%
\begin{table}
 \centering
 \caption[]{Bolometric corrections in several filters for the three UpSco candidates with 
optical and near-infrared spectroscopy (top) and near-infrared spectroscopy only (bottom).
The $\Delta_{BC_{J}}$ and $\Delta_{BC_{K}}$ represent the difference between our measurements
for UpSco L dwarfs and the mean values taken from the polynomial fits for field dwarfs from
\citet{filippazzo15}.
}
 \begin{tabular}{@{\hspace{0mm}}c @{\hspace{1mm}}c @{\hspace{1mm}}c @{\hspace{1mm}}c @{\hspace{1mm}}c @{\hspace{1mm}}c @{\hspace{1mm}}c @{\hspace{1mm}}c @{\hspace{1mm}}c @{\hspace{1mm}}c@{\hspace{0mm}}}
 \hline
 \hline
ID & Name      &  SpT$_{\rm NIR}$ & m$_{\rm bol}$  &  BC$_{J}$ & BC$_{K}$ & BC$_{w1}$ & BC$_{w2}$ & $\Delta_{BC_{J}}$ & $\Delta_{BC_{K}}$ \cr
 \hline
   &          &      &  mag           &   mag     &   mag    &  mag      &   mag     &    mag &  mag \cr
 \hline
 1 & 1559$-$2214 & L1.0 & 19.34 & 2.01 & 3.37 & 4.13 & 4.56 &    0.09 &    0.16 \cr % USco_cand2
 2 & 1613$-$2124 & L1.5 & 19.57 & 2.04 & 3.46 & 4.08 & 4.58 &    0.14 &    0.22 \cr % USco_cand55
 7 & 1609$-$2229 & L3.0 & 19.92 & 1.80 & 3.62 & 4.40 & 4.57 &    0.12 &    0.32 \cr % USco_cand41
 8 & 1607$-$2146 & L4.5 & 19.75 & 1.63 & 3.39 & 4.65 & 5.20 & $-$0.01 &    0.07 \cr % USco_cand32
10 & 1614$-$2211 & L4.5 & 20.18 & 1.72 & 3.51 & 4.61 & 4.89 &    0.08 &    0.19 \cr % USco_cand71
 5 & 1611$-$2215 & L5.0 & 19.75 & 1.50 & 3.69 & 4.55 & 4.98 & $-$0.09 &    0.37 \cr % USco_cand50
 9 & 1604$-$2241 & L6.0 & 20.18 & 1.64 & 3.49 & 4.71 & 4.92 &    0.14 &    0.19 \cr % USco_cand70
13 & 1604$-$2134 & L6.0 & 20.30 & 1.60 & 3.56 & 4.61 & 4.98 &    0.10 &    0.26 \cr % USco_cand75
14 & 1601$-$2212 & L7.0 & 20.97 & 1.40 & 3.68 & 4.67 & 5.45 & $-$0.04 &    0.42 \cr % USco_cand96
15 & 1615$-$2229 & L7.0 & 20.94 & 1.42 & 3.67 & 4.41 & 5.21 & $-$0.01 &    0.40 \cr % USco_cand97
\hline
 1 & 1559$-$2214 & L1.0 & 19.41 & 2.02 & 3.38 & 4.20 & 4.63 &    0.09 &    0.16 \cr % USco_cand2
 2 & 1613$-$2124 & L1.5 & 19.60 & 2.06 & 3.49 & 4.10 & 4.61 &    0.16 &    0.25 \cr % USco_cand55
 7 & 1609$-$2229 & L3.0 & 19.92 & 1.79 & 3.59 & 4.38 & 4.55 &    0.00 &    0.30 \cr % USco_cand41
 8 & 1607$-$2146 & L4.5 & 19.75 & 1.60 & 3.35 & 4.61 & 5.16 &    0.04 &    0.03 \cr % USco_cand32
10 & 1614$-$2211 & L4.5 & 20.17 & 1.73 & 3.50 & 4.60 & 4.88 &    0.09 &    0.18 \cr % USco_cand71
 5 & 1611$-$2215 & L5.0 & 19.75 & 1.51 & 3.68 & 4.54 & 4.97 & $-$0.09 &    0.36 \cr % USco_cand50
 9 & 1604$-$2241 & L6.0 & 20.18 & 1.66 & 3.50 & 4.72 & 4.93 &    0.16 &    0.20 \cr % USco_cand70
13 & 1604$-$2134 & L6.0 & 20.30 & 1.61 & 3.56 & 4.61 & 4.97 &    0.11 &    0.26 \cr % USco_cand75
14 & 1601$-$2212 & L7.0 & 20.95 & 1.37 & 3.66 & 4.65 & 5.43 & $-$0.06 &    0.39 \cr % USco_cand96
15 & 1615$-$2229 & L7.0 & 20.90 & 1.39 & 3.64 & 4.37 & 5.17 & $-$0.04 &    0.37 \cr % USco_cand97
\hline
 \label{tab_USco_GTC_XSH:table_BC}
 \end{tabular}
\end{table}
%

%
%%%%%%%%%%%%%%%%%%%%%%%%%%%%%%%%%%%%%%%%%%%%%%%
%%%%% Figure: BC vs SpType  %%%%%
%%%%%%%%%%%%%%%%%%%%%%%%%%%%%%%%%%%%%%%%%%%%%%%
%
\subsection{Wide companions}
\label{USco_GTC_XSH:wide_comp}

We looked for wide companions cross-matching our list of 15 VISTA candidates with known UpSco members
\citep{luhman12c} within a matching radius of 30 arcsec (corresponding to an upper projected physical
separation of 4350 au at the distance of USco) but none was found. The catalogue of UpSco members
from \citet*{luhman12c} included the most massive members (e.g.\ $\tau$\,Sco; $J$\,=\,3.4 mag) all 
the way down to coolest members known at that time with spectral types around M9/L0 
($J$\,$\sim$\,16.6--17.2 mag). We repeated the same process with the list of 16 L0--L2 dwarfs in 
Table 1 of \citet{lodieu08a} without success. In total, we searched for wide companions around 
29 L dwarfs (2 in common between the aforementioned samples), suggesting a frequency of wide companions 
to early-L dwarfs below 3.4\% at separations larger than $\sim$1.5 arcsec and spectral types between 
B0 and M9/L0, corresponding to $\sim$220 au at the distance
of UpSco. Our conclusions also supplement the low frequency of close companions ($<$7\%) from an
adaptive optic survey of 14 L0--L1 from \citet{lodieu08a} by \citet{biller11}. We conclude
that wide pairs with components separated by more than $\sim$200 au are rare. Only a few systems
have indeed been discovered in USco: USco\,CTIO\,108B \citep[$\sim$670 au;][]{ardila00,bejar08},
1RXS\,J160929.1$-$210524 \citep[$\sim$330 au;][]{lafreniere08,lafreniere10a}, and HIP\,78530B
\citep[$\sim$700 au;][]{lafreniere11}. Our upper limit is consistent with the frequency of $\sim$4\%
of wide (250--1000 au) brown dwarf (0.005--0.040 M$_{\odot}$) companions around higher mass members 
of UpSco \citep[0.2--10 M$_{\odot}$;][]{lafreniere14} and the $\sim$0.6$\pm$0.3\% occurrence of
wide (400--4000 au) down to 0.005 M$_{\odot}$ around M7.5--M9 members \citep{aller13}.

The closest potential system we found cross-matching the above catalogues is located at 69.3 arcsec
or $\sim$10,000 au. It would be composed of the X-ray emitter M1 dwarf 2MASS\,J16063741$-$2108404
(called ``primary'' here) and VISTA\,J16060629$-$2335133 classified as a L0 in the infrared 
\citep{lodieu08a} and M9 in the optical (this work). The proper motion in right ascension and
declination extracted from the UKIDSS GCS catalogue for the primary and secondary are
$-$10.25$\pm$4.11, $-$15.24$\pm$4.11) and ($-$6.60$\pm$2.46, $-$19.21$\pm$2.46) mas/yr, respectively.
The primary has a mass of 0.6 M$_{\odot}$
and has a previously known companion located at 1.279$\pm$0.003 arcsec (position angle of 33.9$\pm$0.3 degrees)
equivalent to $\sim$185 au at the distance of UpSco \citep{koehler00a}. The brightness ratio 
is 0.917$\pm$0.001 in the $K$-band. The catalogue of \citet*{luhman12c} contains 863 B--M members
over a (conservative) area of $\sim$72 square degrees, implying $\sim$0.0033 sources per square arcmin.
The probability of finding one member of UpSco in 4.3 square arcmin is of the order of 41.5\%.
At this stage we can not draw any firm conclusion: the probability indicates that companionship 
might be either true or not. If true, this system composed of a low-mass star and planetary-mass
companion would be of great interest and support early theoretical suggestions that primary stars 
of multiple systems with wide brown dwarf companions tend to be binary themselves 
\citep[e.g.][]{delgado_donate03}.

%
%%%%%%%%%%%%%%%%%%%%%%%%%%%%%%%%%%%%%
%%%%%%%  Conclusions %%%%%%%
%%%%%%%%%%%%%%%%%%%%%%%%%%%%%%%%%%%%%
%
\section*{Conclusions and future work}

We presented a dedicated photometric and spectroscopic follow-up of photometric and astrometric candidates
identified in a deep and wide near-infrared survey of 13.5 square degrees in the central region of UpSco. 
Throughout this work, we argue in favour of a high success rate of our photometric and
astrometric selection, higher than 80\%. Indeed, two sources are rejected with additional
optical photometry among 28 candidates with $i$-band photometric follow-up. Moreover, two candidates
are classified as field dwarfs and another one remains doubtful, suggesting at most three contaminants 
in our near-infrared spectroscopic sample of 15 candidates.
As a consequence, the conclusion drawn earlier in L13 remains valid: the UpSco mass
function in the planetary regime is flat or most likely decreasing.

The main results of our study can be summarised as follows: 
\begin{itemize}
\item [$\bullet$] We derived Sloan $i$-band magnitudes for 25 sources and placed lower limits for
another three substellar candidates in UpSco. 
\item [$\bullet$] We present VLT/X-shooter near-infrared (1.0--2.5 $\mu$m) spectra for 15 candidates 
fainter than $J$\,$\sim$\,17 mag, equivalent to $\sim$10 Jupiter masses at the age ($\sim$5--10 Myr) 
and distance of UpSco.
\item [$\bullet$] We present VLT/X-shooter optical $+$ near-infrared spectra of six late-M dwarf
candidates previously recognised as members and classified in the optical.
\item [$\bullet$] We present GTC/OSIRIS low-resolution optical spectra for all L-type members 
with near-infrared spectra taken with the X-shooter instrument.
\item [$\bullet$] We exploit the WISE mid-infrared survey to extend the current UpSco sequence of
members to fainter magnitudes and lower masses and look for excesses.
\item [$\bullet$] We classify our targets based on their SED in the near-infrared, rejecting only
two of the 15 targets with spectroscopy and cast doubt on another one. This fact demonstrates the strength 
of our photometric and astrometric selections to pick up substellar and planetary-mass members in UpSco.
\item [$\bullet$] We observe a trend towards earlier spectral types in the optical compared to the
infrared for several members classified as M9--L1 in \citet{lodieu08a} and five of our
targets with both optical and infrared spectra. This trend seems to affect sources later than
M9 all the way to late-L dwarfs.
\item [$\bullet$] We detect H$\alpha$ in emission in four L-type sources. Two objects, 
VISTA\,J1611$-$2215 and VISTA\,J1607$-$2146, also have likely mid-infrared flux excesses, thus 
supporting the disk accretion scenario and providing evidence of the persistence of accretion at very 
low masses  and at the age of the UpSco association (5--10 Myr).
\item [$\bullet$] We measure the pseudo-equivalent widths of gravity-sensitive doublets present in 
the near-infrared, whose strength confirms the youth of our candidates.
\item [$\bullet$] We present a wide variety of spectral indices comparing our sample with a sample
of young L dwarfs and several field L0--L9 dwarfs. We show that a few indices (H-cont, CO, CH4-H, FeH-H)
are highly sensitive to gravity (or age) while most of the water-based indices are good proxies
for spectral typing. {\bf{The VO absorption features at $\sim$1.06 $\mu$m and $\sim$1.18 $\mu$m are 
still present in late-L dwarfs at the age of UpSco.}}
\item [$\bullet$] We measure the radial velocities of confirmed members from the X-shooter infrared
spectra. the distribution of velocities is consistent with the mean value reported for higher mass members.
\item [$\bullet$] We infer effective temperatures and gravities fitting the full photometric spectral
energy distribution from 0.8 to 4.5 $\mu$m with different sets of theoretical models. We find that
these physical parameters tend to be underestimated with respect to the parameters derived from the
spectral energy distribution model atmosphere fits to the spectra.
\item [$\bullet$] We derived bolometric corrections for our L dwarf members and found similar values
and trends for the BC$_{J}$ over the L1--L7 range but a slight different tendency for BC$_{K}$ for young L dwarfs 
compared to field L dwarfs.
\item [$\bullet$] We searched for higher mass wide companions to our planetary-mass members and place
an upper limit of 3.4\% on the frequency of systems with projected physical separations greater than 100 au.
\end{itemize}

Our deep survey is currently limited by the depth of the $Y$-band because the final stack is only
0.7 mag deeper in $Y$ than in $J$ (21.2 mag vs 20.5 mag completeness limits) while the colours of
the coolest sources confirmed spectroscopically are in the 1.4--1.8 mag range
(see Fig.\ 3 in L13). Hence, our deep $J$-band might contain some yet-to-be-found 
T-type dwarfs with an age of $\sim$5--10 Myr. 
Assuming that our survey reaches members with masses as low as 5 M$_{\rm Jup}$, we might still be
far from being able to test opacity limit for the fragmentation, i.e.\ the mass at which one object
is unable to contract further because it can not radiate its heat to collapse further
\citep{low76,rees76,boss01}. Indeed, according to the COND03 models \citep{baraffe02},
a 1 M$_{\rm Jup}$ is 4.0 and 5.4 mag fainter in $J$ than a 5 M$_{\rm Jup}$ brown dwarf at 5 and 10 Myr,
respectively. In the future, we might get some hints on the theory of the fragmentation with
upcoming space missions like the James Webb Space Telescope \citep[JWST;][]{clampin08}\footnote{www.jwst.nasa.gov/},
the Wide Field Infrared Survey telescope (WFIRST)\footnote{http://wfirst.gsfc.nasa.gov/}, and
Euclid \citep{mellier16a}\footnote{http://sci.esa.int/euclid/}
as well as the Large Synoptic Survey Telescope from the ground (Ivezic et al.\ 2008)\footnote{www.lsst.org}.

%
%%%%%%%%%%%%%%%%%%%%%%%%%%%%%%%%%%%%%
%%%%%%%  ACKNOWLEDGEMENTS %%%%%%%
%%%%%%%%%%%%%%%%%%%%%%%%%%%%%%%%%%%%%
%
\section*{Acknowledgments}
NL was partly funded by the Ram\'on y Cajal fellowship number 08-303-01-0\@.
NL and VJSB are supported by programme AYA2015-69350-C3-2-P and MRZO by programme
AYA2016-79425-C3-2-P from Spanish Ministry of Economy and Competitiveness (MINECO).

KPR acknowledges CONICYT PAI Concurso Nacional de Inserci\'on en la Academia,
Convocatoria 2016 Folio PAI79160052\@.
NL thanks Kevin Luhman for providing his own spectra of young brown dwarfs.
{\bf{We thank the referee for his/her comments that improved the clarity of the paper.}}

This work is based on observations (programmes GTC4-14A and GTC25-16A; PI Lodieu) 
made with the Gran Telescopio Canarias (GTC), operated on the island of La Palma
in the Spanish Observatorio del Roque de los Muchachos of the Instituto de
Astrof\'isica de Canarias. We are thankful to F.\ Garz\'on and the EMIR consortium
for the development of this instrument on the GTC\@.

Based on observations collected at the European Organisation for Astronomical 
Research in the Southern Hemisphere under ESO programme(s) 095.C-0812(A) and 089-C.0102(ABC).

Based on data from the UKIRT Infrared Deep Sky Survey (UKIDSS). The UKIDSS project 
is defined in \citet{lawrence07} and uses the UKIRT Wide Field Camera \citep[WFCAM;][]{casali07}. 
The photometric system is described in \citet{hewett06} and the calibration is described 
\citet{hodgkin09}. The pipeline processing and science archive are described in 
Irwin et al. (2009, in prep) and \citet{hambly08}.

This publication makes use of VOSA, developed under the Spanish Virtual Observatory project 
supported from the Spanish MICINN through grant AyA2011-24052.

This research has made use of the Simbad and Vizier \citep{ochsenbein00}
databases, operated at the Centre de Donn\'ees Astronomiques de Strasbourg
(CDS), and of NASA's Astrophysics Data System Bibliographic Services (ADS).
This research has also made use of the IRTF spectral library at
http://irtfweb.ifa.hawaii.edu/$\sim$spex/IRTF\_Spectral\_Library/index.html.

%
%%%%%%%%%%%%%%%%%%%%%%%%%%%%%%%%%%%%%%%%%%
%%%%%%%%  Bibliography  %%%%%%%%
%%%%%%%%%%%%%%%%%%%%%%%%%%%%%%%%%%%%%%%%%%
%
\bibliographystyle{mn2e}
\bibliography{../../AA/mnemonic,../../AA/biblio_old}

%
%%%%%%%%%%%%%%%%%%%%%%%%%%%%%%%%%%%%%%%%%%
%%%%%%%%  Appendix  %%%%%%%%
%%%%%%%%%%%%%%%%%%%%%%%%%%%%%%%%%%%%%%%%%%
%
\appendix

\section{Near-infrared spectral indices}
\label{app_USco_GTC_XSH:spectral_indices}

%
%%%%%%%%%%%%%%%%%%%%%%%%%%%%%%%%%%%%%
%%%%% Table: SED fits from SVO %%%%%
%%%%%%%%%%%%%%%%%%%%%%%%%%%%%%%%%%%%%
%
\begin{table*}
\tiny
 \centering
 \caption[]{Values of spectral indices defined in the literature by various authors and computed for
field L dwarfs, young L dwarfs, and our UpSco L-type objects. References for the definitions of the
spectral indices are: \citet{reid01a}, \citet*{tokunaga99}, \citet{burgasser06a}, \citep{geballe02},
\citet{burgasser02}, \citet{allers07}, \citet{mclean03}, \citet{slesnick04}, \citet{allers13},
and \citep{testi01}.
}
 \begin{tabular}{@{\hspace{0mm}}c @{\hspace{0.3mm}}c @{\hspace{0.3mm}}c @{\hspace{0.3mm}}c @{\hspace{0.3mm}}c @{\hspace{0.3mm}}c @{\hspace{0.3mm}}c @{\hspace{0.3mm}}c @{\hspace{0.3mm}}c @{\hspace{0.3mm}}c @{\hspace{0.3mm}}c @{\hspace{0.3mm}}c @{\hspace{0.3mm}}c @{\hspace{0.3mm}}c @{\hspace{0.3mm}}c @{\hspace{0.3mm}}c @{\hspace{0.3mm}}c @{\hspace{0.3mm}}c @{\hspace{0.3mm}}c @{\hspace{0.3mm}}c @{\hspace{0.3mm}}c @{\hspace{0.3mm}}c @{\hspace{0.3mm}}c @{\hspace{0.3mm}}c @{\hspace{0.3mm}}c @{\hspace{0.3mm}}c @{\hspace{0.3mm}}c @{\hspace{0.3mm}}c @{\hspace{0.3mm}}c @{\hspace{0.3mm}}c @{\hspace{0.3mm}}c @{\hspace{0.3mm}}c @{\hspace{0.3mm}}c @{\hspace{0.3mm}}c @{\hspace{0.3mm}}c@{\hspace{0mm}}}
 \hline
 \hline
SpT & H2OA & H2OB & K1 & K2 & H2OJ & CH4J & H2OH & CH4H & H2OK & CH4K & H2O15 & KJ & KH & YJ & HJ & H2Oc & Jcont & Hcont & FeHJ & FeHH & CO & H2OD2 & H2O & H2OD & H2O1 & H2O2 & sHJ & sKJ & sH2OJ & sH2OH1 & sH2OH2 & sH2OK & VOz & Target \cr
 \hline
    &  & &  &  &  & &  & &  &  &  & &  & &  &  &  & &  & &  &  &  & &  & &  &  &  & &  & & & \cr
 \hline
10.0 &  0.705 &   0.827 &  $-$0.885 &   0.043 &   0.961 &   0.909 &   0.855 &   1.110 &   0.969 &   1.103 &   1.194 &   0.366 &   0.590 &   0.768 &   0.670 &   1.544 &   1.068 &   1.007 &   0.942 &   0.908 &   0.775 &   0.883 &   1.122 &   1.073 &   0.684 &   0.887 &   0.368 &   0.894 &   0.060 &   0.257 &   0.376 &   0.033 & 2MASSJ00274197p0503417\_L0b \cr
\hline
 \label{tab_USco_GTC_XSH:table_spec_indices}
 \end{tabular}
\end{table*}
%

%
%%%%%%%%%%%%%%%%%%%%%%%%%%%%%%%%%%%%%%%%%%%%%%%
%%%%% Figure: Spectral Indices vs SpType %%%%%
%%%%%%%%%%%%%%%%%%%%%%%%%%%%%%%%%%%%%%%%%%%%%%%
%
\begin{figure*}
  \centering
  \includegraphics[width=0.49\linewidth, angle=0]{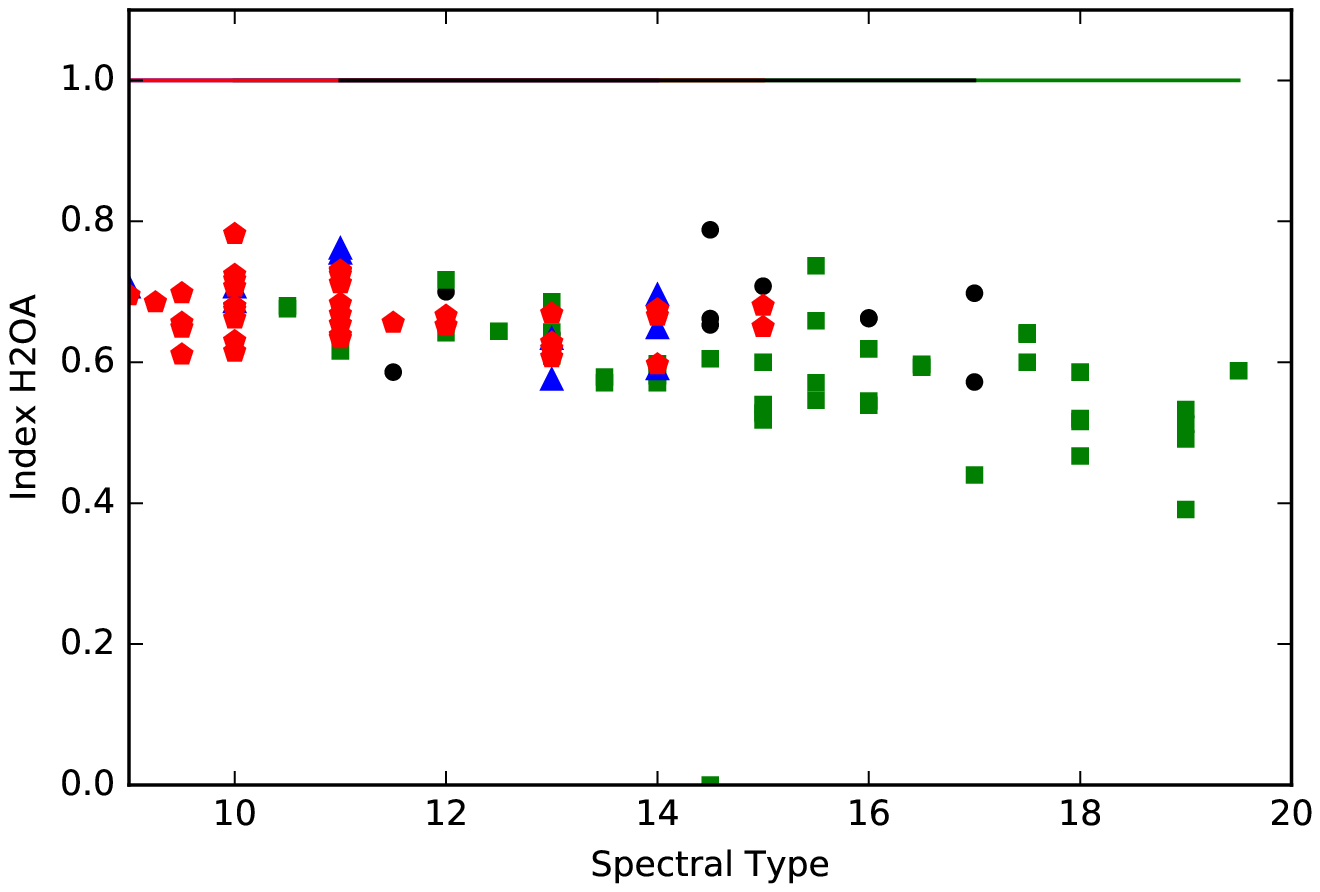}
  \includegraphics[width=0.49\linewidth, angle=0]{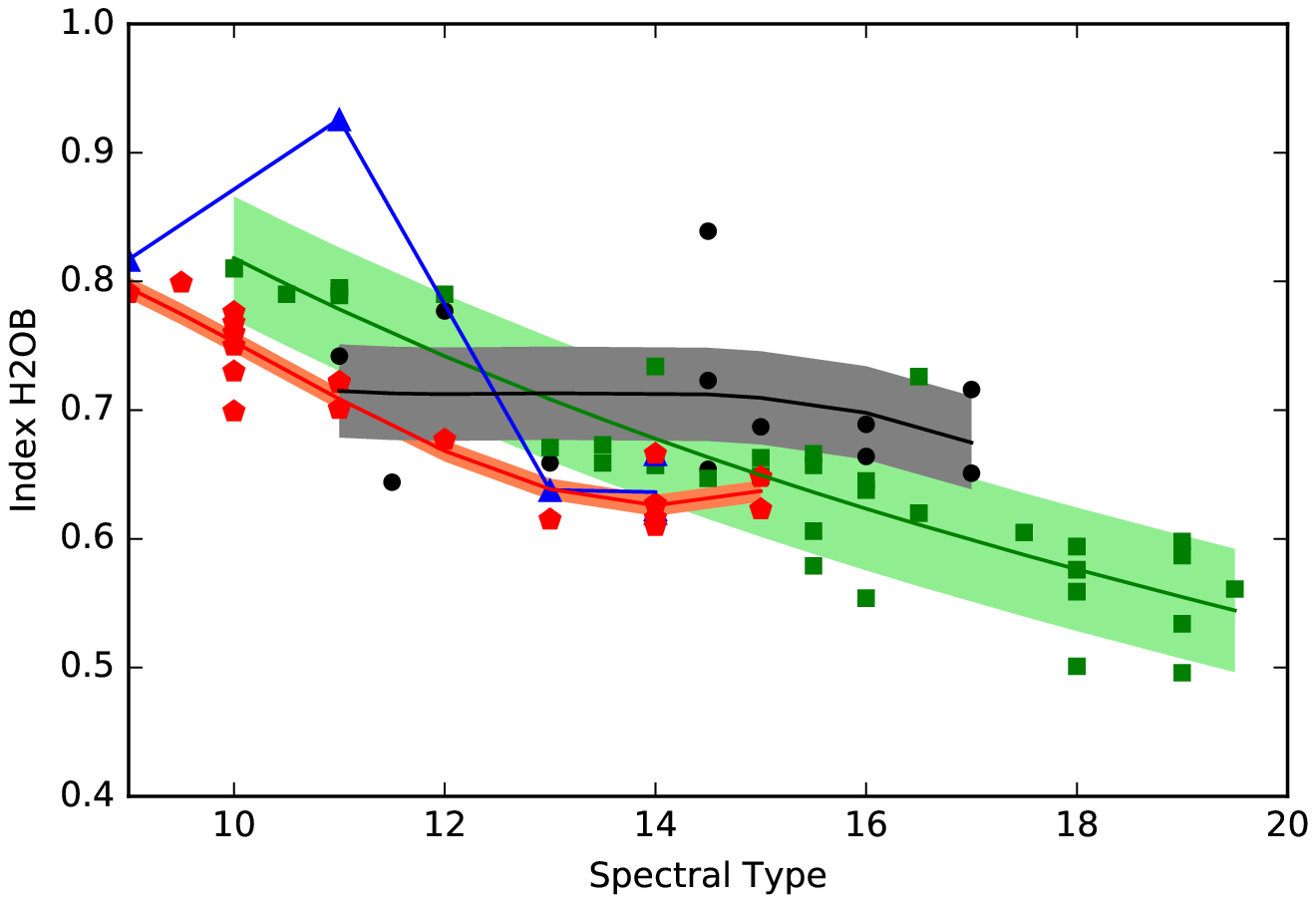}
  \includegraphics[width=0.49\linewidth, angle=0]{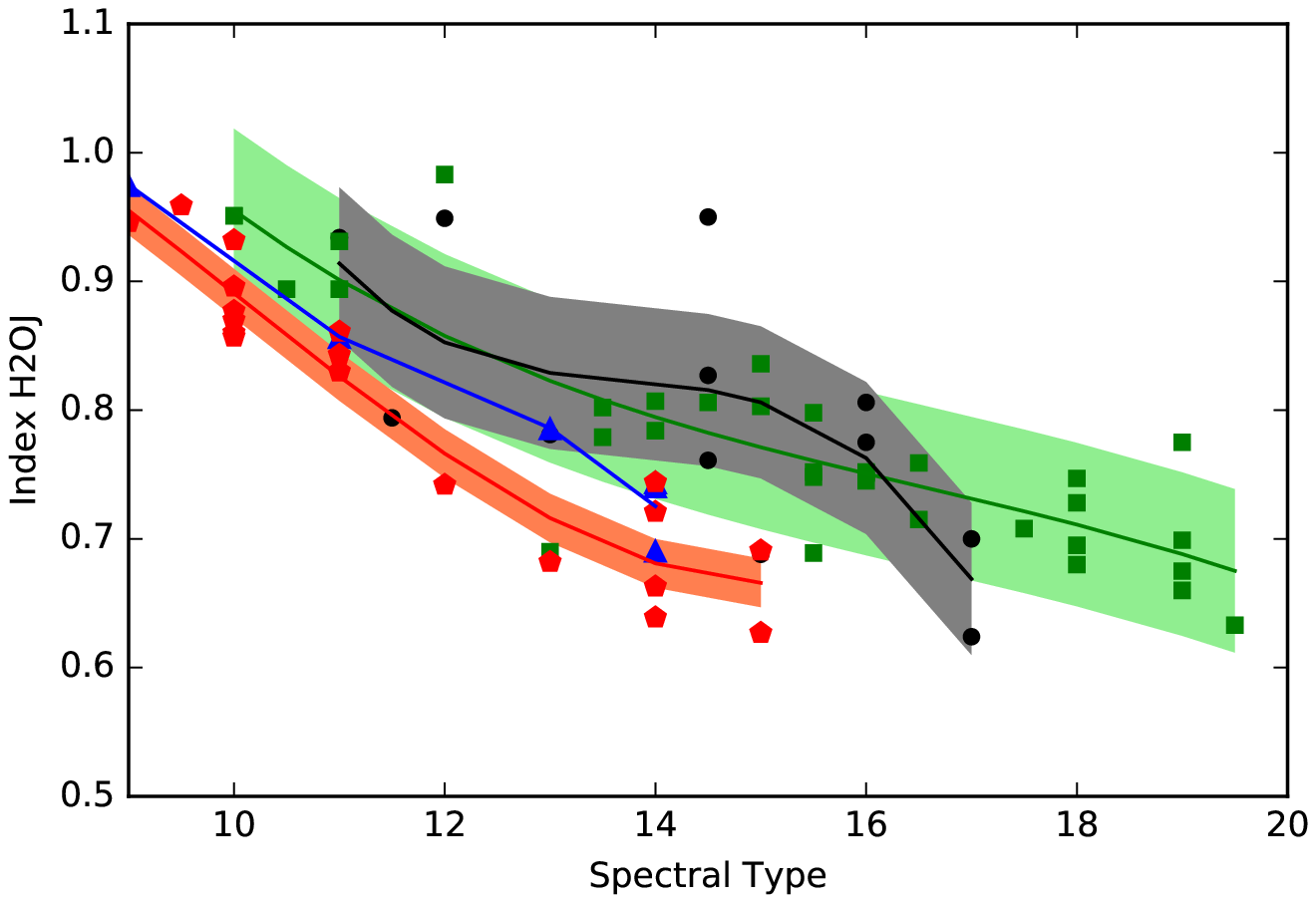}
  \includegraphics[width=0.49\linewidth, angle=0]{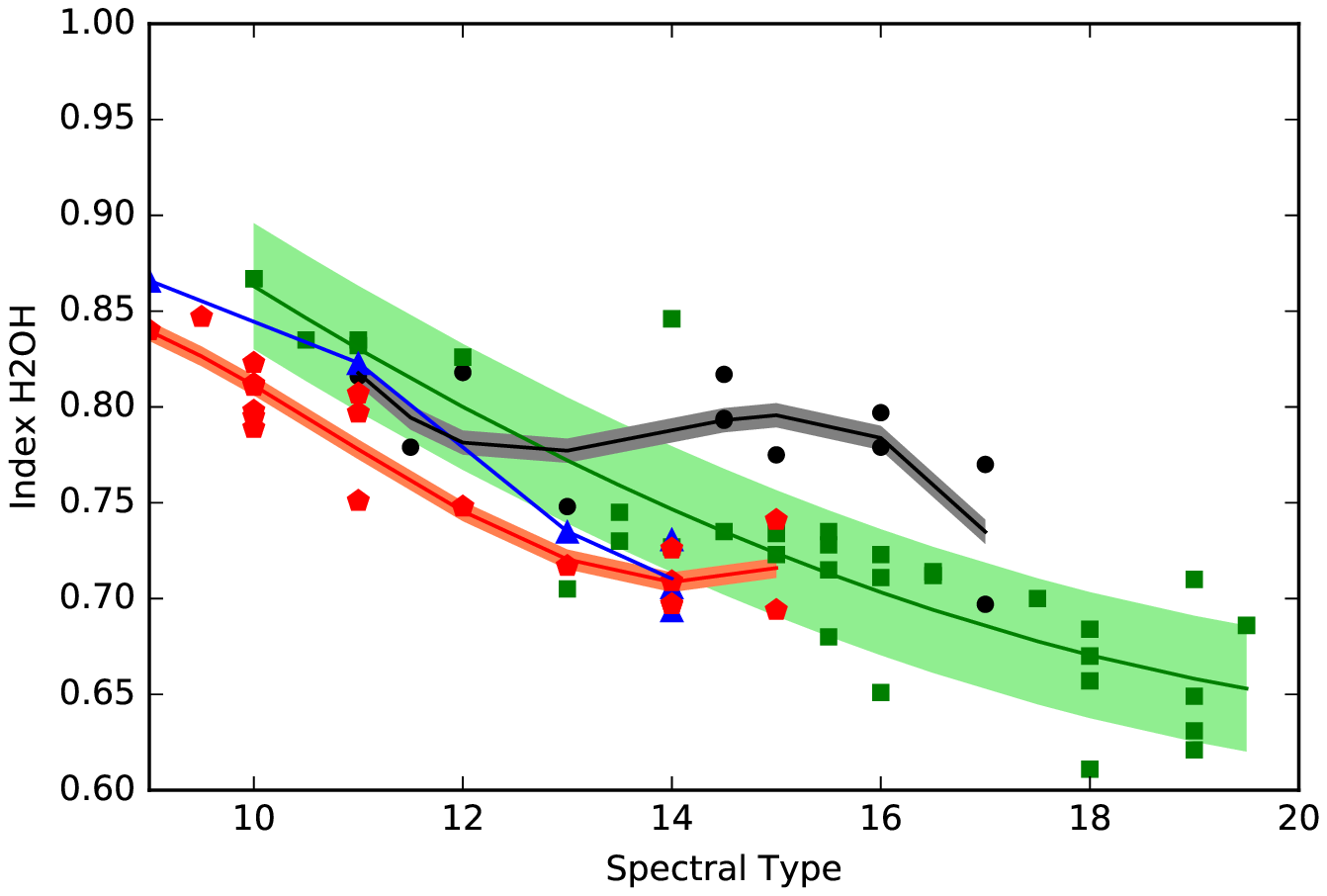}
  \includegraphics[width=0.49\linewidth, angle=0]{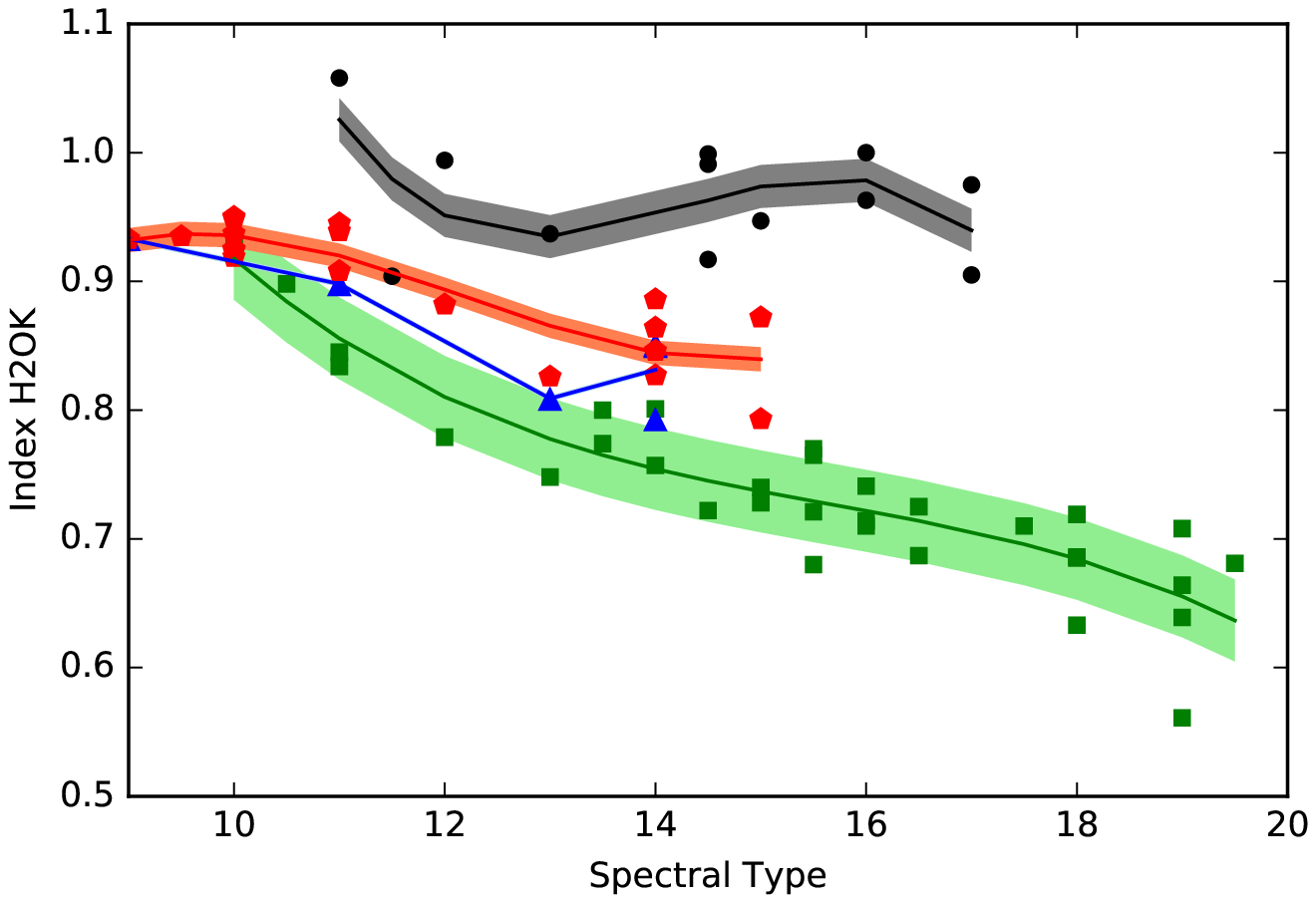}
  \includegraphics[width=0.49\linewidth, angle=0]{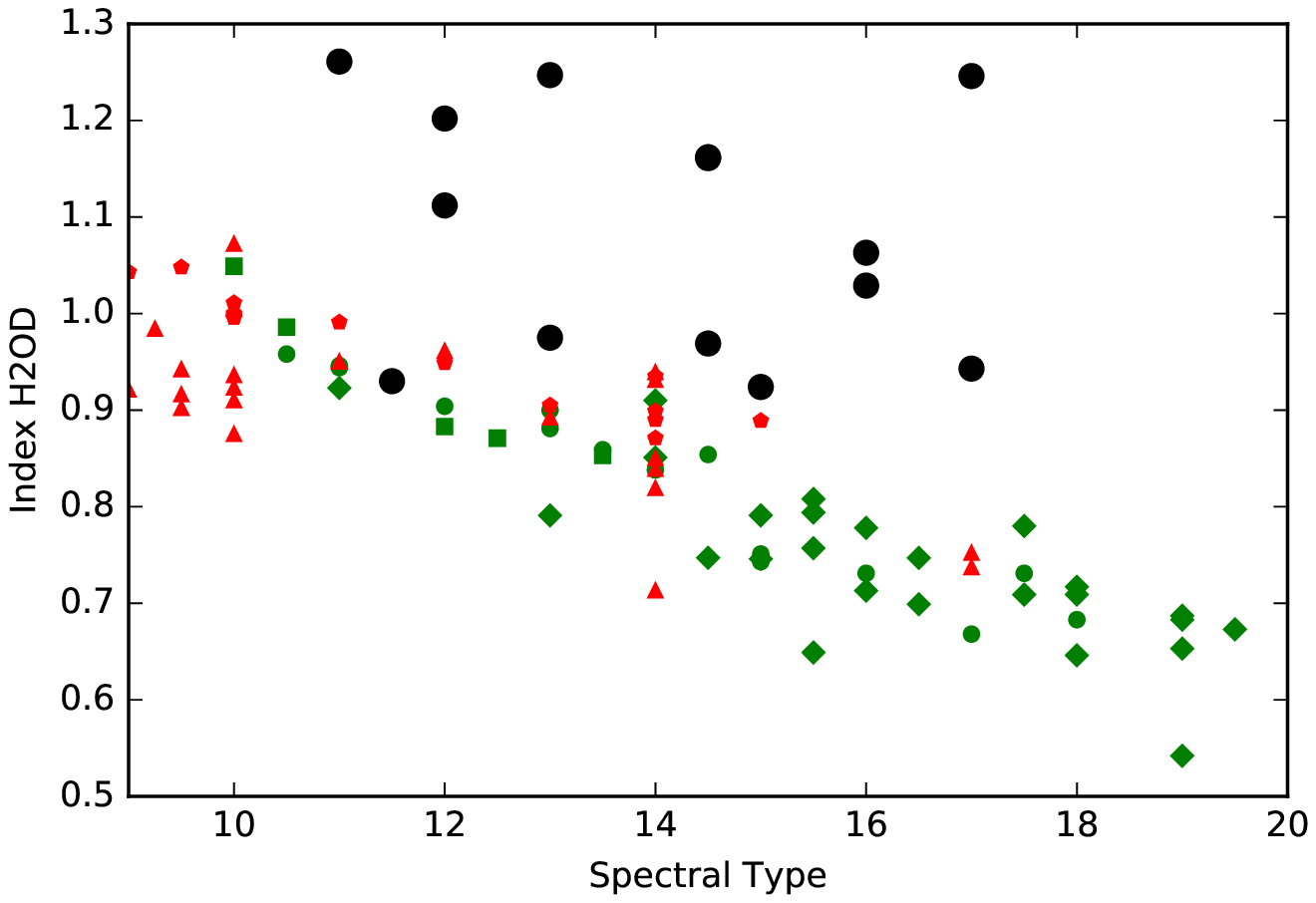}
  \includegraphics[width=0.49\linewidth, angle=0]{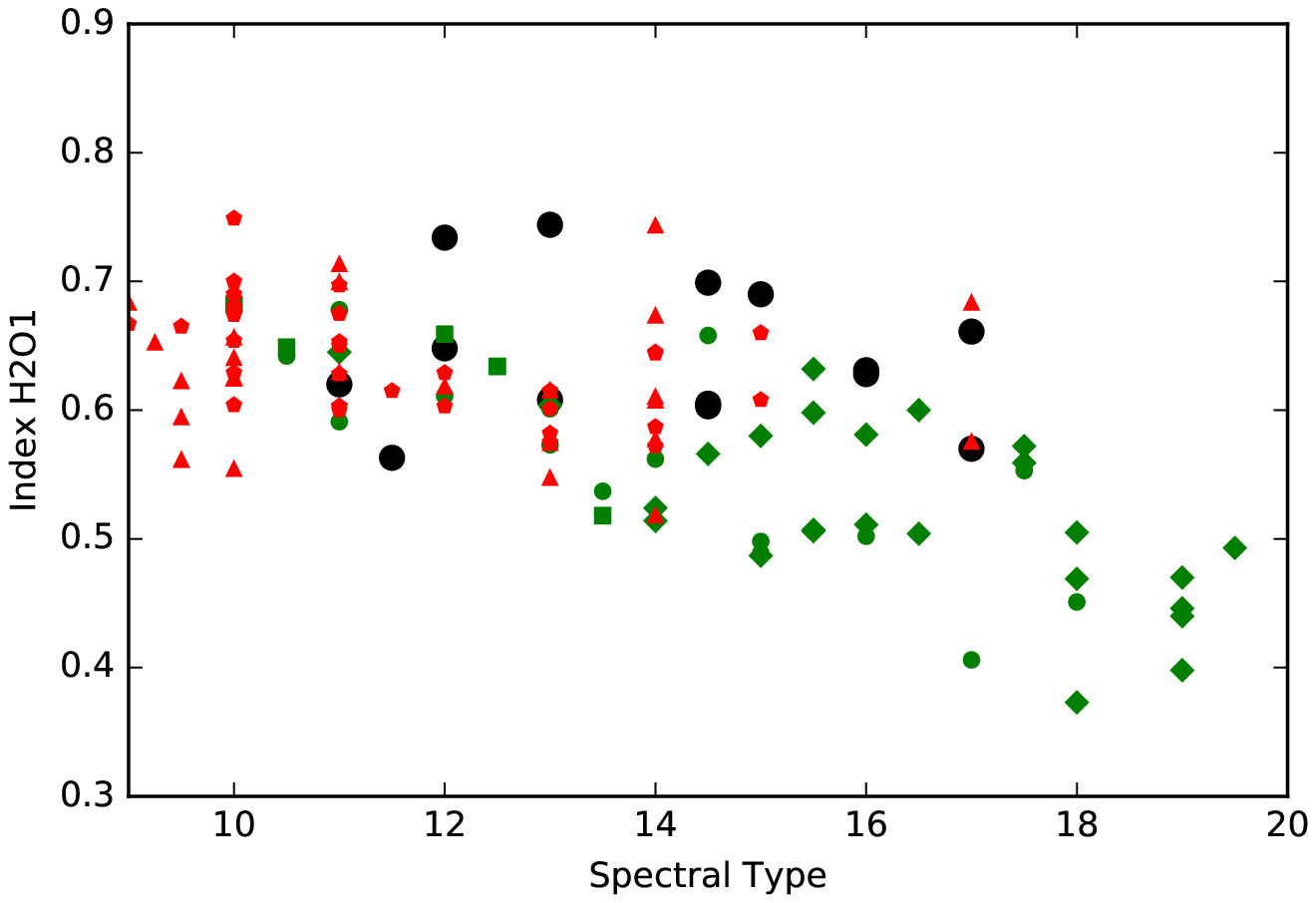}
  \includegraphics[width=0.49\linewidth, angle=0]{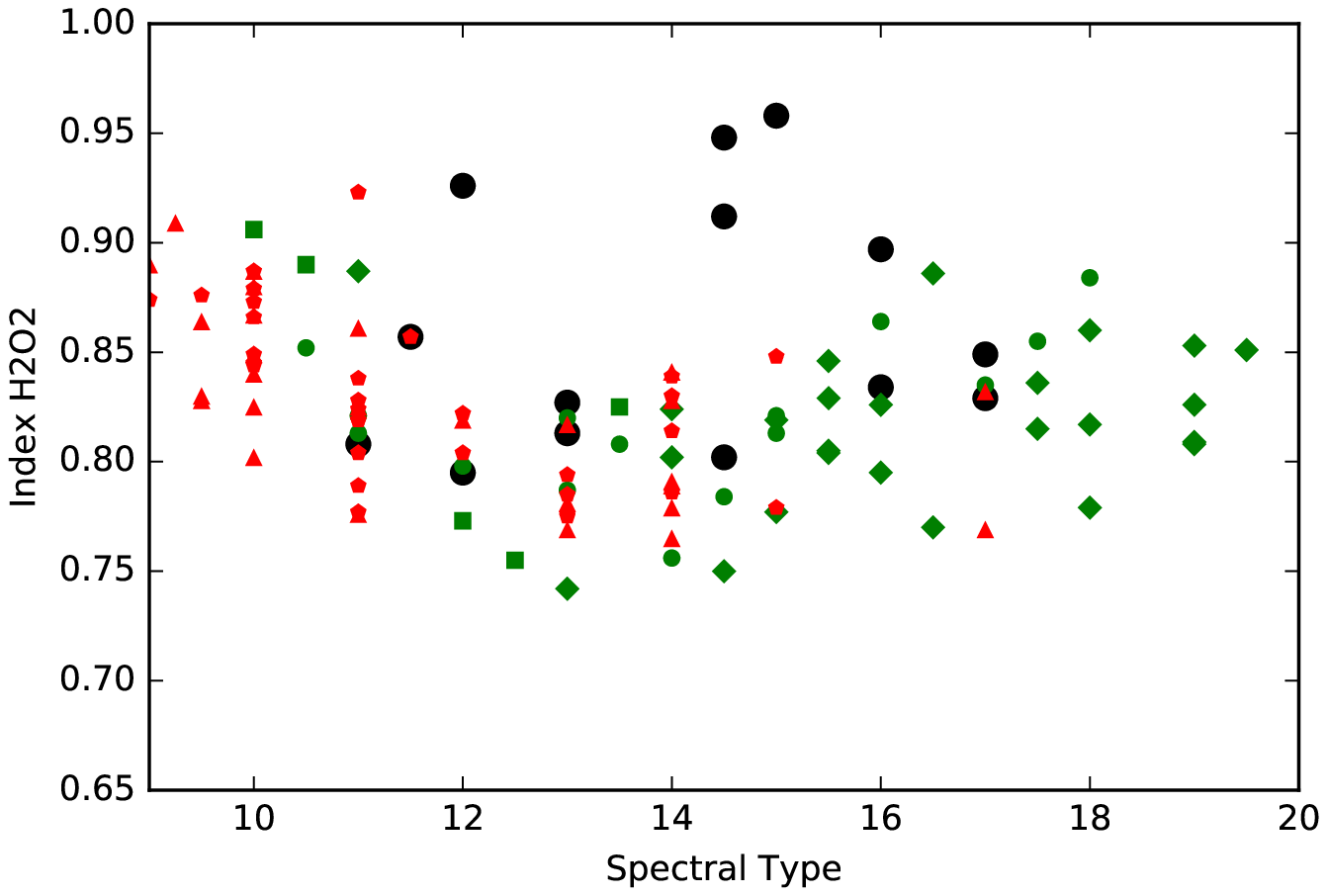}
  \caption{Near-infrared spectral indices defined in the literature vs.\ NIR
spectral types for our UpSco candidates (black dots), young L dwarfs 
\citep[red filled symbols; $\beta$\,=\,triangles; $\gamma$\,=\,pentagons;][]{allers13,bonnefoy14a,gagne15c}
and field L dwarfs \citep[green symbols;][]{reid01a,geballe02,mclean03,knapp04,golimowski04a,chiu06}.
}
  \label{fig_USco_GTC_XSH:plot_spectral_indices_1}
\end{figure*}

\begin{figure*}
  \centering
  \includegraphics[width=0.49\linewidth, angle=0]{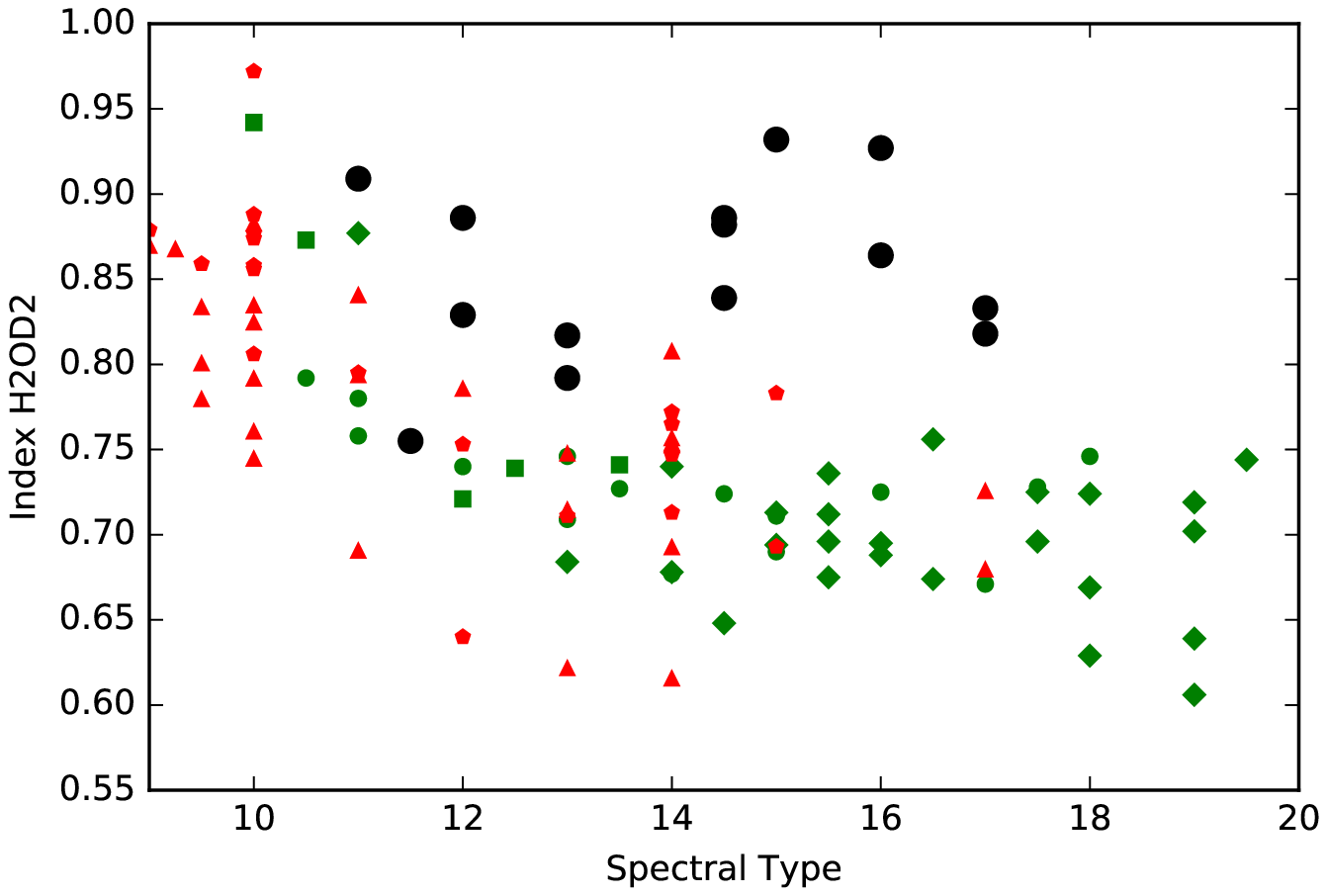}
  \includegraphics[width=0.49\linewidth, angle=0]{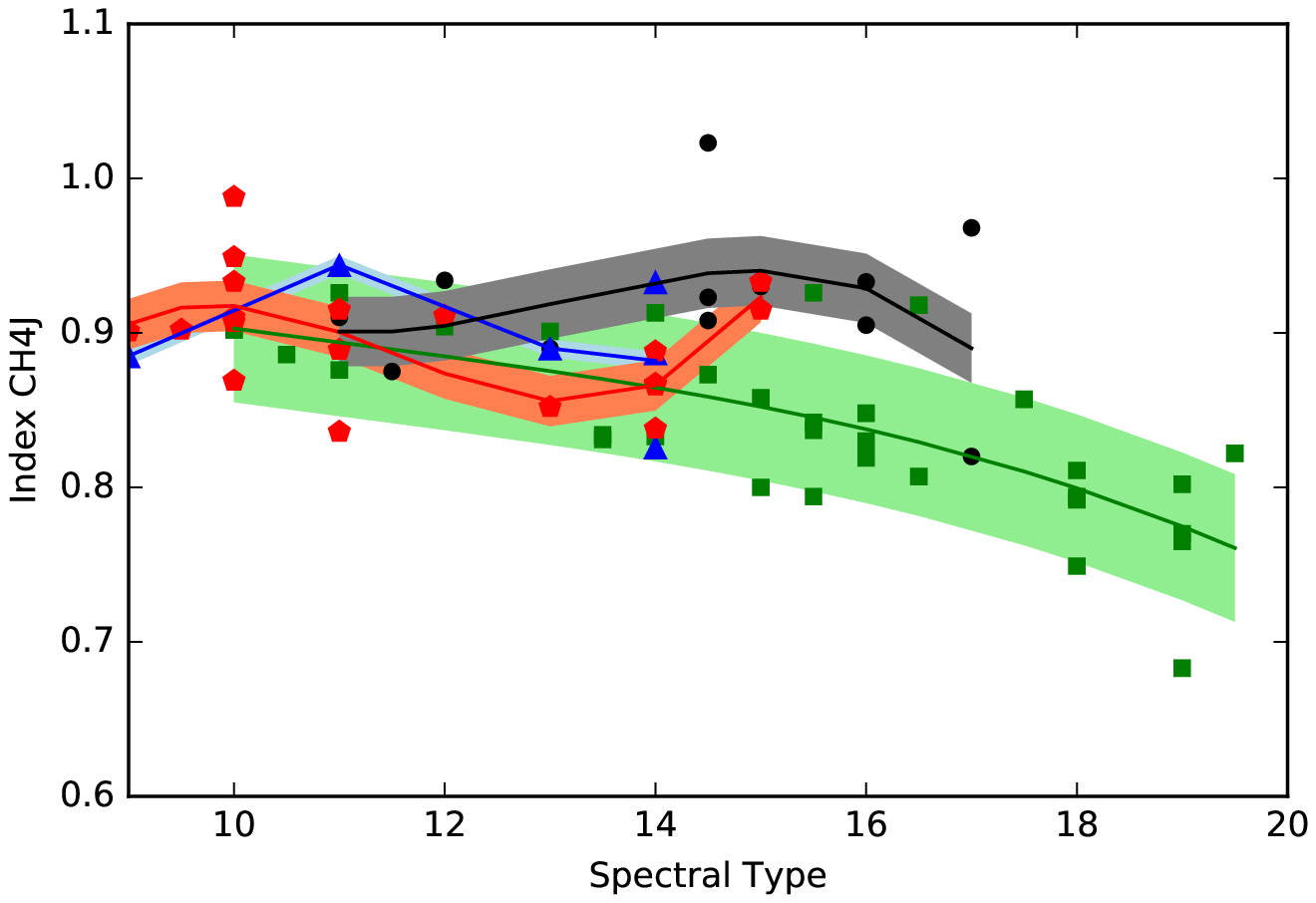}
  \includegraphics[width=0.49\linewidth, angle=0]{plot_indices_USco_XSH_CH4H.ps}
  \includegraphics[width=0.49\linewidth, angle=0]{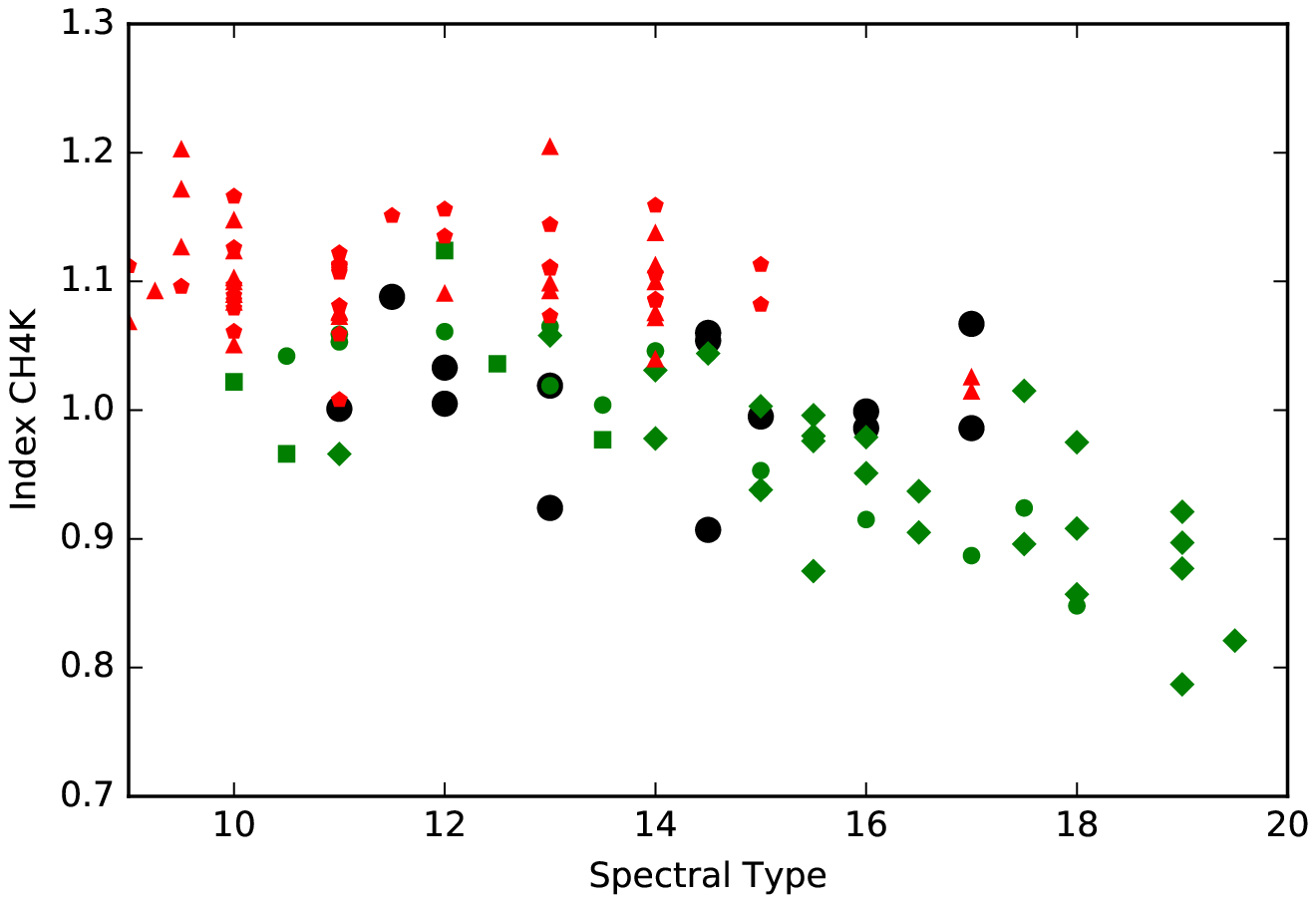}
  \includegraphics[width=0.49\linewidth, angle=0]{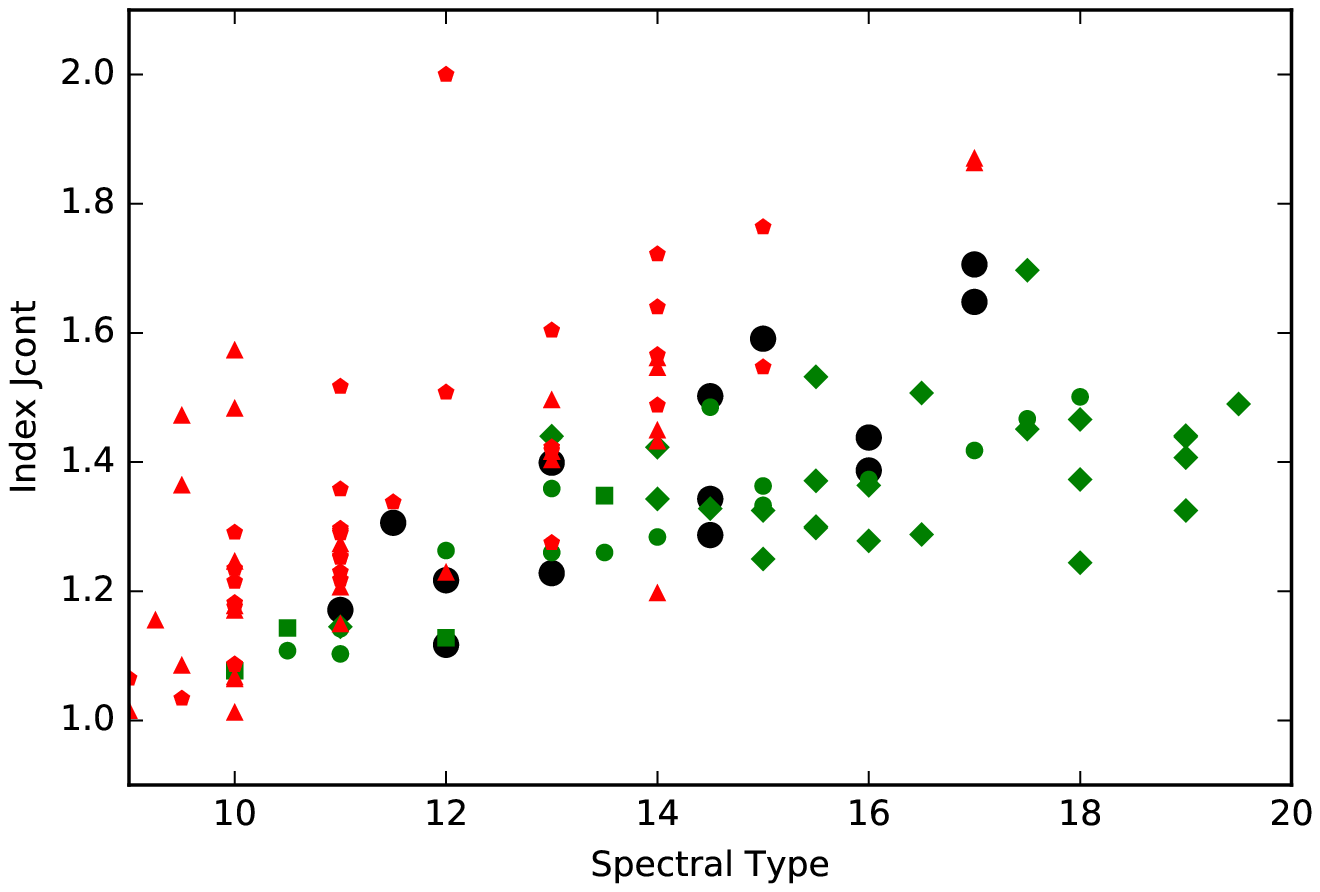}
  \includegraphics[width=0.49\linewidth, angle=0]{plot_indices_USco_XSH_Hcont.ps}
  \includegraphics[width=0.49\linewidth, angle=0]{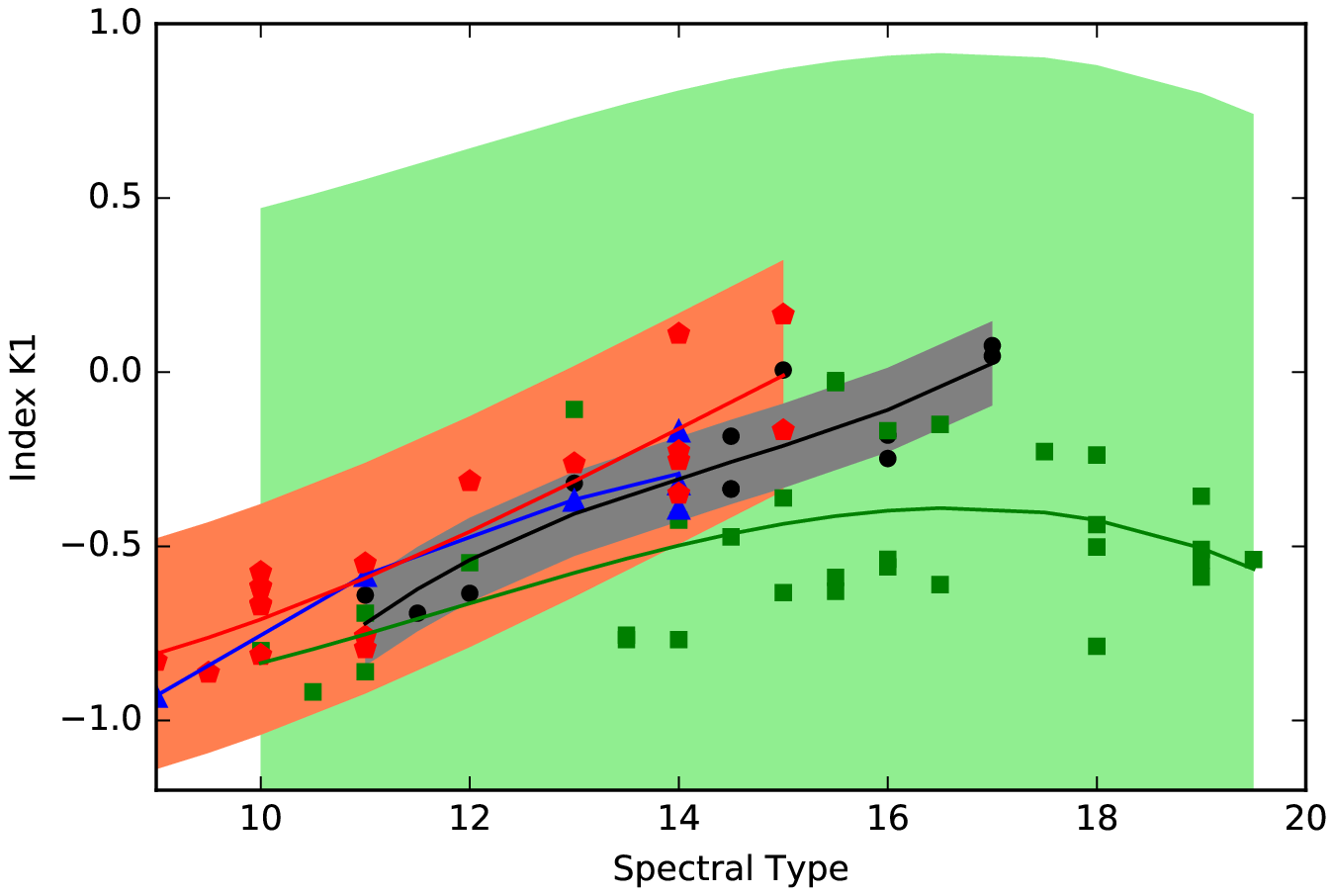}
  \includegraphics[width=0.49\linewidth, angle=0]{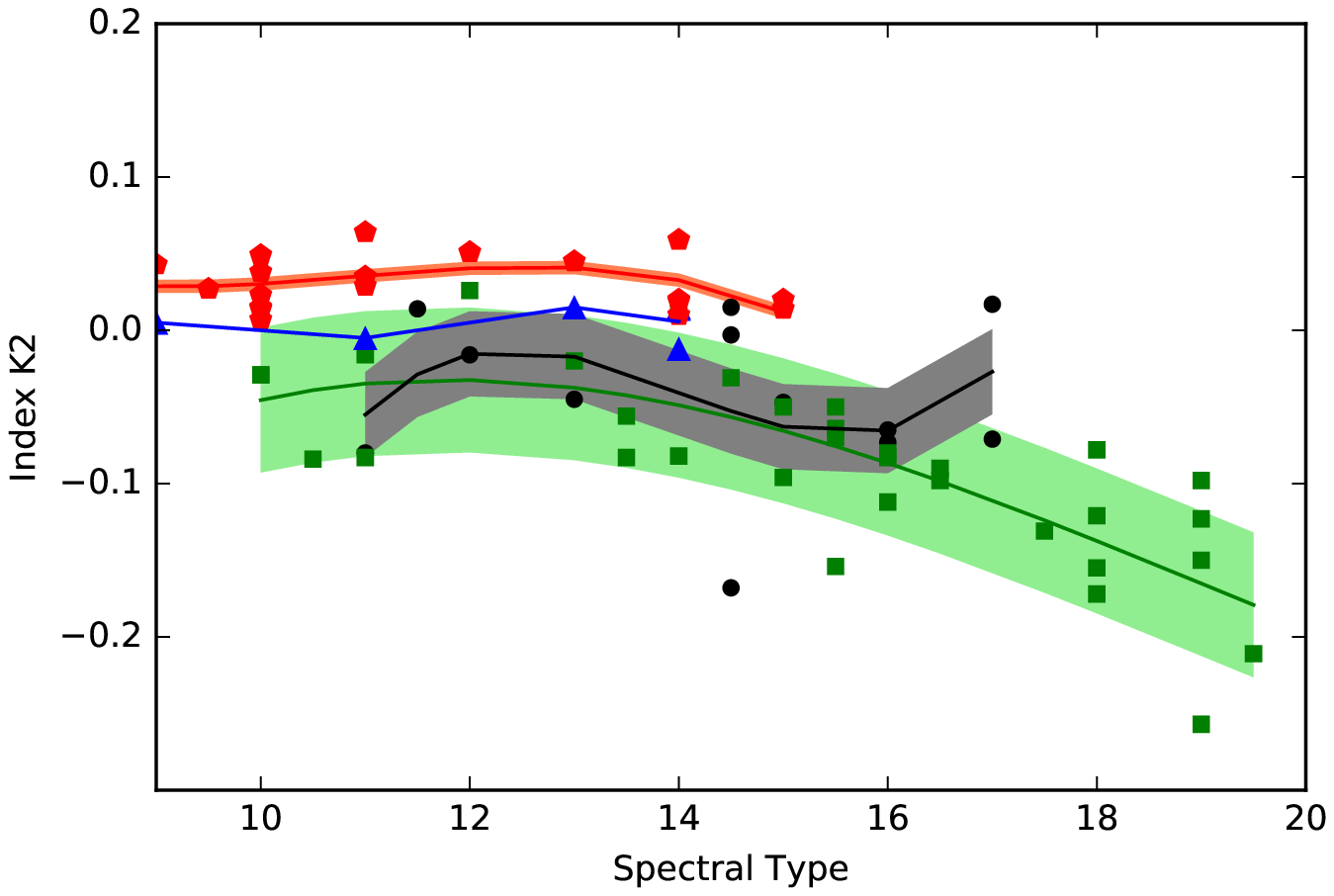}
  \caption{Fig.\ \ref{fig_USco_GTC_XSH:plot_spectral_indices_1} continue}
  \label{fig_USco_GTC_XSH:plot_spectral_indices_2}
\end{figure*}

\begin{figure*}
  \centering
  \caption{Fig.\ \ref{fig_USco_GTC_XSH:plot_spectral_indices_1} continue}
  \includegraphics[width=0.49\linewidth, angle=0]{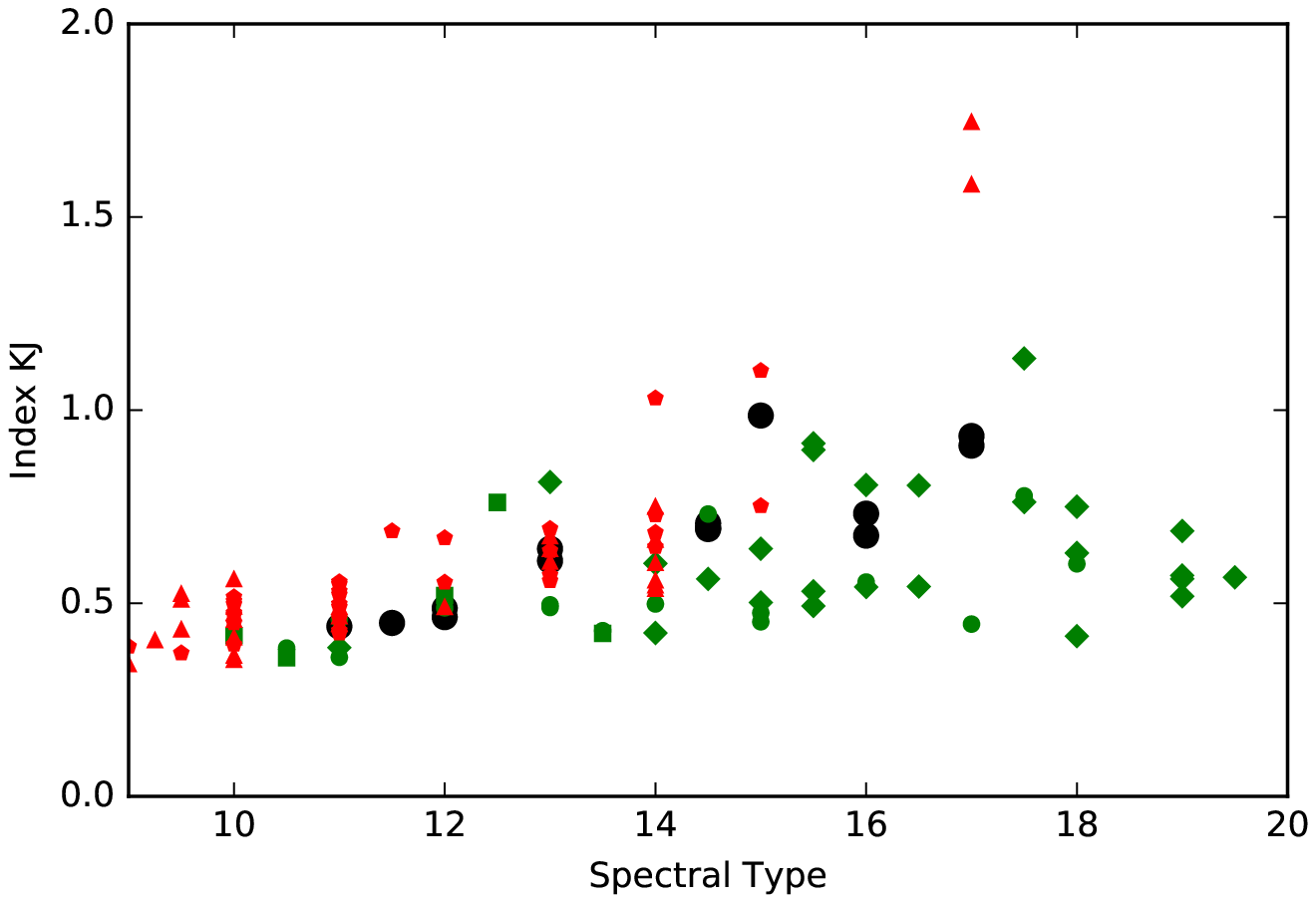}
  \includegraphics[width=0.49\linewidth, angle=0]{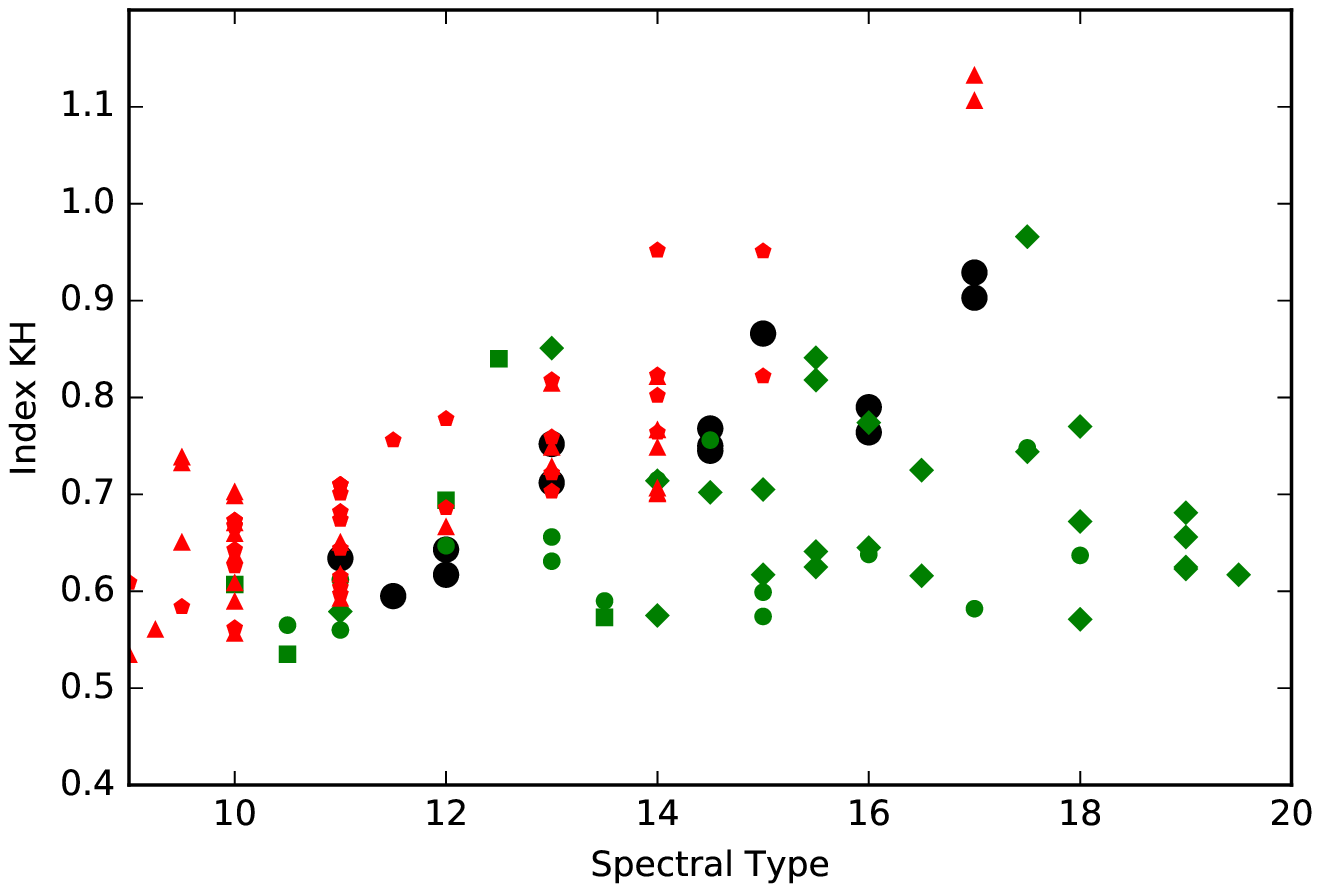}
  \includegraphics[width=0.49\linewidth, angle=0]{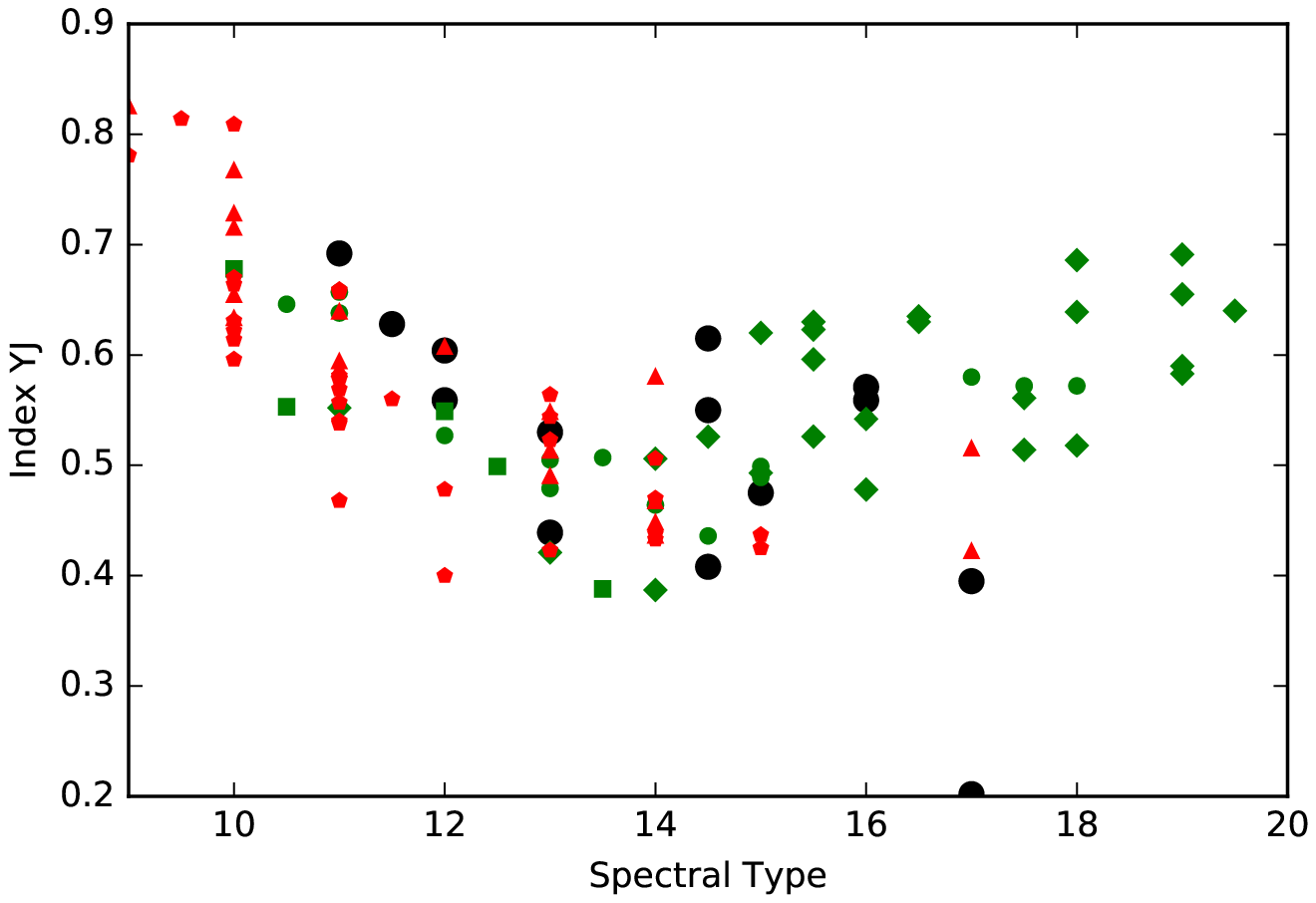}
  \includegraphics[width=0.49\linewidth, angle=0]{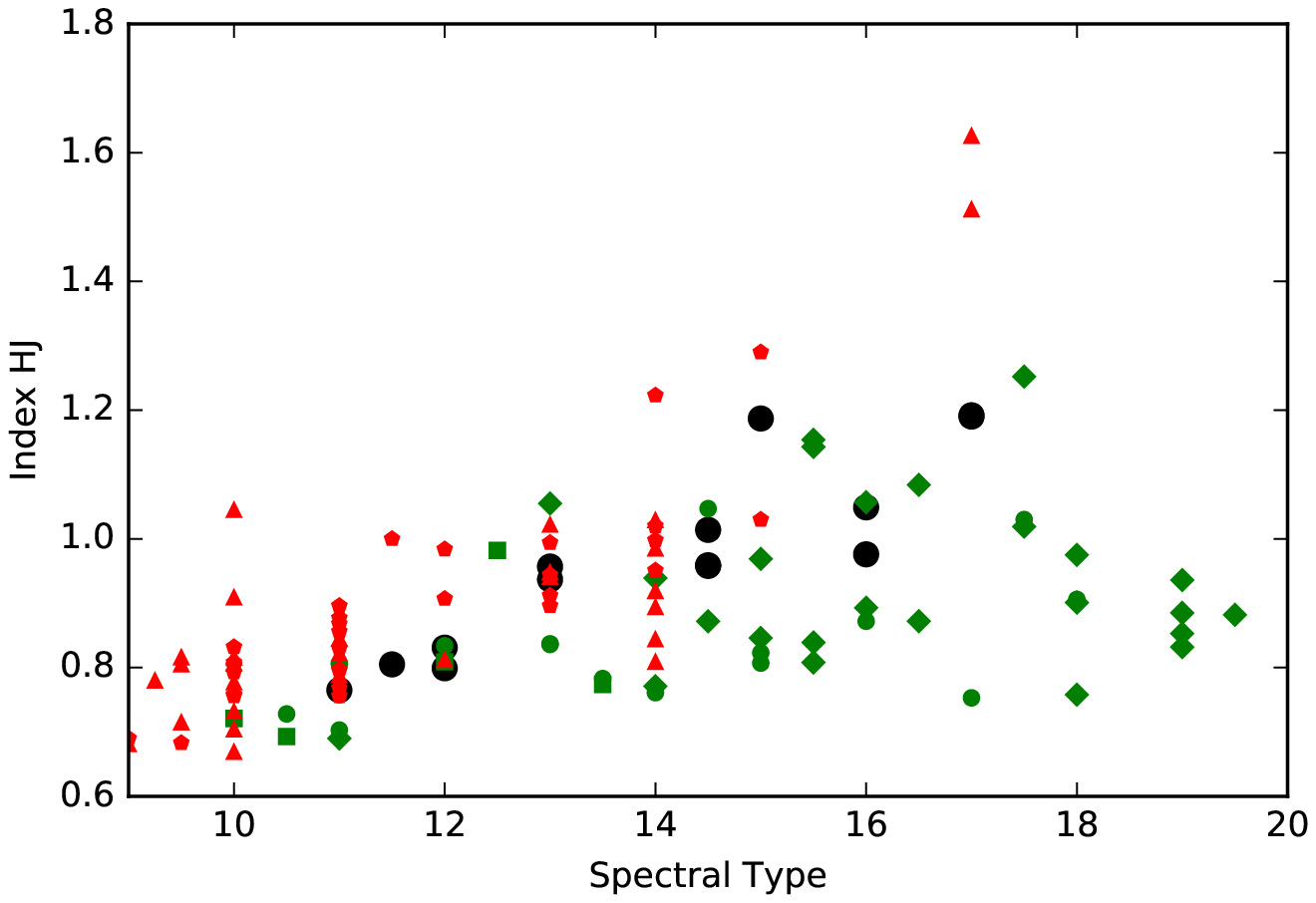}
  \includegraphics[width=0.49\linewidth, angle=0]{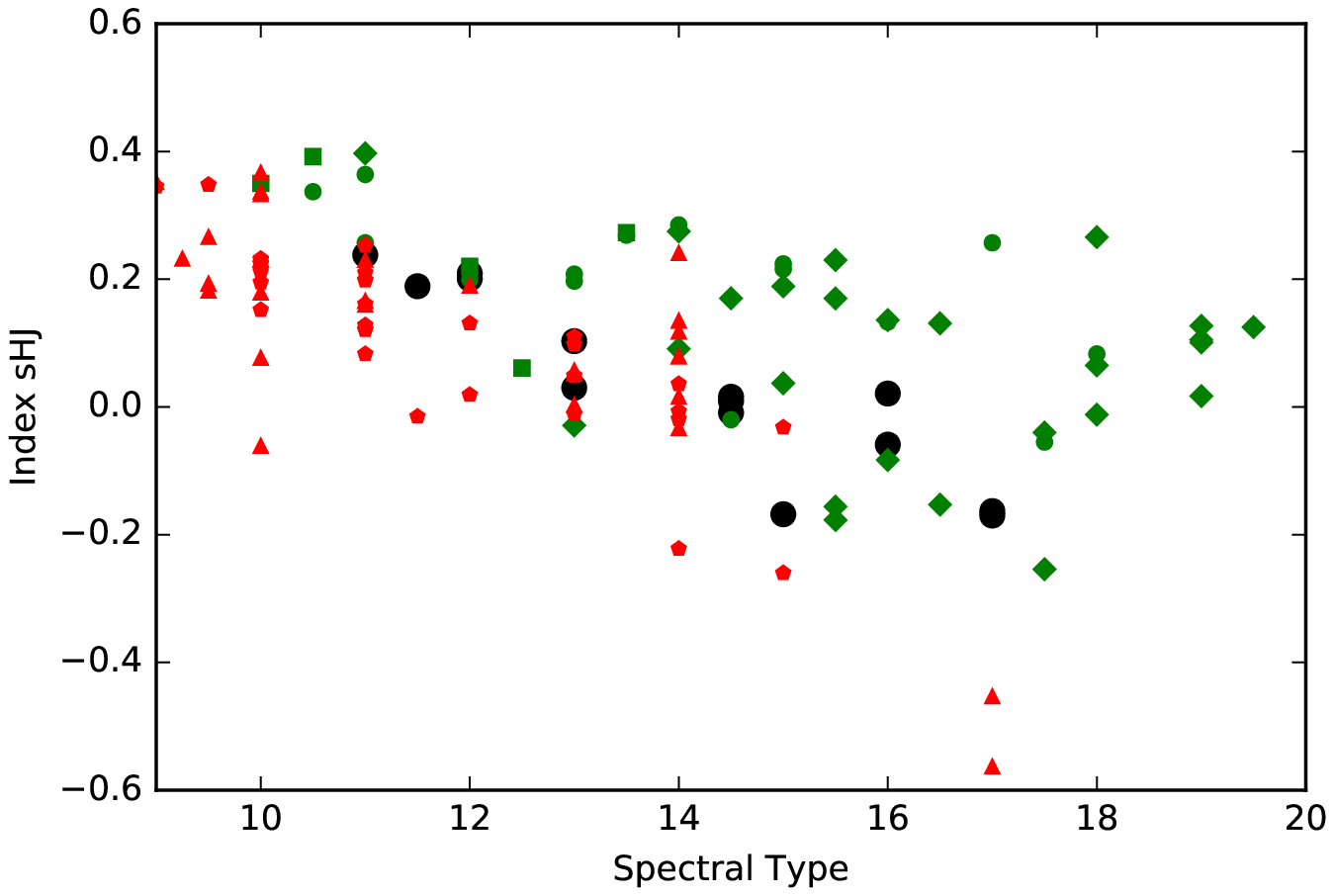}
  \includegraphics[width=0.49\linewidth, angle=0]{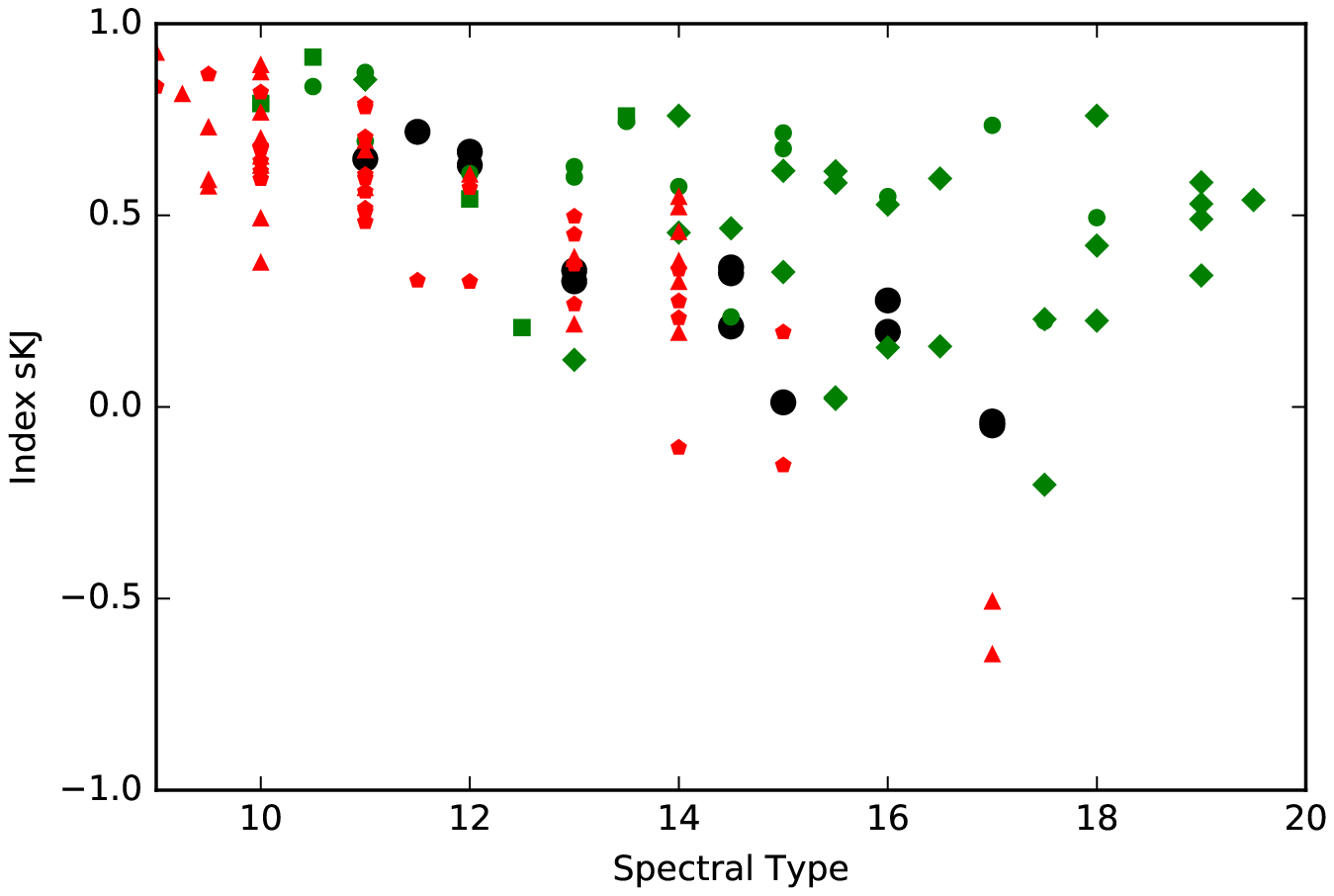}
  \includegraphics[width=0.49\linewidth, angle=0]{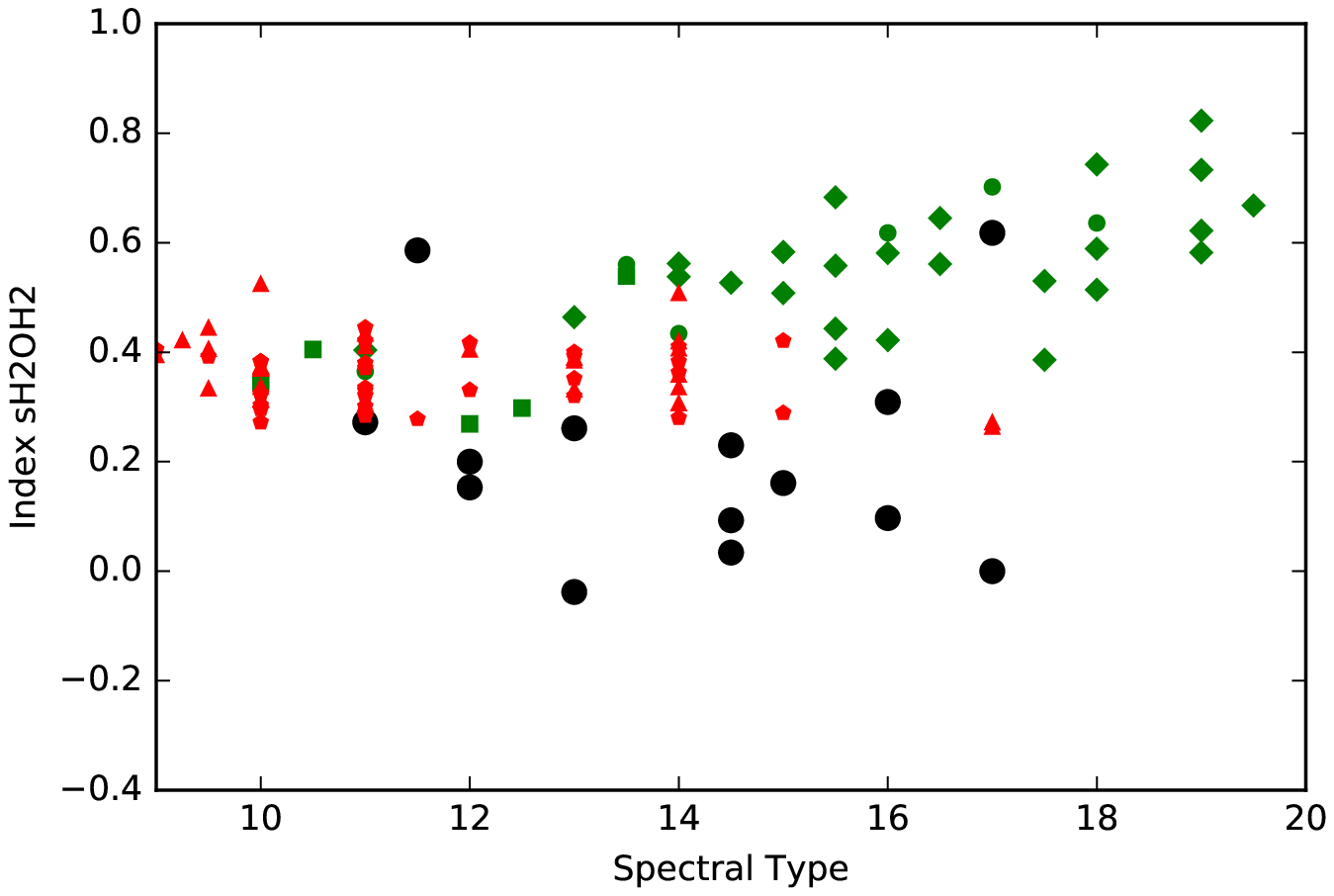}
  \includegraphics[width=0.49\linewidth, angle=0]{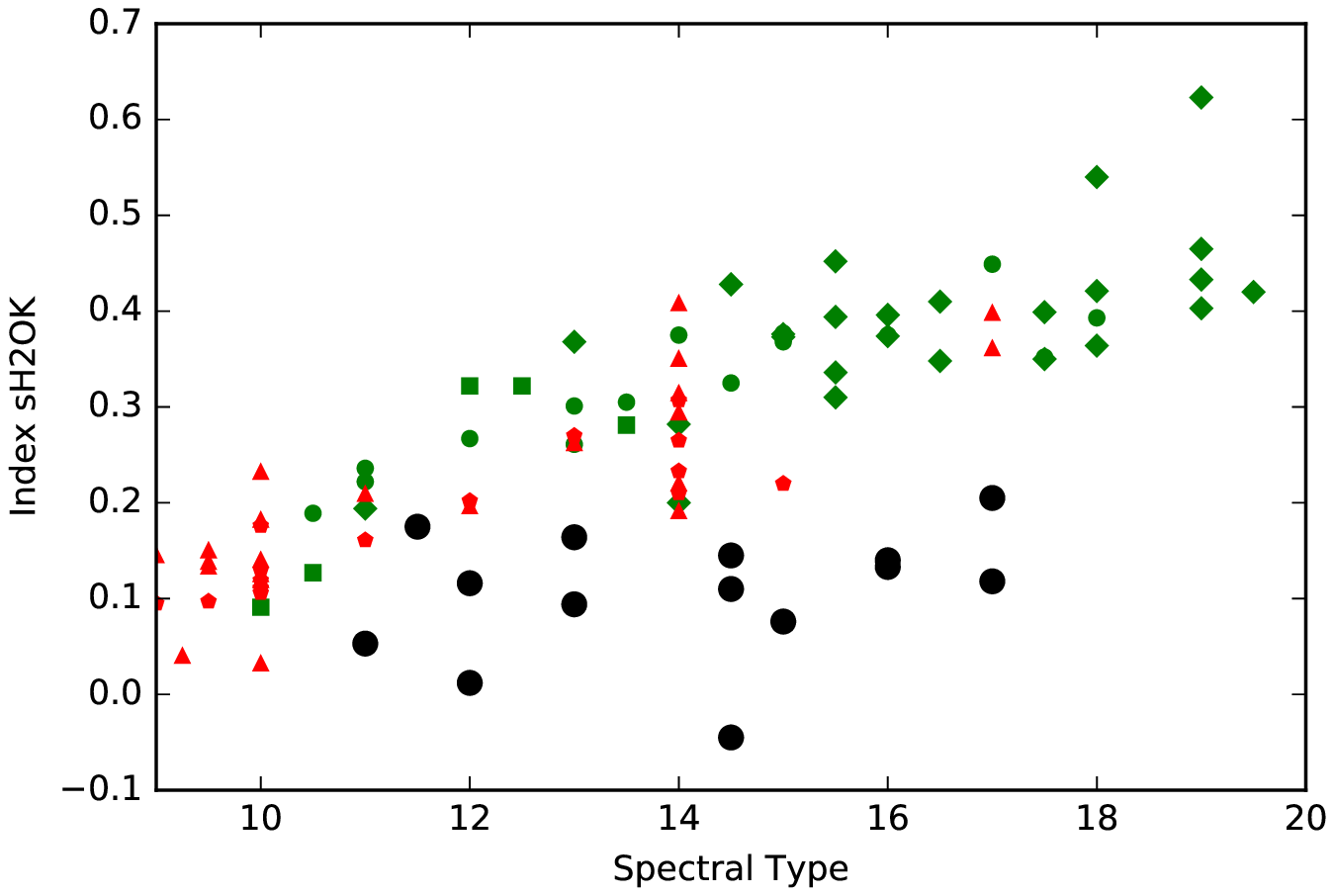}
  \label{fig_USco_GTC_XSH:plot_spectral_indices_3}
\end{figure*}

\begin{figure*}
  \centering
  \includegraphics[width=0.49\linewidth, angle=0]{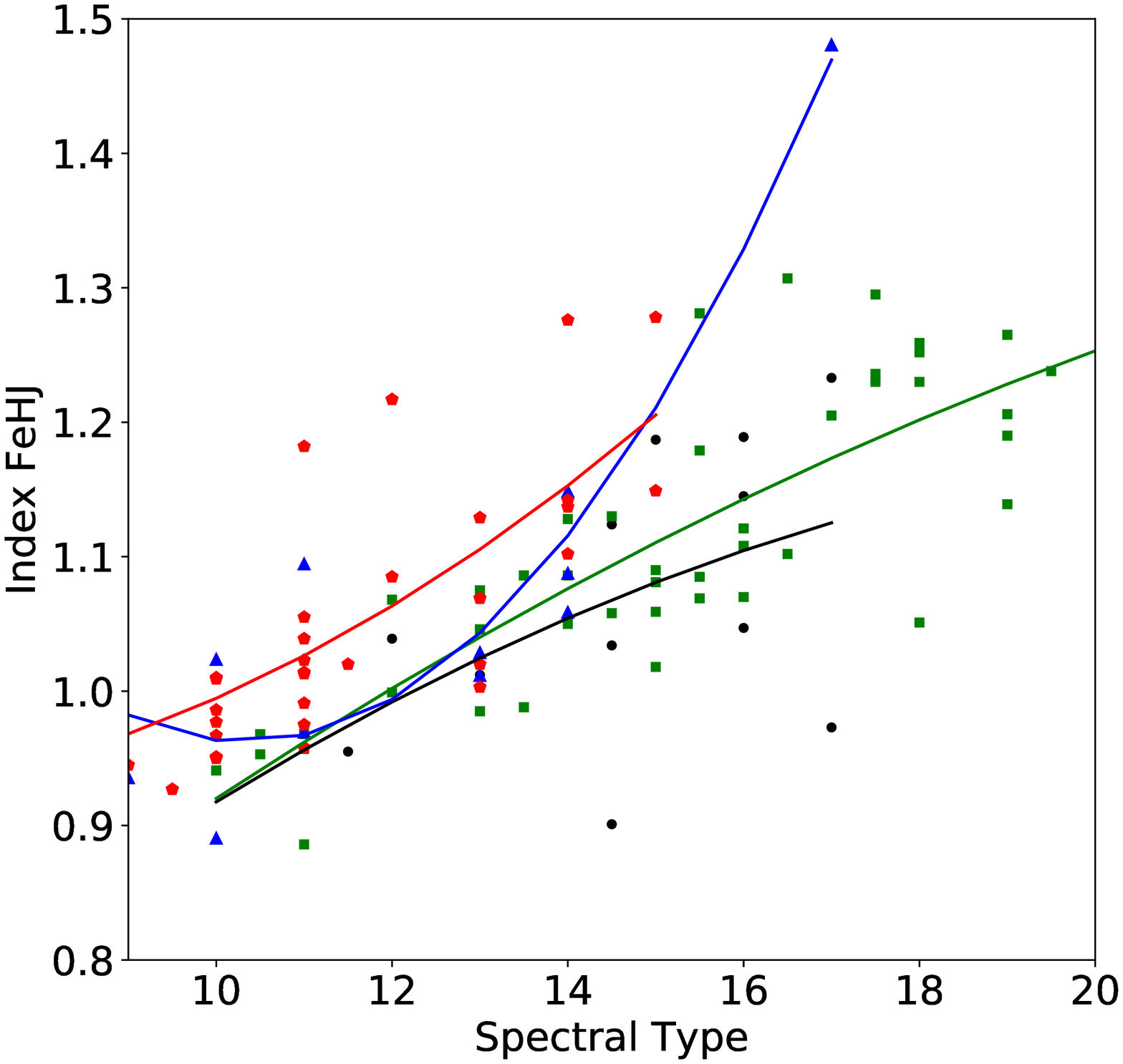}
  \includegraphics[width=0.49\linewidth, angle=0]{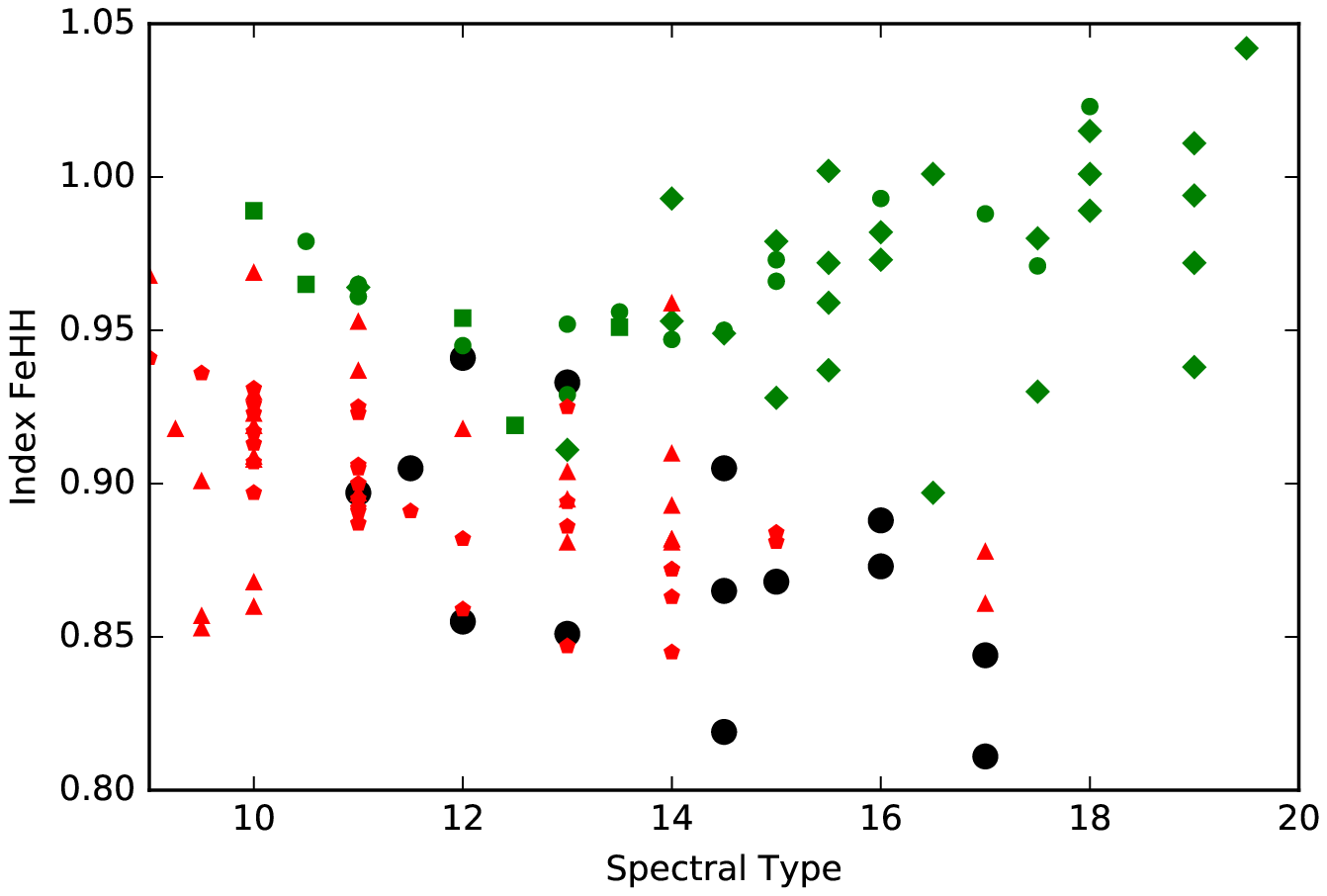}
  \includegraphics[width=0.49\linewidth, angle=0]{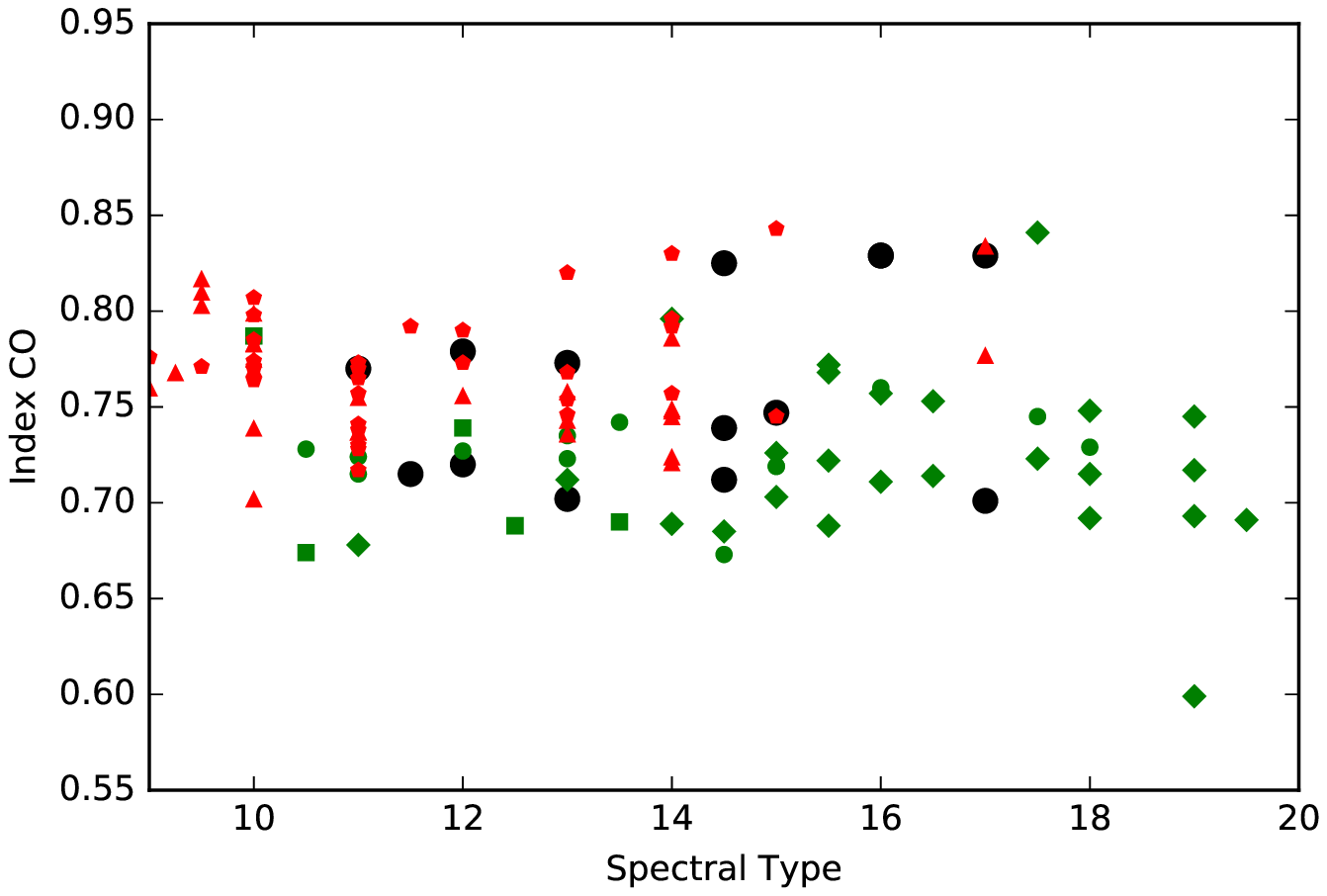}
  \includegraphics[width=0.49\linewidth, angle=0]{plot_indices_USco_XSH_VOz.ps}
  \caption{Fig.\ \ref{fig_USco_GTC_XSH:plot_spectral_indices_1} continue}
  \label{fig_USco_GTC_XSH:plot_spectral_indices_4}
\end{figure*}
%

%%%%%%%%%%%%%%%%%%%%%%%%%%%%%%%%%%%%%%%%%%%

% Don't change these lines
%\bsp    % typesetting comment
%\label{lastpage}
\end{document}